\documentclass[12pt]{article}
\usepackage{amsfonts}
\usepackage[utf8]{inputenc}
\usepackage[english]{babel}
\usepackage{bm}
\usepackage{amsmath}
\usepackage{epsfig}
\usepackage{enumitem}
\usepackage{hyperref}
\usepackage{breqn}

\setlist[enumerate]{%
	wide =0.5\parindent,
	listparindent=0pt%
}

\newcommand{\bmat}{\left(\begin{array}}
	\newcommand{\emat}{\end{array}\right)}

\def\gtrsim{\mathrel{\raise.3ex\hbox{$>$\kern-.75em\lower1ex\hbox{$\sim$}}}}

\def\-{\hphantom{-}}

\def\s2{\frac{1}{\sqrt2}}

\def\mg{m_{3/2}}
\def\mg2{m^2_{3/2}}

\def\Dsl{\,\raise.15ex\hbox{/}\mkern-13.5mu D} 

\def\be{\begin{equation}}
	\def\ee{\end{equation}}
\def\bea{\begin{eqnarray}}
	\def\eea{\end{eqnarray}}

\newcommand{\nn}{\nonumber}

\begin{document}

\pagestyle{plain}

\makeatletter
\@addtoreset{equation}{section}
\makeatother
\renewcommand{\theequation}{\thesection.\arabic{equation}}
\pagestyle{empty}
\begin{center}
\ \

\vskip 2cm

\LARGE{\LARGE\bf The Non-Relativistic Limit of \\ Hohm-Siegel-Zwiebach Theory \\[10mm]}
\vskip 0.5cm
\large{Eric Lescano 
 \\[6mm]}
{\small University of Wroclaw, Faculty of Physics and Astronomy, \\ \small\it  Maksa Borna 9, 50-204 Wroclaw,
Poland}

{\small \verb"eric.lescano@uwr.edu.pl"}\\[2cm]

\small{\bf Abstract} \\[0.5cm]\end{center}

{\small We study the non-relativistic (NR) limit of HSZ theory, a higher-derivative theory of gravity with exact and manifest T-duality invariance. Since the theory can be formulated using the generalized metric formalism, the HSZ Lagrangian remains convergent to all orders in derivatives when taking the NR limit. In this work, we analyze the three-derivative corrections to the symmetry transformations of the fields in the NR case, as well as the terms in the four-derivative action depending on the b-field. Interestingly, the corrections to the metric degrees of freedom cannot be fully trivialized, as in the relativistic case, in order to preserve the convergence of the theory. As HSZ theory interpolates order by order between heterotic and bosonic string theories, the results of this work can be interpreted as a truncation of the four-derivative structure of heterotic supergravity in the NR limit.}

\newpage
\setcounter{page}{1}
\pagestyle{plain}
\renewcommand{\thefootnote}{\arabic{footnote}}
\setcounter{footnote}{0}

\tableofcontents
\newpage
\section{Introduction} \label{SEC:Intro}

Understanding how symmetries constrain higher-derivative interactions has emerged as a central question in high-energy physics. The presence of these contributions in string theory provides a powerful construction where quantum corrections are encoded in higher-derivative terms without violations of unitarity and causality \cite{ME}. For the four- and six-derivative string corrections, T-duality can be promoted to a manifest symmetry, before compactification and for general backgrounds, in the framework of Double Field Theory (DFT) \cite{Siegel:1993xq}–\cite{Hull:2009mi} (see \cite{reviews} for reviews). While recent results show that the manifest form of this symmetry faces obstructions beyond the six-derivative case \cite{Wulff}, at the four-derivative level a two-parameter family of consistent deformations of DFT can be constructed by modifying both the action and the gauge symmetries \cite{TwoParameters}–\cite{Marques:2015vua}. In this sense, the universal structure of four-derivative corrections in DFT embeds those of the bosonic and heterotic string theories \cite{Metsaev:1987zx}, as well as other T-duality invariant theories which are not directly connected to string theory.

Some time ago, Hohm, Siegel, and Zwiebach introduced a higher-derivative theory of gravity that is both exactly and manifestly T-duality invariant \cite{Hohm:2013jaa}, now known as HSZ theory\footnote{See \cite{ReviewHSZ} for a pedagogical introduction.}. HSZ theory is not a string theory, and its spectrum includes a metric, an antisymmetric tensor, and a dilaton; the full theory also contains two massive spin-two ghosts and massive scalars \cite{Hohm:2016lim}. The leading two-derivative theory coincides with ordinary DFT, so HSZ theory represents a non-trivial higher-derivative deformation. HSZ theory is particularly attractive for two main reasons: first, the full Lagrangian is known and finite \cite{Hohm:2013jaa}. After parametrization, the four-derivative structure of the theory coincides with the $Z_2$ odd terms of heterotic supergravity, while the six-derivative structure coincides with the $Z_2$ even correction in bosonic supergravity. Therefore HSZ interpolates between these two formulations; second, HSZ can be expressed in the double geometry in the generalized metric formalism to all orders in derivatives: the double metric \cite{Hohm:2015mka} is identified with the generalized metric plus a specific higher-derivative combination of the generalized metric and dilaton upon integration of the massive fields. This latter property is especially valuable for exploring the non-relativistic (NR) limit of the theory. 

The NR limit of DFT \cite{EandD} consists of a T-duality invariant rewriting of the metric, the B-field and the dilaton under the following expansion
\bea
\label{expansiong}
\hat g_{\mu \nu} & = & h_{\mu \nu} + c^2 \tau_{\mu \nu} \, , \\
\label{expansionb}
\hat B_{\mu \nu} & = & b_{\mu \nu} + c^2 c_{\mu \nu} , \\
\label{expansionphi}
\hat{\Phi} & = & \ln(c) + \varphi \, ,
\eea
where $c$ is an arbitrary parameter that allows us to take the non-relativistic limit in the case $c\rightarrow \infty$ \footnote{For reviews, see \cite{ReviewNR} and references therein.}. Under this expansion the metric tensor has a longitudinal (heavy) and a transversal part (light), while the B-field contains a particular expansion to compensate the two-derivative divergences in the effective supergravity action. 
One long-term goal of the string theory community is to understand the NR limit of the four-derivative structure of bosonic string ($a=b=-{\alpha'}$) and heterotic string ($a=-\alpha'$, $b=0$) according to the action
\bea
S = \int d^Dx \sqrt{-\hat g} e^{-2 \hat \phi} (\hat R + 4 (\partial \hat \phi)^2 - \frac{1}{12} \hat H^2 + L^{(1)}) \, ,
\label{fullA}
\eea
where
\bea
L^{(1)} & = & - \frac{a+b}{8} \Big[  \hat R_{\mu \nu \rho \sigma} \hat R^{\mu \nu \rho \sigma} - \frac12 \hat H^{\mu \nu \rho} \hat H_{\mu \sigma \lambda} \hat R_{\nu \rho}{}^{\sigma \lambda} + \frac{1}{24} H^4 - \frac18 H^2_{\mu \nu} H^{2 \mu \nu} \Big] \nn \\ && + \frac{a-b}{4} H^{\mu \nu \rho} C_{\mu \nu \rho} \, . 
\label{MT}
\eea
Interestingly, the expansions (\ref{expansiong}-\ref{expansionphi}) lead to divergences and, therefore, the NR limit requires new mechanisms to cancel these divergences. In (\ref{MT}) we have introduced the following notation,
\bea
\label{CSdef}
C_{\mu \nu \rho} & = & \hat w_{[\mu}{}^{ab} \partial_{\nu} \hat w_{\rho]ab} + \frac23 \hat w_{[\mu}{}^{ab} \hat w_{\nu b}{}^{c} \hat w_{\rho] ca} \, , \\
H^2_{\mu \nu} & = & \hat H_{\mu}{}^{\rho \sigma} \hat H_{\nu \rho \sigma} \, , \\
H^2 & = & \hat H_{\mu \nu \rho} \hat H^{\mu \nu \rho} \, .
\eea
In this work we will show that the combination $a=-b=-\alpha'$ (HSZ theory) has a finite and convergent Lagrangian once we impose field redefinitions at the supergravity level.  

In \cite{EandD}, it was shown that there exists a tension between the metric formalism and the frame formalism when constructing NR theories, a tension that also appears in the relativistic case, as discussed in \cite{tension}. 
Remarkably, both the generalized metric and the generalized dilaton have convergent NR expansions \cite{EandD}, and thus HSZ theory remains finite to all orders in derivatives. Upon taking the NR limit (and considering the parametrization of the degrees of freedom and the solution to the strong constraints), the study of HSZ provides a useful insight into the long-term goal of computing higher-derivative corrections to bosonic and heterotic supergravity in the NR limit.

This paper is organized as follows: In section \ref{Sec:HSZ} we briefly review the construction of HSZ theory. In section \ref{Sec:NRlimit} we study the NR limit of the theory, imposing a fully metric ansatz for the components of the generalized metric. Here we also analyze the symmetry transformations for the degrees of freedom in the NR case, and finally we present the explicit action for the $b_{\mu \nu}$ contributions. In section \ref{Sec:Redef} we discuss the role of the field redefinitions at the supergravity level, with focus on the idea that the redefinitions should be performed before imposing the NR limit. This point is crucial to describe the supergravity limit of HSZ, and opens the possibility of constructing higher-derivative corrections in heterotic string theory, as discussed in section \ref{Sec:Towards}. Finally, we conclude the work in section \ref{Sec:Conclusions}.

\section{Basics of HSZ Theory} \label{Sec:HSZ}

HSZ theory is defined on a double space equipped with an $O(D,D,\mathbb{R})$ invariant metric $\eta_{M N}$, which raises and lowers the duality indices $M,N,=1,\dots,2D$. All the fields and gauge parameters are restricted by the strong constraint,
\be
\partial_M \dots \partial^M \dots = \partial_M \partial^M \dots = 0 \ . \label{strongconstraint}
\ee

The fundamental fields in HSZ theory are the double metric ${\cal M}_{M N}$ and the generalized dilaton $d$. The double metric is symmetric, but otherwise unconstrained. Under infinitesimal generalized diffeomorphisms parameterized by a generalized vector $\xi^M$, the generalized dilaton transforms as usual
\be
\delta e^{-2d} = \partial_P \left( \xi^P e^{-2d}\right) \label{gaugetransfdilaton}
\ee
but the double metric receives higher-derivative corrections
\be
\delta {\cal M}_{M N} = \widehat {\cal L}_\xi \, {\cal M}_{M N} + {\cal J}^{(1)}({\cal M})_{M N}  + {\cal J}^{(2)} ({\cal M})_{M N} \ , \label{gaugetransfdoublemet}
\ee
where $K_{M N} = 2 \partial_{[M} \xi_{N]}$ and 
\bea
{\cal J}^{(1)}({\cal M})_{M N} &=& - \frac 1 2 \partial_M {\cal M}^{P Q} \partial_P K_{Q N} - \partial_P {\cal M}_{Q M} \partial_N K^{Q P} + (M \leftrightarrow N) \\
{\cal J}^{(2)} ({\cal M})_{M N} &=& - \frac 1 4  \partial_{M} \partial_{K}  {\cal M}^{P Q} \partial_{N P} K_Q{}^K + (M \leftrightarrow N) \ .
\eea

Imposing the strong constraint, the gauge transformations closes
\be
\left[\delta_{\xi_1}, \, \delta_{\xi_2}\right] = \delta_{[\xi_1 ,\, \xi_2]_{\rm C'}} \ ,
\ee
with respect to a deformed C'-bracket
\be
[\xi_1 ,\, \xi_2]_{\rm C'}^M = 2 \xi_{[1}^P \partial_P \xi_{2]}^M - \xi_{[1}^P \partial^M \xi_{2]P} + \partial_P \xi^Q_{[1} \partial^M \partial_Q\xi_{2]}^P \ .
\ee

Interestingly, the gauge transformations (\ref{gaugetransfdoublemet}) do not satisfy the Leibnitz rule and therefore one defines a star-product $\star$ in the following way,
\bea
(T\ \star\ T)_{M N} &=& {T}_{M P} {T}_{N}\,^{P} - \frac{1}{4}\, {\partial}_{M}{{T}_{P Q}}\,  {\partial}_{N}{{T}^{P Q}}\,  + 2\, {\partial}_{N}{{T}_{P}\,^{Q}}\,  {\partial}_{Q}{{T}_{M}\,^{P}}\,  - {\partial}_{P}{{T}_{M}\,^{Q}}\,  {\partial}_{Q}{{T}_{N}\,^{P}}\, \nn\\
 && +\,  \frac{1}{2}\, {\partial}_{M P}{{T}_{Q}\,^{I}}\,  {\partial}_{N I}{{T}^{Q P}}\,  - {\partial}_{N P}{{T}^{Q I}}\,  {\partial}_{Q I}{{T}_{M}\,^{P}}\,  - \frac{1}{8}\, {\partial}_{M P Q}{{T}^{I J}}\,  {\partial}_{N I J}{{T}^{P Q}}\, \nn \\
  &&  +\, \frac{1}{2}\, {T}^{P Q} {\partial}_{P Q}{{T}_{M N}}\,  + \frac{1}{2}\, {\partial}_{M}{{T}_{P}\,^{Q}}\,  {\partial}_{N Q}{{G}^{P}}\,  - \frac{1}{2}\, {\partial}_{N}{{T}^{P Q}}\,  {\partial}_{P Q}{{G}_{M}}\, \nn \\ && -\, \frac{1}{4}\, {\partial}_{M P}{{T}^{Q I}}\,  {\partial}_{N Q I}{{G}^{P}}\,  + \frac{1}{2}\, {G}^{P} {\partial}_{P}{{T}_{M N}}\, -\, {\partial}_{P}{{T}_{M}\,^{Q}}\,  {\partial}_{N Q}{{G}^{P}}\,\nn \\ &&+\, {T}_{M P} {\partial}_{N}{{G}^{P}}\,  - {T}_{M}\,^{P} {\partial}_{P}{{G}_{N}}\, + (M \leftrightarrow N) \vphantom{\frac 1 2} \ ,
\eea
where we used the notation $\partial_{MN\dots}=\partial_{M} \partial_{N}\dots$ \footnote{We use this notation in several parts of this paper. Also $\partial_{\mu\nu}=\partial_{\mu} \partial_{\nu}$ for space-time indices.} and
\bea
G^M(T,d) &=& {\partial}_{N}{{T}^{M N}}\,  - 2\, {T}^{M N} {\partial}_{N}{d}\,  +\, T^{N P} \partial_{N P}{}^M d\label{G} \\
 &&- \frac 1 2 \partial^M \left[\partial_{N P} T^{N P} -4\, \partial_{N} T^{N P} \partial_P d - 2\, T^{N P} \left(\partial_{N P} d - 2 \partial_N d\, \partial_P d\right) \right] \ .\nn
\eea
The star product does satisfy the Leibnitz rule with respect to the gauge transformations,
\be
\delta (T_1 \star T_2) = \delta T_1 \star T_2 + T_1 \star \delta T_2 \ ,
\ee
where $(T_1 \star T_2)_{M N}$, $T_{1 M N}$ and $T_{2 M N}$ transform in the same way as the double metric in (\ref{gaugetransfdoublemet}).

The standard invariant metric contractions also turns out not to be covariant under the deformed gauge transformations. However, the following quantity
\bea
\langle T_1 | T_2 \rangle &=& \frac{1}{2}\, {T_1}_{P Q} {T_2}^{P Q} - {\partial}_{P}{{T_1}_{N}\,^{Q}}\,  {\partial}_{Q}{{T_2}^{N P}}\,  + \frac{1}{4}\, {\partial}_{M N}{{T_1}^{P Q}}\,  {\partial}_{P Q}{{T_2}^{M N}}\,  \\ && - \frac{3}{2}\, {G(T_1)}_{N} {G(T_2)}^{N} + \frac{3}{2}\, {\partial}_{M}{{G(T_1)}^{N}}\,  {\partial}_{N}{{G(T_2)}^{M}}\,  - \frac{3}{2}\, {T_2}_{P}\,^{N} {\partial}_{N}{{G(T_1)}^{P}}\, \nn \\ && - \frac{3}{2}\, {T_1}_{P}\,^{N} {\partial}_{N}{{G(T_2)}^{P}}\, + \frac{3}{4}\, {\partial}_{M}{{T_2}^{N P}}\,  {\partial}_{N P}{{G(T_1)}^{M}}\,  + \frac{3}{4}\, {\partial}_{M}{{T_1}^{N P}}\,  {\partial}_{N P}{{G(T_2)}^{M}}\ , \nn
\eea
transforms as a scalar under the deformed gauge transformations due to the strong constraint
\be
\delta \langle T_1 | T_2 \rangle = \langle \delta T_1 | T_2 \rangle + \langle T_1 | \delta T_2 \rangle = \xi^P \partial_P \langle T_1 | T_2 \rangle \ .
\ee

Using the previous elements now we can write the action of HSZ theory, which is given by
\be
S = \int d^{2D}X \, e^{-2d}\, L \ , \ \ \ \ \ L = \bigg\langle {\cal M}\, \bigg| \, \eta - {\frac 1 6} {\cal M} \star {\cal M} \bigg\rangle \ .\label{ActionDoubleMetric}
 \ee
and it is gauge invariant by construction. Since both the star and inner products are linear in each argument, the action is cubic in powers of the double metric. In order to write the theory in terms of the DFT degrees of freedom one considers the following decomposition of the double metric \cite{Hohm:2015mka}
\be
{\cal M}_{M N} = {\cal H}_{M N} + {\cal F}_{M N} \ ,\label{DecompMHF}
\ee
where ${\cal H}_{M N}$ is the generalized metric of DFT, and ${\cal F}_{M N}$ is a higher order contribution. In \cite{Hohm:2016lim,Hohm:2015mka}, this field was shown to correspond to dynamical massive excitations. However, one can integrate them out ending with an effective action for the massless fields. The result of such a procedure is a functional dependence in terms of the generalized metric and dilaton. For the first order the effective ${\cal F}_{M N}$ field is given by,
\bea
{\cal F}_{MN} &=& \frac{1}{4}\, {\cal H}_{M P}\, {\partial}_{Q}{{\cal H}_{N}\,^{I}}\,  {\partial}_{I}{{\cal H}^{P Q}}\,  - \frac{1}{4}\, {\cal H}_{M}\,^{P} {\partial}_{Q}{\cal H}_{N I}\,  {\partial}_{P}{{\cal H}^{Q I}}\,  - \frac{1}{4}\, {\cal H}_{M}\,^{P} {\partial}_{N}{{\cal H}^{Q I}}\,  {\partial}_{Q}{{\cal H}_{P I}}\, \nn \\ && + \frac{1}{16}\, {\cal H}_{M}\,^{P} {\partial}_{N}{{\cal H}^{Q I}}\,  {\partial}_{P}{{\cal H}_{Q I}}\,  + \frac{1}{8}\, {\cal H}^{P Q} {\partial}_{P}{{\cal H}_{M}\,^{I}}\,  {\partial}_{Q}{{\cal H}_{N I}} + (M \leftrightarrow N) \ . \label{feq}
\eea

Since we would like to explore the NR limit of the four-derivative terms of the HSZ Lagrangian, we will need the following terms of the HSZ Lagrangian,
\bea
L_{\textrm{HSZ}} & = &  L_{\textrm{DFT}} - \frac12 \partial^{N Q}{\mathcal{F}_{N Q}} + 2 \partial^{N}{\mathcal{F}_{N}{}^{Q}} \partial_{Q}{d} - 2 \mathcal{F}^{P Q} \partial_{P}{d} \partial_{Q}{d} - \frac12 \mathcal{F}^{P I} \mathcal{F}_{P}{}^{J} \mathcal{H}_{I J} \nn \\ && + \mathcal{F}^{P I} \mathcal{H}_{P}{}^{J} \partial_{J}{}^{L}{\mathcal{H}_{I L}} - \frac{1}{12} \mathcal{F}^{I J} \mathcal{H}^{K L} \partial_{K L}{\mathcal{H}_{I J}} + \frac12 \mathcal{F}^{P Q} \partial^{J}{\mathcal{H}_{P J}} \partial^{L}{\mathcal{H}_{Q L}}  \nn \\ && - \frac{1}{12} \mathcal{F}^{P Q} \partial^{J}{\mathcal{H}_{P Q}} \partial^{L}{\mathcal{H}_{J L}} - \frac12 \mathcal{F}^{P Q} \partial_{P}{\mathcal{H}^{J L}} \partial_{J}{\mathcal{H}_{Q L}} + \frac{1}{8} \mathcal{F}^{P Q} \partial_{P}{\mathcal{H}^{J L}} \partial_{Q}{\mathcal{H}_{J L}} \nn \\ && - \frac{1}{12} \mathcal{H}^{I J} \mathcal{H}^{K L} \partial_{I J}{\mathcal{F}_{K L}} + \frac12 \mathcal{H}^{I J} \mathcal{H}^{K L} \partial_{I K}{\mathcal{F}_{J L}} + \frac12 \mathcal{H}^{P Q} \partial^{J}{\mathcal{F}_{P}{}^{L}} \partial_{Q}{\mathcal{H}_{J L}} \nn \\ && - \frac{1}{12} \mathcal{H}^{P Q} \partial^{J}{\mathcal{F}_{P Q}} \partial^{L}{\mathcal{H}_{J L}} - \frac12 \mathcal{H}^{P Q} \partial_{P}{\mathcal{F}^{J L}} \partial_{J}{\mathcal{H}_{Q L}} + \frac{1}{12} \mathcal{H}^{P Q} \partial_{P}{\mathcal{F}^{J L}} \partial_{Q}{\mathcal{H}_{J L}} \nn \\ && + \mathcal{H}^{P Q} \partial_{P}{\mathcal{F}_{Q}{}^{J}} \partial^{L}{\mathcal{H}_{J L}} - 2 \mathcal{F}^{P I} \mathcal{H}_{P}{}^{K} \mathcal{H}_{I}{}^{L} \partial_{K L}{d} - 2 \mathcal{F}^{P I} \mathcal{H}_{P}{}^{J} \partial_{J}{\mathcal{H}_{I}{}^{L}} \partial_{L}{d} \nn \\ && - 2 \mathcal{F}^{P I} \mathcal{H}_{P}{}^{J} \partial^{L}{\mathcal{H}_{I L}} \partial_{J}{d} + \frac{1}{6} \mathcal{F}^{I J} \mathcal{H}^{K L} \partial_{K}{\mathcal{H}_{I J}} \partial_{L}{d} - 2 \mathcal{H}^{I J} \mathcal{H}^{K L} \partial_{I}{\mathcal{F}_{J K}} \partial_{L}{d} \nn \\ && + \frac{1}{6} \mathcal{H}^{I J} \mathcal{H}^{K L} \partial_{I}{\mathcal{F}_{K L}} \partial_{J}{d} + 2 \mathcal{F}^{P I} \mathcal{H}_{P}{}^{K} \mathcal{H}_{I}{}^{L} \partial_{K}{d} \partial_{L}{d} + \frac12 \mathcal{F}^{P Q} \partial^{J}{\mathcal{H}_{P}{}^{L}} \partial_{L}{\mathcal{H}_{Q J}} \nn \\ && + 2 \mathcal{F}^{P Q} \partial_{P Q}{d} + \mathcal{O}(\partial^6) \, .
\label{LagHSZ}
\eea
Since ${\cal F}_{M N}$ can be written in terms of ${\cal H}_{M N}$, then HSZ theory can be entirely written in terms of the generalized metric and the generalized dilaton (see Appendix \ref{AppA} for the explicit form of $L^{(4)}$). This is true to all order in derivatives, and therefore HSZ converges upon taking a NR limit to all orders. At the level of the double geometry, this limit consist in taking a particular non-Riemannian solution \cite{NRDFT1}-\cite{NRDFT9} for the generalized metric and dilaton. In this work we will focus on the first non-trivial correction to the action (four-derivative terms) and symmetry transformation (three-derivative terms). 

\section{The non-relativistic limit for HSZ}
\label{Sec:NRlimit}

\subsection{Non-relativistic decomposition in metric formalism}
We start by considering the NR expansion for the generalized metric constructed in \cite{EandD} for a pure metric ansatz,
\bea
{\cal H}_{M N} = {\cal H}^{(0)}_{M N} + {\cal H}^{(-2)}_{M N} \, ,
\eea
where
\begin{align}
    {\cal H}^{(0)}_{M N} &=  
\left(\begin{matrix} h^{\mu \nu} &  - b_{\rho \nu} h^{\rho \mu} - c_{\rho \nu} \tau^{\rho \mu} \\ 
- b_{\rho \mu} h^{\rho \nu} - c_{\rho \mu} \tau^{\rho \nu} &  h_{\mu \nu} + b_{\rho \mu} h^{\rho \sigma} b_{\sigma \nu} + 2 c_{\rho (\mu|} \tau^{\rho \sigma} b_{\sigma |\nu)}  \, \end{matrix} \right) \, ,
\end{align}
and
\bea
   {\cal H}^{(-2)}_{M N} &=  
\left(\begin{matrix} \tau^{\mu \nu} &  -b_{\rho \nu} \tau^{\rho \mu}  \\ 
-b_{\rho \mu} \tau^{\rho \nu} &  b_{\rho \mu} \tau^{\rho \sigma} b_{\sigma \nu} \end{matrix}\right)  \, .
\eea
The components of the previous metric can be thought as an expansion on the metric and B-field of the form,
\bea
\hat g_{\mu \nu} & = & h_{\mu \nu} + c^2 \tau_{\mu \nu} \, , \\
\hat g^{\mu \nu} & = & h^{\mu \nu} + \frac{1}{c^2} \tau^{\mu \nu} \, , \\
\hat B_{\mu \nu} & = & b_{\mu \nu} + c^2 c_{\mu \nu} ,
\eea
where the Newton-Cartan relations in terms of the previous quantities are given by 
\bea
h^{\mu \nu} \tau_{\nu \rho} & = & h^{\mu \nu} c_{\nu \rho} = 0 \, , \\
h_{\mu \nu} \tau^{\nu \rho} & = & 0 \, , \\ 
h_{\mu \nu} h^{\nu \rho} & = & \delta_{\mu}^{\rho} - \tau_{\mu \nu} \tau^{\nu \rho} \, , \label{delta}
\eea
and the antisymmetric $c_{\mu \nu}$ also obeys the following constraint
\bea
c_{\mu \rho} \tau^{\rho \sigma} c_{\sigma \nu} = \tau_{\mu \nu} \, .
\eea
This last constraint is the minimal requirement to have a finite NR expansion in the generalized metric. Moreover, this states that $c_{\mu \nu}$ is gauge invariant under B-shifts. In this work we are using a $c_{\mu \nu}$ field for the divergent part of the B-field expansion since we are interested in a full metric formulation of both supergravity and double geometry. We will discuss about this point in below equation (\ref{csugra}).

On the other hand, the generalized dilaton is expanded considering $\hat{\Phi} = \ln(c) + \varphi$, and therefore it satisfies
\bea
e^{-2 \hat{d}} = e^{-2\hat{\Phi}} \sqrt{-\hat g} = e^{-2 \varphi} \sqrt{f(\tau,h)} = e^{-2 d},
\label{dilaton}
\eea 
where $f(\tau,h) = - \frac{\hat{g}}{c^4}$.

The symmetry transformations of the fundamental fields are
\bea
\label{transfzeroa}
\delta h^{\mu \nu} & = & L_{\xi} \tau^{\mu \nu} + \Delta h^{\mu \nu}(\xi, \zeta) \, , \\
\delta h_{\mu \nu} & = & L_{\xi} \tau_{\mu \nu} - 2 \lambda_{a a'} e_{(\mu}{}^{a'} \tau_{\nu)}{}^{a} + \Delta h_{\mu \nu}(\xi, \zeta) \, , \\
\delta \tau^{\mu \nu} & = & L_{\xi} \tau^{\mu \nu} + 2 \lambda_{a}{}^{a'} e^{(\mu}{}_{a'} \tau^{\nu) a} + \Delta \tau^{\mu \nu}(\xi, \zeta) \, ,  \\
\delta \tau_{\mu \nu} & = & L_{\xi} \tau_{\mu \nu} + \Delta \tau_{\mu \nu}(\xi, \zeta) \, ,  \\
\delta b_{\mu \nu} & = & L_{\xi} b_{\mu \nu} - 2 \epsilon_{a b} \lambda^{a}{}_{a'} \tau_{[\mu}{}^{b} e_{\nu]}{}^{a'} + 2 \partial_{[\mu} \zeta_{\nu]} + \Delta b_{\mu \nu}(\xi, \zeta) \, , \\ 
\delta c_{\mu \nu} & = & L_{\xi} c_{\mu \nu} + \Delta c_{\mu \nu}(\xi, \zeta) \, ,
\label{transfzerob}
\eea
where $\lambda_{a a'}$ is a boost parameter and we have used $\Delta(\xi,\zeta)$ to express a higher-derivative correction to the leading order symmetry transformations. While these higher-order contributions do not deform the boost transformations, they might modify their form due to field redefinitions, as we show in the next section.

\subsection{Three-derivative corrections to the diffeomorphism transformations}
The first order deformation of the gauge transformation of the generalized metric is given by the expressions
\bea
\delta^{(1)}{\cal H}^{(0)}_{M N} &=&  - \frac{1}{4}\, {\partial}_{M}{{\cal H}^{(0)P Q}}\,  {\partial}_{P}{{K}_{Q N}}\,  + \frac{1}{4}\, {\cal H}^{(0)}_{M}\,^{P} {\cal H}^{(0)}_{N}\,^{Q} {\partial}_{P}{{\cal H}^{(0)I J}}\,  {\partial}_{I}{{K}_{J Q}}\, \\ && - \frac{1}{2}\, {\partial}_{P}{{\cal H}^{(0)}_{Q M}}\,  {\partial}_{N}{{K}^{Q P}}\,  + \frac{1}{2}\, {\cal H}^{(0)}_{M}\,^{P} {\cal H}^{(0)}_{N}\,^{Q} {\partial}_{I}{{\cal H}^{(0)}_{J P}}\,  {\partial}_{Q}{{K}^{J I}} + (M \leftrightarrow N) \ ,\nn
\eea
and
\bea
\delta^{(1)}{\cal H}^{(-2)}_{M N} &=&  - \frac{1}{4}\, {\partial}_{M}{{\cal H}^{(-2)P Q}}\,  {\partial}_{P}{{K}_{Q N}}\,  + \frac{1}{4}\, {\cal H}^{(-2)}_{M}\,^{P} {\cal H}^{(0)}_{N}\,^{Q} {\partial}_{P}{{\cal H}^{(0)I J}}\,  {\partial}_{I}{{K}_{J Q}}\, \nn \\ &&
+ \frac{1}{4}\, {\cal H}^{(0)}_{M}\,^{P} {\cal H}^{(-2)}_{N}\,^{Q} {\partial}_{P}{{\cal H}^{(0)I J}}\,  {\partial}_{I}{{K}_{J Q}} + \frac{1}{4}\, {\cal H}^{(0)}_{M}\,^{P} {\cal H}^{(0)}_{N}\,^{Q} {\partial}_{P}{{\cal H}^{(-2)I J}}\,  {\partial}_{I}{{K}_{J Q}}\, \nn \\ &&
- \frac{1}{2}\, {\partial}_{P}{{\cal H}^{(-2)}_{Q M}}\,  {\partial}_{N}{{K}^{Q P}}\,  + \frac{1}{2}\, {\cal H}^{(-2)}_{M}\,^{P} {\cal H}^{(0)}_{N}\,^{Q} {\partial}_{I}{{\cal H}^{(0)}_{J P}}\,  {\partial}_{Q}{{K}^{J I}} \nn \\
&& + \frac{1}{2}\, {\cal H}^{(0)}_{M}\,^{P} {\cal H}^{(-2)}_{N}\,^{Q} {\partial}_{I}{{\cal H}^{(0)}_{J P}}\,  {\partial}_{Q}{{K}^{J I}} + \frac{1}{2}\, {\cal H}^{(0)}_{M}\,^{P} {\cal H}^{(0)}_{N}\,^{Q} {\partial}_{I}{{\cal H}^{(-2)}_{J P}}\,  {\partial}_{Q}{{K}^{J I}} + (M \leftrightarrow N) \, . \nn \\
\eea
From the previous equations we can read some of the corrections to the zeroth order transformations (\ref{transfzeroa})-(\ref{transfzerob}). First, from $\delta^{(1)} {\cal H}^{(0)\mu \nu}$ we can compute
\bea
\label{hinverse}
\Delta h^{\mu \nu} &=& h^{(\mu| \rho} h^{|\nu) \sigma} \partial_{\rho} h^{\alpha \beta} \partial_{\alpha} \partial_{[\beta} \zeta_{\sigma]} - \frac12 h^{(\mu| \rho} (b_{\gamma \sigma} h^{\gamma |\nu)} + c_{\gamma \sigma} \tau^{\gamma |\nu)}) \partial_{\rho} h^{\alpha \beta} \partial_{\alpha} \partial_{\beta} \xi^{\sigma} \nn \\
&& + \frac12 h^{(\mu| \rho} h^{|\nu) \sigma} \partial_{\rho}(b_{\gamma \beta} h^{\gamma \alpha} + c_{\gamma \beta} \tau^{\gamma \alpha}) \partial_{\alpha} \partial_{\sigma} \xi^{\beta} - h^{(\mu| \rho} h^{|\nu) \sigma} \partial_{\alpha}(b_{\gamma \rho} h^{\gamma \beta} + c_{\gamma \rho} \tau^{\gamma \beta}) \partial_{\beta} \partial_{\sigma} \xi^{\alpha} \nn \\
&& 
- (b_{\gamma \rho} h^{\gamma (\mu|} + c_{\gamma \rho} \tau^{\gamma (\mu|}) h^{|\nu) \sigma} \partial_{\alpha} h^{\beta \rho} \partial_{\sigma} \partial_{\beta} \xi^{\alpha} \, . 
\eea
Similarly, from $\delta^{(1)}{\mathcal H}^{(-2) \mu \nu}$ we can obtain the correction to the transformation of $\tau^{\mu \nu}$
\bea
\Delta \tau^{\mu \nu} &=& \tau^{(\mu| \rho} h^{|\nu) \sigma} \partial_{\rho}h^{\alpha \beta} \partial_{\alpha} \partial_{[\beta} \zeta_{\sigma]} + h^{(\mu| \rho} \tau^{|\nu) \sigma} \partial_{\rho}h^{\alpha \beta} \partial_{\alpha} \partial_{[\beta} \zeta_{\sigma]} + h^{(\mu| \rho} h^{|\nu) \sigma} \partial_{\rho} \tau^{\alpha \beta} \partial_{\alpha} \partial_{[\beta} \zeta_{\sigma]} \nn \\ && - \frac12 \tau^{(\mu| \rho} (b_{\epsilon \sigma} h^{\epsilon |\nu)} + c_{\epsilon \sigma} \tau^{\epsilon |\nu)}) \partial_{\rho}h^{\alpha \beta} \partial_{\alpha} \partial_{\beta} \xi^{\sigma} - \frac12 h^{(\mu| \rho} b_{\pi \sigma} \tau^{\pi |\nu)} \partial_{\rho} h^{\alpha \beta} \partial_{\alpha} \partial_{\beta} \xi^{\sigma} \nn \\ && - \frac12 h^{(\mu| \rho} ( b_{\epsilon \sigma} h^{\epsilon |\nu)} + c_{\epsilon \sigma} \tau^{\epsilon |\nu)}) \partial_{\rho} \tau^{\alpha \beta} \partial_{\alpha} \partial_{\beta} \xi^{\sigma} + \frac12 h^{(\mu| \rho} h^{|\nu) \sigma} \partial_{\rho}(b_{\pi \beta} \tau^{\pi \alpha})  \partial_{\alpha} \partial_{\sigma} \xi^{\beta} \nn \\ && + \frac12 \tau^{(\mu| \rho} h^{|\nu) \sigma} \partial_{\rho} (b_{\epsilon \beta} h^{\epsilon \alpha} + c_{\epsilon \beta} \tau^{\epsilon \alpha}) \partial_{\alpha} \partial_{\sigma} \xi^{\beta} + \frac12 h^{(\mu| \rho} \tau^{|\nu) \sigma} \partial_{\rho} (b_{\epsilon \beta} h^{\epsilon \alpha)} + c_{\epsilon \beta} \tau^{\epsilon \alpha)}) \partial_{\alpha} \partial_{\sigma} \xi^{\beta} \nn \\ &&  - \tau^{(\mu| \rho} h^{|\nu) \sigma} \partial_{\alpha} (b_{\epsilon \rho} h^{\epsilon \beta} + c_{\epsilon \rho} \tau^{\epsilon \beta}) \partial_{\beta} \partial_{\sigma} \xi^{\alpha} - h^{(\mu| \rho} \tau^{|\nu) \sigma} \partial_{\alpha} (b_{\epsilon \rho} h^{\epsilon \beta} + c_{\epsilon \rho} \tau^{\epsilon \beta}) \partial_{\beta} \partial_{\sigma} \xi^{\alpha} \nn \\
&& - h^{(\mu| \rho} h^{|\nu) \sigma} \partial_{\alpha}(b_{\pi \rho} \tau^{\pi \beta})  \partial_{\beta} \partial_{\sigma} \xi^{\alpha} - b_{\pi \rho} \tau^{\pi (\mu|} h^{|\nu)\sigma} \partial_{\alpha} h^{\beta \rho} \partial_{\sigma} \partial_{\beta} \xi^{\alpha} \nn \\ && - (b_{\epsilon \rho} h^{\epsilon (\mu|} + c_{\epsilon \rho} \tau^{\epsilon (\mu|}) \tau^{|\nu)\sigma} \partial_{\alpha}h^{\beta \rho} \partial_{\sigma} \partial_{\beta}\xi^{\alpha} - (b_{\epsilon \rho} h^{\epsilon (\mu|} + c_{\epsilon \rho} \tau^{\epsilon (\mu|}) h^{|\nu)\sigma} \partial_{\alpha} \tau^{\beta \rho} \partial_{\sigma} \partial_{\beta}\xi^{\alpha} \, . \nn \\ 
\eea
At this point we observe that the corrections $\Delta h^{\mu \nu}$ and $\Delta \tau^{\mu \nu}$ involves both diffeomorphisms and b-shifts,
\bea
\Delta h^{\mu \nu} & = & \Delta_{\xi} h^{\mu \nu} + \Delta_{\zeta} h^{\mu \nu} \, , \\
\Delta \tau^{\mu \nu} & = & \Delta_{\xi} \tau^{\mu \nu} + \Delta_{\zeta} \tau^{\mu \nu} \, .
\eea
The last pieces of the previous transformations, involving $\zeta_{\mu}$, can be eliminated considering a field redefinition. Explicitly,
\bea
\tilde h^{\mu \nu} & = & h^{\mu \nu} - \frac12 h^{(\mu| \rho} h^{|\nu) \sigma} \partial_{\rho} h^{\alpha \beta} \partial_{\alpha} b_{\beta \sigma} \, , \\
\tilde \tau^{\mu \nu} & = & \tau^{\mu \nu} - \frac12 (\tau^{(\mu| \rho} h^{|\nu) \sigma} \partial_{\rho}h^{\alpha \beta} + h^{(\mu| \rho} \tau^{|\nu) \sigma} \partial_{\rho}h^{\alpha \beta} + h^{(\mu| \rho} h^{|\nu) \sigma} \partial_{\rho} \tau^{\alpha \beta}) \partial_{\alpha} b_{\beta \sigma} \, . 
\label{redeftauup}
\eea
The remanent $\Delta h^{\mu \nu}$ and $\Delta \tau^{\mu \nu}$ cannot be eliminated with field redefinitions. The reader can easily prove that given the following factorization,
\bea
\Delta_{\xi} h^{\mu \nu} & = & h_{1 \beta}^{\mu \nu \sigma \alpha} \partial_{\sigma} \partial_{\alpha} \xi^{\beta} \, , \\
\Delta_{\xi} \tau^{\mu \nu} & = & \tau_{1 \beta}^{\mu \nu \sigma \alpha} \partial_{\sigma} \partial_{\alpha} \xi^{\beta} \, ,
\eea
both $h_{1 \beta}^{\mu \nu \sigma \alpha}$ and $\tau_{1 \beta}^{\mu \nu \sigma \alpha}$ transform in a non-covariant fashion. This behavior is inherited from the relativistic limit of the theory.

The redefinition (\ref{redeftauup}) is not boost invariant, and therefore $\tilde \tau^{\mu \nu}$ receives a correction in its boost transformation given by,
\bea
\Delta_{\lambda} \tilde \tau^{\mu \nu} & = & - \frac12 \big(\lambda_{a}{}^{a'} e^{(\mu}{}_{a'} \tau^{\rho) a} \ h^{\nu \sigma} \partial_{\rho}h^{\alpha \beta} + h^{\mu \rho} \lambda_{a}{}^{a'} e^{(\nu}{}_{a'} \tau^{\sigma) a} \partial_{\rho}h^{\alpha \beta} + h^{\mu \rho} h^{\nu \sigma} \partial_{\rho} (\lambda_{a}{}^{a'} e^{(\alpha}{}_{a'} \tau^{\beta) a}) \big) \partial_{\alpha} b_{\beta \sigma} 
\nn \\ &&
+ \frac12 (\tau^{\mu \rho} h^{\nu \sigma} \partial_{\rho}h^{\alpha \beta} + h^{\mu \rho} \tau^{\nu \sigma} \partial_{\rho}h^{\alpha \beta} + h^{\mu \rho} h^{\nu \sigma} \partial_{\rho} \tau^{\alpha \beta}) \partial_{\alpha}(\epsilon_{a b} \lambda^{a}{}_{a'} \tau_{[\beta}{}^{b} e_{\sigma]}{}^{a'})
\nn \\ && + (\mu \leftrightarrow \nu) \, . \eea
Now we move to the analysis of the other components of the generalized metric. The expresion for $\delta^{(1)}{\mathcal H}^{(0)}_{\mu}{}^{\nu}$ is given by
\bea
&& \delta^{(1)}{\mathcal H}^{(0)}_{\mu}{}^{\nu} = - \frac14 \partial_{\mu} h^{\rho \sigma} \partial_{\rho} \partial_{\sigma} \xi^{\nu} - \frac12 \partial_{\rho} h^{\sigma \nu} \partial_{\mu} \partial_{\sigma} \xi^{\rho} 
\nn \\ &&
- \frac14 (b_{\epsilon \mu} h^{\epsilon \rho} + c_{\epsilon \mu} \tau^{\epsilon \rho}) \Big[ (b_{\pi \sigma} h^{\pi \nu} + c_{\pi \sigma} \tau^{\pi \nu}) \partial_{\rho} h^{\beta \alpha} \partial_{\alpha} \partial_{\beta} \xi^{\sigma} 
+ 2 h^{\nu \sigma} \partial_{\rho}h^{\beta \alpha} \partial_{\alpha} \partial_{[\beta} \zeta_{\sigma]} \nn \\ && - 2 h^{\nu \sigma} \partial_{\rho}(b_{\pi \beta} h^{\pi \alpha} + c_{\pi \beta} \tau^{\pi \alpha}) \partial_{\alpha} \partial_{\sigma} \xi^{\beta} 
- 2 h^{\nu \sigma} \partial_{\beta}(b_{\pi \rho} h^{\pi \alpha} + c_{\pi \rho} \tau^{\pi \alpha}) \partial_{\sigma} \partial_{\alpha}\xi^{\beta} \Big] 
\nn \\ &&
+ \frac12 (h_{\mu \rho} + b_{\epsilon \mu} h^{\epsilon \pi} b_{\pi \rho} + 2 c_{\epsilon (\mu|} \tau^{\epsilon \pi} b_{\pi |\rho)}) h^{\nu \sigma} \partial_{\beta} h^{\alpha \rho} \partial_{\sigma}\partial_{\alpha}\xi^{\beta} 
\nn \\ &&
+ \frac14 h^{\nu \rho} (h_{\mu \sigma} + b_{\epsilon \mu} h^{\epsilon \pi} b_{\pi \sigma} + 2 c_{\epsilon (\mu|} \tau^{\epsilon \pi} b_{\pi |\sigma)}) \partial_{\rho} h^{\beta \alpha} \partial_{\alpha} \partial_{\beta} \xi^{\sigma}
\nn \\ &&
+ \frac12 (b_{\epsilon \mu} h^{\epsilon \sigma} + c_{\epsilon \mu} \tau^{\epsilon \sigma}) \Big[ \frac12 h^{\nu \rho} \partial_{\rho}(b_{\pi \beta} h^{\pi \alpha} + c_{\pi \beta} \tau^{\pi \alpha}) \partial_{\alpha} \partial_{\sigma} \xi^{\beta} 
- h^{\nu \rho} \partial_{\rho} h^{\beta \alpha} \partial_{\alpha [\beta} \zeta_{\sigma]} \nn \\ && - h^{\nu \rho} \partial_{\alpha} (b_{\pi \rho} h^{\pi \beta} + c_{\pi \rho} \tau^{\pi \beta}) \partial_{\sigma} \partial_{\beta} \xi^{\alpha}
+ (b_{\pi \rho} h^{\pi \beta} + c_{\pi \rho} \tau^{\pi \beta}) \partial_{\alpha}h^{\rho \beta} \partial_{\sigma} \partial_{\beta} \xi^{\alpha} \Big] \, .
\eea
From the definition of this component we get the relation,
\bea
\label{exb}
\delta^{(1)}{\mathcal H}^{(0)}_{\mu}{}^{\nu} + b_{\psi \mu} \Delta h^{\psi \nu} + c_{\psi \mu} \Delta \tau^{\psi \nu} = -\Delta b_{\psi \mu} h^{\psi \nu} - \Delta c_{\psi \mu} \tau^{\psi \nu} \, ,
\eea
where the full LHS is known. We can then factorize $h^{\psi \nu}$ and $\tau^{\psi \nu}$ of this sum and in this way to obtain part of the transformations of $c_{\mu \nu}$ and $b_{\mu \nu}$.  However, some of the terms cannot be directly factorized in this way, and in those cases we need to invoke the relation (\ref{delta}). Explicitly,
\bea
\label{exc}
&& - \Delta c_{\psi \mu} \tau^{\psi \nu} =    - \frac14 \partial_{\mu} h^{\rho \sigma} \partial_{\rho} \partial_{\sigma} \xi^{\epsilon} \tau_{\pi \epsilon} \tau^{\pi \nu} - \frac12 \partial_{\rho} h^{\sigma \epsilon} \partial_{\mu} \partial_{\sigma} \xi^{\rho} \tau_{\pi \epsilon} \tau^{\pi \nu} \nn \\ && - \frac14 (b_{\epsilon \mu} h^{\epsilon \rho} + c_{\epsilon \mu} \tau^{\epsilon \rho}) c_{\pi \sigma} \tau^{\pi \nu} \partial_{\rho} h^{\beta \alpha} \partial_{\alpha} \partial_{\beta} \xi^{\sigma}
+ \frac12 (b_{\epsilon \mu} h^{\epsilon \sigma} + c_{\epsilon \mu} \tau^{\epsilon \sigma}) c_{\pi \rho} \tau^{\pi \nu} \partial_{\alpha}h^{\rho \beta} \partial_{\sigma} \partial_{\beta} \xi^{\alpha}
\nn \\ &&
- b_{\psi \mu} \Big[ \frac14 h^{\psi \rho} c_{\gamma \sigma} \tau^{\gamma \nu} \partial_{\rho} h^{\alpha \beta} \partial_{\alpha} \partial_{\beta} \xi^{\sigma}
+ \frac12 c_{\gamma \rho} \tau^{\gamma \nu} h^{\psi \sigma} \partial_{\alpha} h^{\beta \rho} \partial_{\sigma} \partial_{\beta} \xi^{\alpha} \Big] \nn \\ &&
+ c_{\psi \mu} \Big[ - \frac12 \tau^{(\psi| \rho} c_{\epsilon \sigma} \tau^{\epsilon |\nu)} \partial_{\rho}h^{\alpha \beta} \partial_{\alpha} \partial_{\beta} \xi^{\sigma} - c_{\epsilon \rho} \tau^{\epsilon (\psi|} \tau^{|\nu)\sigma} \partial_{\alpha}h^{\beta \rho} \partial_{\sigma} \partial_{\beta}\xi^{\alpha}\Big] \, ,
\eea
and
\bea
\label{exb}
&&  -\Delta b_{\psi \mu} h^{\psi \nu}  = - \frac14 \partial_{\mu} h^{\rho \sigma} \partial_{\rho} \partial_{\sigma} \xi^{\epsilon} h_{\pi \epsilon} h^{\pi \nu} - \frac12 \partial_{\rho} h^{\sigma \epsilon} \partial_{\mu} \partial_{\sigma} \xi^{\rho} h_{\pi \epsilon} h^{\pi \nu} 
\nn \\ &&
- \frac14 (b_{\epsilon \mu} h^{\epsilon \rho} + c_{\epsilon \mu} \tau^{\epsilon \rho}) \Big[ b_{\pi \sigma} h^{\pi \nu} \partial_{\rho} h^{\beta \alpha} \partial_{\alpha} \partial_{\beta} \xi^{\sigma} 
+ 2 h^{\nu \sigma} \partial_{\rho}h^{\beta \alpha} \partial_{\alpha} \partial_{[\beta} \zeta_{\sigma]} \nn \\ && - 2 h^{\nu \sigma} \partial_{\rho}(b_{\pi \beta} h^{\pi \alpha} + c_{\pi \beta} \tau^{\pi \alpha}) \partial_{\alpha} \partial_{\sigma} \xi^{\beta}
- 2 h^{\nu \sigma} \partial_{\beta}(b_{\pi \rho} h^{\pi \alpha} + c_{\pi \rho} \tau^{\pi \alpha}) \partial_{\sigma} \partial_{\alpha}\xi^{\beta} \Big] 
\nn \\ &&
+ \frac12 (h_{\mu \rho} + b_{\epsilon \mu} h^{\epsilon \pi} b_{\pi \rho} + 2 c_{\epsilon (\mu|} \tau^{\epsilon \pi} b_{\pi \rho)}) h^{\nu \sigma} \partial_{\beta} h^{\alpha \rho} \partial_{\sigma}\partial_{\alpha}\xi^{\beta} 
\nn \\ &&
+ \frac14 h^{\nu \rho} (h_{\mu \sigma} + b_{\epsilon \mu} h^{\epsilon \pi} b_{\pi \sigma} + 2 c_{\epsilon (\mu|} \tau^{\epsilon \pi} b_{\pi \sigma)}) \partial_{\rho} h^{\beta \alpha} \partial_{\alpha} \partial_{\beta} \xi^{\sigma} 
\nn \\ &&
+ \frac12 (b_{\epsilon \mu} h^{\epsilon \sigma} + c_{\epsilon \mu} \tau^{\epsilon \sigma}) \Big[ \frac12 h^{\nu \rho} \partial_{\rho}(b_{\pi \beta} h^{\pi \alpha} + c_{\pi \beta} \tau^{\pi \alpha}) \partial_{\alpha} \partial_{\sigma} \xi^{\beta} 
- h^{\nu \rho} \partial_{\rho} h^{\beta \alpha} \partial_{\alpha [\beta} \zeta_{\sigma]} \nn \\ && - h^{\nu \rho} \partial_{\alpha} (b_{\pi \rho} h^{\pi \beta} + c_{\pi \rho} \tau^{\pi \beta}) \partial_{\sigma} \partial_{\beta} \xi^{\alpha}
+ b_{\pi \rho} h^{\pi \nu} \partial_{\alpha}h^{\rho \beta} \partial_{\sigma} \partial_{\beta} \xi^{\alpha} \Big]
\nn \\ &&
+ b_{\psi \mu} \Big[ h^{(\psi| \rho} h^{|\nu) \sigma} \partial_{\rho} h^{\alpha \beta} \partial_{\alpha} \partial_{[\beta} \zeta_{\sigma]} - \frac12 h^{(\psi| \rho} b_{\gamma \sigma} h^{\gamma |\nu)} \partial_{\rho} h^{\alpha \beta} \partial_{\alpha} \partial_{\beta} \xi^{\sigma} - \frac14 h^{\nu \rho} c_{\gamma \sigma} \tau^{\gamma \psi} \partial_{\rho} h^{\alpha \beta} \partial_{\alpha} \partial_{\beta} \xi^{\sigma} \nn \\
&& + \frac12 h^{(\psi| \rho} h^{|\nu) \sigma} \partial_{\rho}(b_{\gamma \beta} h^{\gamma \alpha} + c_{\gamma \beta} \tau^{\gamma \alpha}) \partial_{\alpha} \partial_{\sigma} \xi^{\beta} - h^{(\psi| \rho} h^{|\nu) \sigma} \partial_{\alpha}(b_{\gamma \rho} h^{\gamma \beta} + c_{\gamma \rho} \tau^{\gamma \beta}) \partial_{\beta} \partial_{\sigma} \xi^{\alpha} \nn \\
&& 
- b_{\gamma \rho} h^{\gamma (\psi|} h^{|\nu) \sigma} \partial_{\alpha} h^{\beta \rho} \partial_{\sigma} \partial_{\beta} \xi^{\alpha} - \frac12 c_{\gamma \rho} \tau^{\gamma \psi} h^{\nu \sigma} \partial_{\alpha} h^{\beta \rho} \partial_{\sigma} \partial_{\beta} \xi^{\alpha} \Big] 
\nn \\ &&
+ c_{\psi \mu} \Big[ \frac12 \tau^{\psi \rho} h^{\nu \sigma} \partial_{\rho}h^{\alpha \beta} \partial_{\alpha} \partial_{[\beta} \zeta_{\sigma]} + \frac12 h^{\nu \rho} \tau^{\psi \sigma} \partial_{\rho}h^{\alpha \beta} \partial_{\alpha} \partial_{[\beta} \zeta_{\sigma]} \nn \\ && - \frac14 \tau^{\psi \rho} b_{\epsilon \sigma} h^{\epsilon \nu} \partial_{\rho}h^{\alpha \beta} \partial_{\alpha} \partial_{\beta} \xi^{\sigma} - \frac14 h^{\nu \rho} b_{\pi \sigma} \tau^{\pi \psi} \partial_{\rho} h^{\alpha \beta} \partial_{\alpha} \partial_{\beta} \xi^{\sigma} - \frac14 h^{\nu \rho} c_{\epsilon \sigma} \tau^{\epsilon \psi} \partial_{\rho} \tau^{\alpha \beta} \partial_{\alpha} \partial_{\beta} \xi^{\sigma} 
\nn \\ &&
+ \frac14 \tau^{\psi \rho} h^{\nu \sigma} \partial_{\rho} (b_{\epsilon \beta} h^{\epsilon \alpha} + c_{\epsilon \beta} \tau^{\epsilon \alpha}) \partial_{\alpha} \partial_{\sigma} \xi^{\beta} + \frac14 h^{\nu \rho} \tau^{\psi \sigma} \partial_{\rho} (b_{\epsilon \beta} h^{\epsilon \alpha} + c_{\epsilon \beta} \tau^{\epsilon \alpha}) \partial_{\alpha} \partial_{\sigma} \xi^{\beta} \nn \\ &&  - \frac12 \tau^{\psi \rho} h^{\nu \sigma} \partial_{\alpha} (b_{\epsilon \rho} h^{\epsilon \beta} + c_{\epsilon \rho} \tau^{\epsilon \beta}) \partial_{\beta} \partial_{\sigma} \xi^{\alpha} - \frac12 h^{\nu \rho} \tau^{\psi \sigma} \partial_{\alpha} (b_{\epsilon \rho} h^{\epsilon \beta} + c_{\epsilon \rho} \tau^{\epsilon \beta}) \partial_{\beta} \partial_{\sigma} \xi^{\alpha} \nn \\ && - \frac12 b_{\pi \rho} \tau^{\pi \psi} h^{\nu \sigma} \partial_{\alpha} h^{\beta \rho} \partial_{\sigma} \partial_{\beta} \xi^{\alpha} - \frac12 b_{\epsilon \rho} h^{\epsilon \nu} \tau^{\psi \sigma} \partial_{\alpha}h^{\beta \rho} \partial_{\sigma} \partial_{\beta}\xi^{\alpha} - \frac12 c_{\epsilon \rho} \tau^{\epsilon \psi} h^{\nu \sigma} \partial_{\alpha} \tau^{\beta \rho} \partial_{\sigma} \partial_{\beta}\xi^{\alpha}\Big] \, . \nn \\
\eea
The full correction for $\Delta c_{\mu \nu}$ is given by
\bea
\Delta c_{\mu \nu} = \tau^{\rho \sigma} \Delta c_{\rho [\nu|}  \tau_{\sigma |\mu]} - c_{\rho [\nu|} \Delta h^{\rho \sigma} h_{\sigma |\mu]} \, ,
\eea
where we have used the fact that $c_{\mu \nu} h^{\nu \rho}=0$. Once again we can split the transformation as
\bea
\Delta c_{\mu \nu} & = & \Delta_{\xi} c_{\mu \nu} + \Delta_{\zeta} c_{\mu \nu} \, . 
\eea
The reader might quickly verify that the former vanishes, and therefore the correction is purely deforming the diffeomorphism transformation. Particularly, 
\bea
\Delta_{\xi} c_{\rho \sigma} & = & - \frac18 \tau_{\nu \rho}  \Big[ \partial_{\sigma} h^{\kappa \delta} \partial_{\kappa} \partial_{\delta} \xi^{\epsilon} \tau_{\pi \epsilon} \tau^{\pi \nu} + 2 \partial_{\kappa} h^{\delta \epsilon} \partial_{\sigma} \partial_{\delta} \xi^{\kappa} \tau_{\pi \epsilon} \tau^{\pi \nu} \nn \\ && + (b_{\epsilon \sigma} h^{\epsilon \kappa} + c_{\epsilon \sigma} \tau^{\epsilon \kappa}) c_{\pi \delta} \tau^{\pi \nu} \partial_{\kappa} h^{\beta \alpha} \partial_{\alpha} \partial_{\beta} \xi^{\delta}
- 2 (b_{\epsilon \sigma} h^{\epsilon \delta} + c_{\epsilon \sigma} \tau^{\epsilon \delta}) c_{\pi \kappa} \tau^{\pi \nu} \partial_{\alpha}h^{\kappa \beta} \partial_{\delta} \partial_{\beta} \xi^{\alpha}
\nn \\ &&
+ b_{\psi \sigma} (h^{\psi \kappa} c_{\gamma \delta} \tau^{\gamma \nu} \partial_{\kappa} h^{\alpha \beta} \partial_{\alpha} \partial_{\beta} \xi^{\delta}
+ 2 c_{\gamma \kappa} \tau^{\gamma \nu} h^{\psi \delta} \partial_{\alpha} h^{\beta \kappa} \partial_{\delta} \partial_{\beta} \xi^{\alpha} ) \nn \\ &&
+ 2 c_{\psi \sigma} (  \tau^{(\psi| \kappa} c_{\epsilon \delta} \tau^{\epsilon |\nu)} \partial_{\kappa}h^{\alpha \beta} \partial_{\alpha} \partial_{\beta} \xi^{\delta} + 2 c_{\epsilon \kappa} \tau^{\epsilon (\psi|} \tau^{|\nu)\delta} \partial_{\alpha}h^{\beta \kappa} \partial_{\delta} \partial_{\beta}\xi^{\alpha}) \Big] \nn \\ && 
+ \frac18 c_{\mu \sigma} \Big[ h^{\nu \psi} c_{\gamma \sigma} \tau^{\gamma \mu} \partial_{\psi} h^{\alpha \beta} \partial_{\alpha} \partial_{\beta} \xi^{\sigma} + 2 c_{\gamma \psi} \tau^{\gamma \mu} h^{\nu \delta} \partial_{\alpha} h^{\beta \psi} \partial_{\delta} \partial_{\beta} \xi^{\alpha} \Big] h_{\nu \rho} - (\rho \leftrightarrow \sigma) \, . \nn
\\ \label{ctransf}
\eea
One may try to eliminate it with field redefinitions. For doing so we factorize the second derivative of the diffeomorphism parameter as
\bea
\Delta c_{\rho \sigma} = c_{1}^{\delta}{}_{[\rho| \kappa} \partial_{|\sigma]} \partial_{\delta} \xi^{\kappa} + c_{2}^{\delta \beta}{}_{\alpha[\rho \sigma]} \partial_{\delta} \partial_{\beta} \xi^{\alpha}  \, ,  
\eea
where $c_{1}^{\delta}{}_{[\rho \kappa}$ and $c_{2}^{\delta \beta}{}_{\alpha[\rho \sigma]}$ are given by
\bea
c_{1}^{\delta}{}_{\rho \kappa} & = & - \frac12 \tau_{\nu \rho} \partial_{\kappa} h^{\delta \nu} \, , \\
c_{2}^{\delta \beta}{}_{\alpha [\rho \sigma]} & = & - \frac14 \tau_{\alpha [\rho}  \partial_{\sigma]} h^{\delta \beta}  - \frac14 \tau_{\nu [\rho|} (b_{\epsilon |\sigma]} h^{\epsilon \kappa} + c_{\epsilon |\sigma]} \tau^{\epsilon \kappa}) c_{\pi \alpha} \tau^{\pi \nu} \partial_{\kappa} h^{\beta \delta}
\nn \\ && + \frac12 \tau_{\nu [\rho|} (b_{\epsilon |\sigma]} h^{\epsilon \delta} + c_{\epsilon |\sigma]} \tau^{\epsilon \delta}) c_{\pi \kappa} \tau^{\pi \nu} \partial_{\alpha}h^{\kappa \beta}
\nn \\ &&
- \frac14 \tau_{\nu [\rho|} b_{\psi |\sigma]} h^{\psi \kappa} c_{\gamma \alpha} \tau^{\gamma \nu} \partial_{\kappa} h^{\delta \beta}
- \frac12 \tau_{\nu [\rho|} b_{\psi |\sigma]} c_{\gamma \kappa} \tau^{\gamma \nu} h^{\psi \delta} \partial_{\alpha} h^{\beta \kappa} \nn \\ &&
- \frac12 \tau_{\nu [\rho|} c_{\psi |\sigma]}  \tau^{(\psi| \kappa} c_{\epsilon \alpha} \tau^{\epsilon |\nu)} \partial_{\kappa}h^{\delta \beta} - \tau_{\nu [\rho|} c_{\psi |\sigma]} c_{\epsilon \kappa} \tau^{\epsilon (\psi|} \tau^{|\nu)\delta} \partial_{\alpha}h^{\beta \kappa} \nn \\ && 
+ \frac14 c_{\mu [\sigma|} h^{\nu \psi} c_{\gamma \alpha} \tau^{\gamma \mu} \partial_{\psi} h^{\delta \beta} h_{\nu |\rho]} + \frac12 c_{\mu [\sigma|} c_{\gamma \psi} \tau^{\gamma \mu} h^{\nu \delta} \partial_{\alpha} h^{\beta \psi} h_{\nu |\rho]} \, .
\eea
Inspecting the transformation properties of the previous objects, we can easily see that $c_{1}^{\delta}{}_{\rho \kappa}$ transforms covariantly, while some of the terms of $c_{2}^{\delta \beta}{}_{\alpha[\rho \sigma]}$ transforms non-covariantly. This indicates that (\ref{ctransf}) cannot be removed with field redefinitions and it is a non-ambiguous deformation of the diffeomorphism transformation of $c_{\rho \sigma}$. Explicitly, the non-ambiguous terms in the transformation are given by
\bea
\Delta_{\xi} c_{\rho \sigma}|_{\textrm{non-ambiguous}} & = & \Big[ - \frac14 \tau_{\alpha [\rho}  \partial_{\sigma]} h^{\delta \beta}  - \frac14 \tau_{\nu [\rho|} (b_{\epsilon |\sigma]} h^{\epsilon \kappa} + c_{\epsilon |\sigma]} \tau^{\epsilon \kappa}) c_{\pi \alpha} \tau^{\pi \nu} \partial_{\kappa} h^{\beta \delta}
\nn \\ &&
- \frac14 \tau_{\nu [\rho|} b_{\psi |\sigma]} h^{\psi \kappa} c_{\gamma \alpha} \tau^{\gamma \nu} \partial_{\kappa} h^{\delta \beta}
- \frac12 \tau_{\nu [\rho|} c_{\psi |\sigma]}  \tau^{(\psi| \kappa} c_{\epsilon \alpha} \tau^{\epsilon |\nu)} \partial_{\kappa}h^{\delta \beta} 
\nn \\ && 
+ \frac14 c_{\mu [\sigma|} h^{\nu \psi} c_{\gamma \alpha} \tau^{\gamma \mu} \partial_{\psi} h^{\delta \beta} h_{\nu |\rho]} \Big] \partial_{\delta} \partial_{\beta} \xi^{\alpha} \, . \label{gsc} 
\eea
The previous transformation is the equivalent to a Green-Schwarz mechanism for diffeomorphism for the $c_{\mu \nu}$ field. Now we would like to find the same mechanism, but for the $b_{\mu \nu}$ field. The full transformation of the $b_{\mu \nu}$ field is given by
\bea
\Delta b_{\mu \nu} = \Delta b_{\rho [\nu|} h^{\rho \sigma} h_{\sigma |\mu]} + \Delta b_{\rho [\nu|} \tau^{\rho \sigma} \tau_{\sigma |\mu]} \, .
\eea
The second piece of the previous transformation must be read from $\delta^{(1)} {\cal H}^{-2}_{\mu}{}^{\nu}$ using the relation,
\bea
 \delta^{(1)} {\cal H}^{(-2)}_{\nu}{}^{\sigma} = - \Delta b_{\rho \nu} \tau^{\rho \sigma}  - b_{\rho \nu} \Delta \tau^{\rho \sigma} \, . 
\eea
Therefore, the transformation can be written as
\bea
\Delta b_{\mu \nu} = \Delta b_{\rho [\nu|} h^{\rho \sigma} h_{\sigma |\mu]} - \delta^{(1)} {\cal H}^{(-2)}_{[\nu|}{}^{\rho} \tau_{\rho |\mu]} - b_{\rho [\nu|} \Delta \tau^{\rho \sigma} \tau_{\sigma |\mu]} \, .
\eea
To explicitly obtain the previous transformation we first compute,
\bea
&& \delta^{(1)}{\cal H}^{(-2)}_{\nu}{}^{\rho} \tau_{\rho \mu} =  \Big[ - \frac{1}{4}\, {\partial}_{\nu}{{\tau}^{\tau \sigma}}\,  {\partial}_{\tau}{\partial_{\sigma} \xi^{\rho}}\,  \,
+ \frac{1}{4}\, b_{\delta \nu} \tau^{\delta \sigma} c_{\lambda \gamma} \tau^{\lambda \rho} {\partial}_{\sigma}{h}^{\alpha \beta}\,  {\partial}_{\alpha}\partial_{\beta} \xi^{\gamma} \nn \\ &&
- \frac{1}{2}\, (b_{\lambda \nu} h^{\lambda \sigma} + c_{\lambda \nu} \tau^{\lambda \sigma}) {\tau}^{\rho \gamma} {\partial}_{\sigma}{{h}^{\alpha \beta}}\,  {\partial}_{\alpha}{\partial_{[\beta} \zeta_{\gamma]}}\, 
+ \frac{1}{4}\, (b_{\lambda \nu} h^{\lambda \sigma} + c_{\lambda \nu} \tau^{\lambda \sigma}) \tau^{\delta \rho} b_{\delta \gamma} {\partial}_{\sigma}{h}^{\alpha \beta}\,  {\partial}_{\alpha}\partial_{\beta} \xi^{\gamma} \nn \\ && -\frac{1}{4}\, (b_{\lambda \nu} h^{\lambda \sigma} + c_{\lambda \nu} \tau^{\lambda \sigma}) {\tau}^{\rho \gamma} {\partial}_{\sigma}(b_{\epsilon \beta} h^{\epsilon \alpha} + c_{\epsilon \beta} \tau^{\epsilon \alpha})\,  {\partial}_{\alpha}{\partial_{\gamma} \xi^{\beta}} \,
+ \frac{1}{4}\, (b_{\lambda \nu} h^{\lambda \sigma} + c_{\lambda \nu} \tau^{\lambda \sigma}) c_{\epsilon \gamma} \tau^{\epsilon \rho} {\partial}_{\sigma}{{\tau}^{\alpha \beta}}\,  {\partial}_{\alpha}\partial_{\beta} \xi^{\gamma}  \nn \\ &&
+ \frac{1}{2}\, (b_{\lambda \nu} h^{\lambda \gamma} + c_{\lambda \nu} \tau^{\lambda \gamma}) {\tau}^{\rho \sigma} {\partial}_{\alpha}{(b_{\epsilon \gamma} h^{\epsilon \beta} + c_{\epsilon \gamma} \tau^{\epsilon \beta})}\,  {\partial}_{\sigma}\partial_{\beta} \xi^{\alpha}  - \frac{1}{2}\, 
{\tau}^{\rho \sigma} (b_{\lambda \nu} h^{\lambda \gamma} + c_{\lambda \nu} \tau^{\lambda \gamma}) 
\partial_{\sigma}{h}^{\alpha \beta}\,  \partial_{\alpha} \partial_{[\beta} \zeta_{\gamma]}\, 
\nn \\ &&
+ \frac{1}{2}\, (h_{\nu \gamma} + b_{\lambda \nu} h^{\lambda \tau} b_{\tau \gamma} + 2 c_{\lambda (\nu|} \tau^{\lambda \tau} b_{\tau |\gamma)}) {\tau}^{\rho \sigma} {\partial}_{\alpha}{{h}^{\beta \gamma}}\,  {\partial}_{\sigma}\partial_{\beta} \xi^{\alpha} \nn \\ &&
 + \frac{1}{4}\, 
{\tau}^{\rho \sigma} (h_{\nu \gamma} + b_{\lambda \nu} h^{\lambda \tau} b_{\tau \gamma} + 2 c_{\lambda (\nu|} \tau^{\lambda \tau} b_{\tau |\gamma)}) 
\partial_{\sigma}{h}^{\alpha \beta}\,  \partial_{\alpha} \partial_{\beta} \xi^{\gamma}\,
 \nn \\ &&
 - \frac{1}{4}\, 
{\tau}^{\rho \sigma} (b_{\lambda \nu} h^{\lambda \sigma} + c_{\lambda \nu} \tau^{\lambda \gamma}) 
\partial_{\sigma}(b_{\epsilon \beta} h^{\epsilon \alpha} + c_{\epsilon \beta} \tau^{\epsilon \alpha})\,  \partial_{\alpha} \partial_{\gamma} \xi^{\beta}\,
 \,
 \nn \\ && + \frac{1}{2}\, {\tau}^{\rho \gamma} (b_{\lambda \nu} h^{\lambda \sigma} + c_{\lambda \nu} \tau^{\lambda \sigma}) {\partial}_{\alpha}(b_{\epsilon \gamma} h^{\epsilon \beta} + c_{\epsilon \gamma} \tau^{\epsilon \beta})\,  {\partial}_{\sigma}\partial_{\beta} \xi^{\alpha}  \nn \\ && + \frac{1}{2}\, \tau^{\delta \rho} b_{\delta \gamma} (b_{\lambda \nu} h^{\lambda \sigma} + c_{\lambda \nu} \tau^{\lambda \sigma}) {\partial}_{\alpha}{{h}^{\beta \gamma}}\,  {\partial}_{\sigma}\partial_{\beta} \xi^{\alpha}
 + \frac{1}{2}\, c_{\lambda \gamma} \tau^{\lambda \rho} b_{\delta \nu} \tau^{\delta \sigma} {\partial}_{\alpha}{{h}^{\beta \gamma}}\,  {\partial}_{\sigma}\partial_{\beta} \xi^{\alpha}  \nn \\ &&
+ \frac{1}{2}\, c_{\lambda \gamma} \tau^{\lambda \rho} (b_{\lambda \nu} h^{\lambda \sigma} + c_{\lambda \nu} \tau^{\lambda \sigma}) {\partial}_{\alpha}{{\tau}^{\beta \gamma}}\,  {\partial}_{\sigma}\partial_{\beta} \xi^{\alpha}
- \frac{1}{2}\, \partial_{\sigma}{\tau}^{\tau \rho}\,  {\partial}_{\nu} \partial_{\tau} \xi^{\sigma}\, \Big] \tau_{\rho \mu} \, .
\eea
At this point we can explicitly write the corrections to the leading order b-shifts. They are given by,
\bea
&& \Delta_{\zeta} b_{\mu \nu} 
=  \frac14
\Big[ (b_{\epsilon \nu} h^{\epsilon \rho} + c_{\epsilon \nu} \tau^{\epsilon \rho}) h^{\sigma \delta} \partial_{\rho}h^{\beta \alpha} 
+ (b_{\epsilon \nu} h^{\epsilon \delta} + c_{\epsilon \nu} \tau^{\epsilon \delta})
h^{\sigma \rho} \partial_{\rho} h^{\beta \alpha} \nn \\ &&
- 2 b_{\psi \nu} h^{(\psi| \rho} h^{|\sigma) \delta} \partial_{\rho} h^{\alpha \beta}
- \tau^{\psi \rho} h^{\sigma \delta} \partial_{\rho}h^{\alpha \beta} c_{\psi \nu} - h^{\sigma \rho} \tau^{\psi \delta} \partial_{\rho}h^{\alpha \beta} c_{\psi \nu} \Big] \partial_{\alpha} \partial_{[\beta} \zeta_{\delta]} h_{\sigma \mu} \, \nn \\ && - \frac{1}{4}\, \Big[(b_{\lambda \nu} h^{\lambda \sigma} + c_{\lambda \nu} \tau^{\lambda \sigma}) {\tau}^{\rho \gamma} +\, 
{\tau}^{\rho \sigma} (b_{\lambda \nu} h^{\lambda \gamma} + c_{\lambda \nu} \tau^{\lambda \gamma}) \Big]
\partial_{\sigma}{h}^{\alpha \beta}\,  \partial_{\alpha} \partial_{[\beta} \zeta_{\gamma]} \tau_{\rho \mu} \nn \\ && - \frac14 b_{\rho \nu} (\tau^{\sigma \gamma} h^{\rho \delta} \partial_{\gamma}h^{\alpha \beta} \partial_{\alpha} \partial_{[\beta} \zeta_{\delta]} + h^{\rho \gamma} \tau^{\sigma \delta} \partial_{\gamma}h^{\alpha \beta} \partial_{\alpha} \partial_{[\beta} \zeta_{\delta]}) \tau_{\sigma \mu}
- (\mu \leftrightarrow \nu) \, .
\eea
We can easily see that all these terms can be eliminated using a field redefinition for the b-field of the form
\bea
\tilde b_{\mu \nu} & = &  b_{\mu \nu} - \frac14
\Big[ (b_{\epsilon [\nu|} h^{\epsilon \rho} + c_{\epsilon [\nu|} \tau^{\epsilon \rho}) h^{\sigma \delta} \partial_{\rho}h^{\beta \alpha}  + (b_{\epsilon [\nu|} h^{\epsilon \delta} + c_{\epsilon [\nu|} \tau^{\epsilon \delta})
h^{\sigma \rho} \partial_{\rho} h^{\beta \alpha} \nn \\ && 
- 2 b_{\psi [\nu|} h^{(\psi| \rho} h^{|\sigma) \delta} \partial_{\rho} h^{\alpha \beta}  
- \tau^{\psi \rho} h^{\sigma \delta} \partial_{\rho}h^{\alpha \beta} c_{\psi [\nu|} - h^{\sigma \rho} \tau^{\psi \delta} \partial_{\rho}h^{\alpha \beta} c_{\psi [\nu|} \Big] \partial_{\alpha} b_{\beta \delta} h_{\sigma |\mu]} \, \nn \\ && 
- \frac{1}{4}\, \Big[ (b_{\lambda [\nu|} h^{\lambda \sigma} + c_{\lambda [\nu|} \tau^{\lambda \sigma}) {\tau}^{\rho \gamma} +\, 
{\tau}^{\rho \sigma} (b_{\lambda [\nu|} h^{\lambda \gamma} + c_{\lambda [\nu|} \tau^{\lambda \gamma}) \Big] 
\partial_{\sigma}{h}^{\alpha \beta}\,  \partial_{\alpha} b_{\beta \gamma} \tau_{\rho |\mu]} \nn \\ && - \frac14 b_{\rho [\nu|} (\tau^{\sigma \gamma} h^{\rho \delta} \partial_{\gamma}h^{\alpha \beta} \partial_{\alpha} b_{\beta \delta} + h^{\rho \gamma} \tau^{\sigma \delta} \partial_{\gamma}h^{\alpha \beta} \partial_{\alpha} b_{\beta \delta}) \tau_{\sigma |\mu]} \, .
\eea

As in the relativistic case, the correction to the diffeomorphisms transformations, $\Delta_{\xi} b_{\mu \nu}$, cannot be eliminated by field redefinitions. This correction is given by
\bea
\Delta_{\xi} b_{\mu \nu} & = & \Delta_{\xi} b_{\psi [\nu|} h^{\psi \chi} h_{\chi |\mu]} - \delta^{(1)} {\cal H}^{(-2)}_{[\nu|}{}^{\rho} \tau_{\rho |\mu]} - b_{\rho [\nu|} \Delta \tau^{\rho \sigma} \tau_{\sigma |\mu]} 
\nn \\ 
& = & b_{1}^{\delta}{}_{[\mu| \kappa} \partial_{|\nu]} \partial_{\delta} \xi^{\kappa} + b_{2}^{\alpha \beta}{}_{\omega[\mu \nu]} \partial_{\alpha} \partial_{\beta} \xi^{\omega} \, ,
\label{GSb}
\eea
where in the last equality we consider, once more, a factorization for the derivatives of the diffeomorphism parameter. For the latter we find,
\bea
b_{1}^{\delta}{}_{\mu \kappa} = \frac12 \partial_{\kappa} h^{\delta \rho}  h_{\mu \rho}  - 2 \, \partial_{\kappa}{\tau}^{\delta \rho}\,  \, \tau_{\mu \rho} \, .
\eea
This quantity transforms non-covariantly and therefore it cannot be removed from the b-field correction. The same happens with $b_{2}^{\alpha \beta}{}_{\omega[\mu \nu]}$, which also transforms non-covariantly. Due to the extension of this object, we report it in the appendix B. The corrections given in (\ref{GSb}) are the equivalent of the Green-Schwarz for diffeomorphisms for the b-field.

As happens for $\tilde \tau^{\mu \nu}$, the new $\tilde b_{\mu \nu}$ receives a correction to its boost transformation. This correction is given by,
\bea
\Delta_{\lambda} \tilde b_{\mu \nu} & = & \frac14
\Big[ b_{\epsilon \nu} h^{\epsilon \rho} h^{\sigma \delta} \partial_{\rho}h^{\beta \alpha} + c_{\epsilon \nu} \tau^{\epsilon \rho} h^{\sigma \delta} \partial_{\rho}h^{\beta \alpha}  + b_{\epsilon \nu} h^{\epsilon \delta} h^{\sigma \rho} \partial_{\rho} h^{\beta \alpha} + c_{\epsilon \nu} \tau^{\epsilon \delta}
h^{\sigma \rho} \partial_{\rho} h^{\beta \alpha} \nn \\ && 
- 2 b_{\psi \nu} h^{(\psi| \rho} h^{|\sigma) \delta} \partial_{\rho} h^{\alpha \beta}  
- \tau^{\psi \rho} h^{\sigma \delta} \partial_{\rho}h^{\alpha \beta} c_{\psi \nu} - h^{\sigma \rho} \tau^{\psi \delta} \partial_{\rho}h^{\alpha \beta} c_{\psi \nu} \Big] \partial_{\alpha} b_{\beta \delta} \lambda_{a a'} e_{(\sigma}{}^{a'} \tau_{\mu)}{}^{a} \, \nn \\ &&
+ \frac14
\Big[ b_{\epsilon \nu} h^{\epsilon \rho} h^{\sigma \delta} \partial_{\rho}h^{\beta \alpha} + c_{\epsilon \nu} \tau^{\epsilon \rho} h^{\sigma \delta} \partial_{\rho}h^{\beta \alpha}  + b_{\epsilon \nu} h^{\epsilon \delta} h^{\sigma \rho} \partial_{\rho} h^{\beta \alpha} + c_{\epsilon \nu} \tau^{\epsilon \delta}
h^{\sigma \rho} \partial_{\rho} h^{\beta \alpha} \nn \\ && 
- 2 b_{\psi \nu} h^{(\psi| \rho} h^{|\sigma) \delta} \partial_{\rho} h^{\alpha \beta}  
- \tau^{\psi \rho} h^{\sigma \delta} \partial_{\rho}h^{\alpha \beta} c_{\psi \nu} - h^{\sigma \rho} \tau^{\psi \delta} \partial_{\rho}h^{\alpha \beta} c_{\psi \nu} \Big] \partial_{\alpha} (\epsilon_{a b} \lambda^{a}{}_{a'} \tau_{[\beta}{}^{a} e_{\delta]}{}^{a'}) h_{\sigma \mu} \, \nn \\ &&
- \frac18
\Big[ -2 \epsilon_{a b} \lambda^{a}{}_{a'} \tau_{[\epsilon}{}^{b} e_{\nu]}{}^{a'} h^{\epsilon \rho} h^{\sigma \delta} \partial_{\rho}h^{\beta \alpha} + c_{\epsilon \nu}  \lambda_{a}{}^{a'} e^{\rho}{}_{a'} \tau^{\epsilon a} h^{\sigma \delta} \partial_{\rho}h^{\beta \alpha}   -2 \epsilon_{a b} \lambda^{a}{}_{a'} \tau_{[\epsilon}{}^{b} e_{\nu]}{}^{a'} h^{\epsilon \delta} h^{\sigma \rho} \partial_{\rho} h^{\beta \alpha} \nn \\ && + c_{\epsilon \nu} \lambda_{a}{}^{a'} e^{\delta}{}_{a'} \tau^{\epsilon a}
h^{\sigma \rho} \partial_{\rho} h^{\beta \alpha} + 4 \epsilon_{a b} \lambda^{a}{}_{a'} \tau_{[\psi}{}^{b} e_{\nu]}{}^{a'}) h^{(\psi| \rho} h^{|\sigma) \delta} \partial_{\rho} h^{\alpha \beta}  
- 2 \lambda_{a}{}^{a'} e^{(\psi}{}_{a'} \tau^{\rho) a} h^{\sigma \delta} \partial_{\rho}h^{\alpha \beta} c_{\psi \nu} \nn \\ && - 2 h^{\sigma \rho} \lambda_{a}{}^{a'} e^{(\psi}{}_{a'} \tau^{\delta) a} \partial_{\rho}h^{\alpha \beta} c_{\psi \nu} \Big] \partial_{\alpha} b_{\beta \delta} h_{\sigma \mu} \, 
\nn \\ &&
+ \frac{1}{4}\, \Big[ b_{\lambda \nu} h^{\lambda \sigma} \tau^{\rho \gamma} + c_{\lambda \nu} \tau^{\lambda \sigma} {\tau}^{\rho \gamma} +\, 
{\tau}^{\rho \sigma} b_{\lambda \nu} h^{\lambda \gamma} + {\tau}^{\rho \sigma} c_{\lambda \nu} \tau^{\lambda \gamma} \Big] 
\partial_{\sigma}{h}^{\alpha \beta}\,  \partial_{\alpha} (\epsilon_{a b} \lambda^{a}{}_{a'} \tau_{[\beta}{}^{b} e_{\gamma]}{}^{a'}) \tau_{\rho \mu} 
\nn \\ && 
- \frac{1}{8}\, \Big[ b_{\lambda \nu} h^{\lambda \sigma} \lambda_{a}{}^{a'} e^{\gamma}{}_{a'} \tau^{\rho a} + c_{\lambda \nu} \tau^{\lambda \sigma} \lambda_{a}{}^{a'} e^{\gamma}{}_{a'} \tau^{\rho a} \nn \\ && +\, 
{\tau}^{\rho \sigma} \epsilon_{a b} \lambda^{a}{}_{a'} \tau_{\nu}{}^{b} e_{\lambda}{}^{a'} h^{\lambda \gamma} + {\tau}^{\rho \sigma} c_{\lambda \nu} \lambda_{a}{}^{a'} e^{\gamma}{}_{a'} \tau^{\lambda a} \Big] 
\partial_{\sigma}{h}^{\alpha \beta}\,  \partial_{\alpha} b_{\beta \gamma} \tau_{\rho \mu} 
\nn \\ && 
- \frac{1}{8}\, \Big[ \epsilon_{a b} \lambda^{a}{}_{a'} \tau_{\nu}{}^{b} e_{\lambda}{}^{a'} h^{\lambda \sigma} \tau^{\rho \gamma} + c_{\lambda \nu} \lambda_{a}{}^{a'} e^{\sigma}{}_{a'} \tau^{\lambda a} {\tau}^{\rho \gamma} \nn \\ && +\, 
 \lambda_{a}{}^{a'} e^{\sigma}{}_{a'} \tau^{\rho a} b_{\lambda \nu} h^{\lambda \gamma} +  \lambda_{a}{}^{a'} e^{\sigma}{}_{a'} \tau^{\rho a} c_{\lambda \nu} \tau^{\lambda \gamma} \Big] 
\partial_{\sigma}{h}^{\alpha \beta}\,  \partial_{\alpha} b_{\beta \gamma} \tau_{\rho \mu} 
\nn \\ && 
- \frac18 \epsilon_{a b} \lambda^{a}{}_{a'} \tau_{\nu}{}^{b} e_{\rho}{}^{a'} (\tau^{\sigma \gamma} h^{\rho \delta} \partial_{\gamma}h^{\alpha \beta} \partial_{\alpha} b_{\beta \delta} + h^{\rho \gamma} \tau^{\sigma \delta} \partial_{\gamma}h^{\alpha \beta} \partial_{\alpha} b_{\beta \delta}) \tau_{\sigma \mu} 
\nn \\ &&
- \frac18 b_{\rho \nu} \Big( \tau^{\sigma \gamma} h^{\rho \delta} \partial_{\gamma}h^{\alpha \beta} \partial_{\alpha} (-2 \epsilon_{a b} \lambda^{a}{}_{a'} \tau_{[\beta}{}^{b} e_{\delta]}{}^{a'}) + h^{\rho \gamma} \tau^{\sigma \delta} \partial_{\gamma}h^{\alpha \beta} \partial_{\alpha} (-2 \epsilon_{a b} \lambda^{a}{}_{a'} \tau_{[\beta}{}^{b} e_{\delta]}{}^{a'}) \Big) \tau_{\sigma \mu} 
\nn \\ &&
- \frac18 b_{\rho \nu} \Big(  \lambda_{a}{}^{a'} e^{\gamma}{}_{a'} \tau^{\sigma a} h^{\rho \delta} \partial_{\gamma}h^{\alpha \beta} \partial_{\alpha} b_{\beta \delta} + h^{\rho \gamma} \lambda_{a}{}^{a'} e^{\delta}{}_{a'} \tau^{\sigma a} \partial_{\gamma}h^{\alpha \beta} \partial_{\alpha} b_{\beta \delta} \Big) \tau_{\sigma \mu} 
- (\mu \leftrightarrow \nu ) \, .
\eea
Finally, moving to $\Delta h_{\mu \nu}$ and $\Delta \tau_{\mu \nu}$. These transformations can be extracted from 
\bea
\Delta h_{\mu \nu} &=& \delta^{(1)} {\cal H}^{(0)}_{\mu \nu} - 2 \Delta b_{\rho (\mu|} h^{\rho \sigma} b_{\sigma |\nu)} - b_{\rho \mu} \Delta h^{\rho \sigma} b_{\sigma \nu} \nn \\ && - 2 \Delta c_{\rho (\mu|} \tau^{\rho \sigma} b_{\sigma |\nu)} - 2  c_{\rho (\mu|} \Delta \tau^{\rho \sigma} b_{\sigma |\nu)} - 2 c_{\rho (\mu|} \tau^{\rho \sigma} \Delta b_{\sigma |\nu)} \, ,
 \eea
 and
\bea
\Delta \tau_{\mu \nu}= 2 \Delta c_{(\mu| \rho} \tau^{\rho \sigma} c_{\sigma |\nu)} + c_{\mu \rho} \Delta \tau^{\rho \sigma} c_{\sigma \nu} \, ,  
\eea
respectively. Since these transformations effectively depend on $\Delta b_{\mu \nu}$ and $\Delta c_{\mu \nu}$, it is a straightforward computation to prove that these quantities also contain non-ambiguous corrections to their diffeomorphisms transformations. Similarly to $\tilde \tau^{\mu \nu}$ and $\tilde b_{\mu \nu}$, the new field $\tilde h_{\mu \nu}$ will inherit new higher-derivative boost corrections. Due to the systematic of the computation, we do not report these corrections explicitly. We observe that in the relativistic case all the higher-derivative correction in the transformation of the metric and its inverse can be completely eliminated using field redefinitions (i.e. $\Delta g_{\mu \nu}=\Delta g^{\mu \nu}=0$), while in the NR limit this is not possible. We will discuss about this important feature of the NR limit of HSZ theory in the section \ref{Sec:Redef}. 

Another interesting point about the way of choosing the fundamental fields of the theory is related to $c_{\mu \nu}$. While at the two derivative level one can choose this field to be \cite{NSNS},
\bea
c_{\mu \nu} = - \tau_{\mu}{}^{a} \tau_{\nu}{}^{b} \epsilon_{a b} \, ,
\label{csugra}
\eea

the inclusion of higher-derivative terms in HSZ implies
\bea
\Delta c_{\mu \nu} = - \Delta \tau_{\mu}{}^{a} \tau_{\nu}{}^{b} \epsilon_{a b} - \tau_{\mu}{}^{a} \Delta \tau_{\nu}{}^{b} \epsilon_{a b} \, ,
\eea
which is not satisfied by the $c_{\mu \nu}$ field. However, it should be possible to find a field redefinition which allows one to fully write the HSZ theory in vielbein formalism, as in \cite{Marques:2015vua}, for the relativistic case. A subtle point here is that the vielbein formulation of HSZ in the relativistic limit contains a non-trivial transformation for the Lorentz transformation of the generalized metric, instead of a deformation of the generalized diffeomorphism. For this reason, we expect a similar behavior when analyzing the NR limit: the higher-derivative transformation (\ref{transfzeroa})-(\ref{transfzerob}) should turn into non-trivial deformations of the double Lorentz transformations, where the decomposition (\ref{csugra}) should be consistent with the symmetry rules of the formulation.   

\subsection{NR HSZ Lagrangian up to four-derivatives terms}

Our goal is to compute the Lagrangian of HSZ up to four derivatives.  The two-derivative Lagrangian $L^{(2)}$ for HSZ matches the DFT NR Lagrangian, 
\bea
\label{DFTaction}
L^{(2)}_{DFT} & = & \frac54 \partial_{\mu}{h_{\nu \rho}} \partial_{\sigma}{h^{\nu \rho}} h^{\mu \sigma} - \frac14 \partial_{\mu}{c_{\nu \rho}} \partial_{\sigma}{c_{\gamma \epsilon}} \tau^{\nu \gamma} \tau^{\rho \epsilon} h^{\mu \sigma} - \frac12 \partial_{\mu}{\tau^{\nu \rho}} \partial_{\sigma}{c_{\nu \gamma}} \tau^{\gamma \epsilon} c_{\rho \epsilon} h^{\mu \sigma} \nn \\ && - \frac14 \partial_{\mu}{\tau^{\nu \rho}} \partial_{\sigma}{\tau^{\gamma \epsilon}} c_{\nu \gamma} c_{\rho \epsilon} h^{\mu \sigma} - \frac12 \partial_{\mu}{h_{\nu \rho}} \partial_{\sigma}{h^{\mu \nu}} h^{\rho \sigma} - \frac12 \partial_{\mu}{c_{\nu \rho}} \partial_{\sigma}{h^{\mu \nu}} \tau^{\rho \gamma} \tau^{\sigma \epsilon} c_{\gamma \epsilon} \nn \\ && + \frac12 \partial_{\mu}{\tau^{\nu \rho}} \partial_{\sigma}{h^{\mu \gamma}} \tau^{\sigma \epsilon} c_{\nu \gamma} c_{\rho \epsilon} - \frac12 \partial_{\mu}{c_{\nu \rho}} \partial_{\sigma}{c_{\gamma \epsilon}} \tau^{\mu \gamma} \tau^{\nu \epsilon} h^{\rho \sigma} - \frac12 \partial_{\mu}{\tau^{\nu \rho}} \partial_{\nu}{c_{\sigma \gamma}} \tau^{\sigma \epsilon} c_{\rho \epsilon} h^{\mu \gamma} \nn \\ && - \frac12 \partial_{\mu}{c_{\nu \rho}} \partial_{\sigma}{h^{\nu \gamma}} \tau^{\mu \epsilon} \tau^{\rho \sigma} c_{\gamma \epsilon} + 4 \partial_{\mu \nu}{\varphi} h^{\mu \nu} + \partial_{\mu \nu}{h^{\rho \sigma}} h_{\rho \sigma} h^{\mu \nu} \nn \\ && + \partial_{\mu \nu}{\tau^{\rho \sigma}} \tau_{\rho \sigma} h^{\mu \nu} + \partial_{\mu}{\tau_{\nu \rho}} \partial_{\sigma}{\tau^{\nu \rho}} h^{\mu \sigma} + 4 \partial_{\mu}{\varphi} \partial_{\nu}{h^{\mu \nu}} + \partial_{\mu}{h^{\mu \nu}} \partial_{\nu}{h^{\rho \sigma}} h_{\rho \sigma} + \partial_{\mu}{\tau^{\nu \rho}} \partial_{\sigma}{h^{\mu \sigma}} \tau_{\nu \rho} \nn \\ && - 4 \partial_{\mu}{\varphi} \partial_{\nu}{\varphi} h^{\mu \nu} - 2 \partial_{\mu}{\varphi} \partial_{\nu}{h^{\rho \sigma}} h_{\rho \sigma} h^{\mu \nu} - 2 \partial_{\mu}{\varphi} \partial_{\nu}{\tau^{\rho \sigma}} \tau_{\rho \sigma} h^{\mu \nu} - \frac14 \partial_{\mu}{h^{\nu \rho}} \partial_{\sigma}{h^{\gamma \epsilon}} h_{\nu \rho} h_{\gamma \epsilon} h^{\mu \sigma} 
\nn \\ && - \frac12 \partial_{\mu}{\tau^{\nu \rho}} \partial_{\sigma}{h^{\gamma \epsilon}} \tau_{\nu \rho} h_{\gamma \epsilon} h^{\mu \sigma} - \frac14 \partial_{\mu}{\tau^{\nu \rho}} \partial_{\sigma}{\tau^{\gamma \epsilon}} \tau_{\nu \rho} \tau_{\gamma \epsilon} h^{\mu \sigma} - \partial_{\mu \nu}{h^{\mu \nu}} 
\nn \\ &&
+ \frac32 h_{\mu \nu \rho} \partial_{\sigma}{h^{\nu \gamma}} \tau^{\rho \epsilon} c_{\gamma \epsilon} h^{\mu \sigma} - \frac14 h_{\mu \nu \rho} h_{\sigma \gamma \epsilon} h^{\mu \sigma} h^{\nu \gamma} h^{\rho \epsilon} - \frac32 h_{\mu \nu \rho} \partial_{\sigma}{c_{\gamma \epsilon}} \tau^{\nu \gamma} h^{\mu \sigma} h^{\rho \epsilon} \, ,
\eea
where $h_{\mu \nu \rho} = 3 \partial_{[\mu}b_{\nu \rho]}$ is the curvature of the b-field. 

The four-derivative corrections in the setup under consideration are given by the following contributions,
\bea
L^{(4)}_{\textrm{NR-HSZ}} & = & - \frac12 \partial^{N Q}{\mathcal{F}^{(0)}_{N Q}} + 2 \partial^{N}{\mathcal{F}^{(0)}_{N}{}^{Q}} \partial_{Q}{d} - 2 \mathcal{F}^{(0)P Q} \partial_{P}{d} \partial_{Q}{d} - \frac12 \mathcal{F}^{(0)P I} \mathcal{F}^{(0)}_{P}{}^{J} \mathcal{H}^{(0)}_{I J} \nn \\ && +
\mathcal{F}^{(0)P I} \mathcal{H}^{(0)}_{P}{}^{J} \partial_{J}{}^{L}{\mathcal{H}^{(0)}_{I L}} - \frac{1}{12} \mathcal{F}^{(0)I J} \mathcal{H}^{(0)K L} \partial_{K L}{\mathcal{H}^{(0)}_{I J}} + \frac12 \mathcal{F}^{(0)P Q} \partial^{J}{\mathcal{H}^{(0)}_{P J}} \partial^{L}{\mathcal{H}^{(0)}_{Q L}}  \nn \\ && 
- \frac{1}{12} \mathcal{F}^{(0)P Q} \partial^{J}{\mathcal{H}^{(0)}_{P Q}} \partial^{L}{\mathcal{H}^{(0)}_{J L}} - \frac12 \mathcal{F}^{(0)P Q} \partial_{P}{\mathcal{H}^{(0)J L}} \partial_{J}{\mathcal{H}^{(0)}_{Q L}} + \frac{1}{8} \mathcal{F}^{(0)P Q} \partial_{P}{\mathcal{H}^{(0)J L}} \partial_{Q}{\mathcal{H}^{(0)}_{J L}} \nn \\ && 
- \frac{1}{12} \mathcal{H}^{(0) I J} \mathcal{H}^{(0) K L} \partial_{I J}{\mathcal{F}^{(0)}_{K L}} + \frac12 \mathcal{H}^{(0)I J} \mathcal{H}^{(0)K L} \partial_{I K}{\mathcal{F}^{(0)}_{J L}} + \frac12 \mathcal{H}^{(0) P Q} \partial^{J}{\mathcal{F}^{(0)}_{P}{}^{L}} \partial_{Q}{\mathcal{H}^{(0)}_{J L}} \nn \\ && 
- \frac{1}{12} \mathcal{H}^{(0) P Q} \partial^{J}{\mathcal{F}^{(0)}_{P Q}} \partial^{L}{\mathcal{H}^{(0)}_{J L}} 
- \frac12 \mathcal{H}^{(0)P Q} \partial_{P}{\mathcal{F}^{(0)J L}} \partial_{J}{\mathcal{H}^{(0)}_{Q L}} 
+ \frac{1}{12} \mathcal{H}^{(0)P Q} \partial_{P}{\mathcal{F}^{(0)J L}} \partial_{Q}{\mathcal{H}^{(0)}_{J L}} \nn \\ && 
+ \mathcal{H}^{(0)P Q} \partial_{P}{\mathcal{F}^{(0)}_{Q}{}^{J}} \partial^{L}{\mathcal{H}^{(0)}_{J L}} - 2 \mathcal{F}^{(0)P I} \mathcal{H}^{(0)}_{P}{}^{K} \mathcal{H}^{(0)}_{I}{}^{L} \partial_{K L}{d} - 2 \mathcal{F}^{(0)P I} \mathcal{H}^{(0)}_{P}{}^{J} \partial_{J}{\mathcal{H}^{(0)}_{I}{}^{L}} \partial_{L}{d} \nn \\ && 
- 2 \mathcal{F}^{(0)P I} \mathcal{H}^{(0)}_{P}{}^{J} \partial^{L}{\mathcal{H}^{(0)}_{I L}} \partial_{J}{d} + \frac{1}{6} \mathcal{F}^{(0)I J} \mathcal{H}^{(0)K L} \partial_{K}{\mathcal{H}^{(0)}_{I J}} \partial_{L}{d} - 2 \mathcal{H}^{(0)I J} \mathcal{H}^{(0)K L} \partial_{I}{\mathcal{F}^{(0)(0)}_{J K}} \partial_{L}{d} \nn \\ && 
+ \frac{1}{6} \mathcal{H}^{(0)I J} \mathcal{H}^{(0)K L} \partial_{I}{\mathcal{F}^{(0)}_{K L}} \partial_{J}{d} + 2 \mathcal{F}^{(0)P I} \mathcal{H}^{(0)}_{P}{}^{K} \mathcal{H}^{(0)}_{I}{}^{L} \partial_{K}{d} \partial_{L}{d} + \frac12 \mathcal{F}^{(0)P Q} \partial^{J}{\mathcal{H}^{(0)}_{P}{}^{L}} \partial_{L}{\mathcal{H}^{(0)}_{Q J}} \nn \\ && 
+ 2 \mathcal{F}^{(0)P Q} \partial_{P Q}{d}\, .
\label{piece1} 
\eea
After parametrization, the full $L^{(4)}_{\textrm{NR-HSZ}}$ Lagrangian can be schematically written as
\bea
L^{(4)}_{\textrm{NR-HSZ}} & = & \partial_{\mu} b_{\nu \rho} T^{\mu \nu \rho}(h,\tau) + \partial_{\mu} \partial_{\nu} b_{\rho \sigma} T^{\mu \nu \rho \sigma}(h,\tau) 
+ \partial_{\mu} \partial_{\nu} \partial_{\rho} b_{\sigma \gamma} T^{\mu \nu \rho \sigma \gamma}(h,\tau) \nn \\ && + \partial_{\mu} b_{\nu \rho} \partial_{\sigma} b_{\gamma \epsilon} \partial_{\lambda} b_{\alpha \beta} T^{\mu \nu \rho \sigma \gamma \epsilon \lambda \alpha \beta}(h,\tau) + L(c) \, ,  
\eea
where $L(c)$ are the c-field contributions of the Lagrangian. The explicit computation of the full Lagrangian is a hard computational challenge, due to the extension of it and the lack of well-defined curvatures for the anomalous transformations of all the fields of the theory. We report in the appendix C the contributions given by $T^{\mu \nu \rho}(h.\tau)$, $T^{\mu \nu \rho \sigma}(h,\tau)$, $T^{\mu \nu \rho \sigma \gamma}(h,\tau)$ and $T^{\mu \nu \rho \sigma \gamma \epsilon \lambda \alpha \beta}(h,\tau)$. With a suitable rewriting of these contributions plus partial integration, we expect higher-derivative corrections in the curvature of the b-field, and similar higher-order curvatures for the remanent fields.   

\section{Field redefinitions {\it before} taking the NR limits}
\label{Sec:Redef}
In the relativistic case, the supergravity limit of HSZ up to four-derivatives is given by Chern-Simons terms $\Omega_{\mu \nu \rho}$ that correct the three-form field strength of the Kalb-Ramond field \cite{Hohm:2014eba}
 \be
\tilde H_{\mu \nu \rho} = 3\, \partial_{[\mu} \tilde B_{\nu \rho]} + 3 \tilde \Omega_{\mu \nu \rho} \ , \ \ \ \ \tilde \Omega_{\mu \nu \rho} = \tilde \Gamma_{[\mu| \sigma}^\delta \partial_{|\nu} \tilde \Gamma_{\rho] \delta}^\sigma + \frac 2 3
\tilde \Gamma_{[\mu | \sigma}^\delta \tilde \Gamma_{|\nu| \lambda}^\sigma \tilde \Gamma_{|\rho] \delta}^\lambda \ . \label{Hhat}
\ee

The state of the art is that to first order, the gauge invariant action of HSZ theory is the following
\be
S = \int d^Dx \sqrt{-\tilde g} e^{-2 \tilde \phi}\left(\tilde R + 4 (\nabla \tilde \phi)^2  - \frac 1 {12} \tilde H^2 \right) \ , \label{UsmanBartonAction}
\ee
where the $\tilde B$-field has a Green-Schwarz transformation given by
\be
\delta \tilde B_{\mu \nu} = \partial_{[\mu|} \partial_\rho \xi^\sigma \tilde \Gamma_{|\nu]\sigma}^\rho \ . \label{GS}
\ee

As first glance, the NR limit cannot be taken in this Lagrangian since the product $H^{\mu \nu \rho} \Omega_{\mu \nu \rho}$ contains divergences up to $c^6$. Therefore, one can impose a first order non-covariant field redefinition {\it before} taking the NR limit. 

One possible choice for the family of field redefinitions that makes HSZ theory convergent at the supergravity limit up to four-derivative contributions are
\bea
\tilde g_{\mu \nu} &=& \hat g_{\mu \nu} - \Delta_{1} g_{\mu \nu} \, , \label{DeltaG}\\
\tilde B_{\mu \nu} &=& \hat B_{\mu \nu} - \Delta_{1} B_{\mu \nu}  \label{DeltaB} , \\
\tilde \phi &=& \hat \phi - \Delta_{1} \phi \, ,
\label{DeltaP}
\eea
where
\bea
\Delta_1 g_{\mu \nu} &=&  - \frac{1}{4}\, {\partial}_{\mu}{{\hat B}_{\rho \sigma}}\,  {\partial}_{\gamma}{{\hat g}_{\nu \epsilon}}\,  {\hat g}^{\rho \gamma} {\hat g}^{\sigma \epsilon} - \frac{1}{4}\, {\partial}_{\rho}{{\hat B}_{\mu \sigma}}\,  {\partial}_{\gamma}{{\hat g}_{\nu \epsilon}}\,  {\hat g}^{\rho \epsilon} {\hat g}^{\sigma \gamma} \nn \\ && + \frac{1}{4}\, {\partial}_{\rho}{{\hat B}_{\mu \sigma}}\,  {\partial}_{\gamma}{{\hat g}_{\nu \epsilon}}\,  {\hat g}^{\rho \gamma} {\hat g}^{\sigma \epsilon} - \frac{1}{4}\, {\partial}_{\rho}{{\hat B}_{\mu \sigma}}\,  {\partial}_{\nu}{{\hat g}_{\gamma \epsilon}}\,  {\hat g}^{\rho \gamma} {\hat g}^{\sigma \epsilon} + (\mu \leftrightarrow \nu) \, , \nn\\
\Delta_1 B_{\mu \nu} &=& \frac{1}{4}\, {\partial}_{\mu}{{\hat g}_{\rho \sigma}}\,  {\partial}_{\gamma}{{\hat g}_{\nu \epsilon}}\,  {\hat g}^{\rho \gamma} {\hat g}^{\sigma \epsilon} - (\mu \leftrightarrow \nu) \ , \\
\Delta_{1} \hat \phi & = & - \frac 1 4 \log \frac {\Delta_{1} \hat g}{\hat g} \, . 
\label{FieldRedef}
\eea
Considering the previous field redefinitions, the supergravity limit of HSZ theory coincides with the results coming from a double geometry in the generalized metric approach and, therefore, its Lagrangian converges. According to the results in \cite{EandDiego}, we expect the same behavior at the next order in derivatives (six-derivative case). We summarize the previous statements in the following diagram,

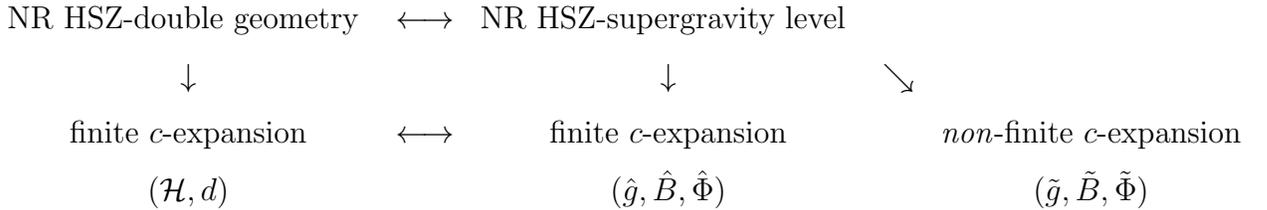
\begin{align}
    \begin{array}{ccccc}
      \textrm{NR HSZ-double geometry }   &  \longleftrightarrow & \textrm{NR HSZ-supergravity level }& & \\
       \downarrow  & & \downarrow & \searrow & \\
       \text{finite $c$-expansion} &  \longleftrightarrow & \text{finite $c$-expansion} & & \text{\textit{non}-finite $c$-expansion} \\
        ({\mathcal{H}},{d}) & & (\hat{g},\hat{B},\hat{\Phi}) & & (\tilde{g},\tilde{B},\tilde{\Phi})
    \end{array} \nonumber
\end{align}
The results of this work indicate that including higher-derivative corrections in bosonic, heterotic or type II string theory might need field redefinitions before taking the NR limit. The simplest case of study could be the bosonic or heterotic string, which contain four-derivative terms in their $\alpha'$ expansion. For this reason, we will discuss this case in the next section. 

\section{Towards the $\alpha'$-corrections in NR heterotic and bosonic string}
\label{Sec:Towards}
One of the most intriguing aspects of the HSZ theory lies in its structure as an order-by-order interpolation between the heterotic and bosonic string theories, truncated by a discrete $\mathbb{Z}_2$ symmetry transformation, as described in \cite{EandDiego}. This interpolation offers a unified higher-derivative framework that captures features of both string theories within a single formalism, which also is T-duality invariant. At the four-derivative level, the theory incorporates the $Z_2$ odd terms of heterotic string theory and a Green–Schwarz mechanism for generalized diffeomorphisms. Importantly, in the NR limit, this leads to the specific form of the corrections of the symmetry transformations of the degrees of freedom of the theory, presented in equations (\ref{transfzeroa})–(\ref{transfzerob}). The higher-derivative corrections in the boost transformations of the new redefined fields $\tilde h_{\mu \nu}$, $\tilde \tau^{\mu \nu}$ and $\tilde b_{\mu \nu}$ are a distinctive feature from the HSZ formulation in its NR limit.     
When the gauge field of the heterotic supergravity is included by considering a Yang-Mills interaction in the Lagrangian and Chern-Simons terms, the NR limit admits two distinct $c$-expansions, both of which are constructed to be invariant under T-duality. On the one hand, one can adopt a direct expansion of the fundamental fields—namely, the metric, the $B$-field, and a suitably chosen expansion of the non-Abelian gauge field $A_{\mu}{}^{i}$. This method was developed in \cite{BRomano}, where the corresponding Buscher rules \cite{Buscher} were also derived. While this formulation is consistent and captures essential features of the NR dynamics, it lacks a manifest T-duality invariance in terms of a DFT metric \footnote{Relaxing the constraint ${\cal H}^2=1$ could be a potential solution to incorporate this formulation in a relaxed form of DFT.}. A similar behavior appears in all the formulations of string theory at order ${\alpha'}^3$, in the relativistic case, as discussed in \cite{Wulff}. Apparently, a similar problem appears at the two-derivative level for the NR heterotic string.  

To overcome this limitation, a second formulation was introduced in \cite{EandD}, in which the fields are embedded directly in an $O(D,D+N)$ generalized metric. This embedding requires a specific expansion of the $A$-field that is compatible with the constraint ${\cal H}^2=1$ of DFT, thereby preserving the fact that the generalized metric is an element of the duality group. This approach was further developed in \cite{Enew}, where the explicit form of the heterotic supergravity was inspected. The two formulations are expected to be related through field redefinitions at the level of two-derivative supergravity and none of them seems to be suited for systematically including higher-derivative corrections. In particular, the presence of Riem$^2$ terms in the four-derivative Lagrangian seems to indicate divergences that will require field redefinitions in both approaches. 

For the second-order corrections, a new study of the HSZ theory to that order could be very useful. HSZ theory contains all the $\mathbb{Z}_2$-even contributions of the bosonic string in the relativistic setting, as explicitly demonstrated in \cite{EandDiego}. Extending the analysis of this paper to include six-derivative corrections would be a natural next step, potentially allowing for the construction of a consistent truncation of the NR limit of the bosonic string. Such an extension would deepen our understanding of the structure of NR string theories at higher orders.

In addition to these theoretical implications, our findings could have applications for exploring cosmological solutions within higher-derivative contributions. In particular, these results are useful in the study of non-relativistic cosmologies, where NR limits of string theory could provide a simplified framework to test string dualities as in \cite{DOT}, upon exploring the treatment of matter with higher-derivative terms \cite{matter} in the NR limit.

In this work, we have focused on the NR formulation of the HSZ theory within the generalized metric formalism, consistent with the original framework introduced by Hohm, Siegel, and Zwiebach. However, it is important to emphasize that HSZ theory also admits an equivalent description in terms of generalized fluxes, where the generalized metric transforms as a tensor under generalized diffeomorphisms but transforms non-trivially under double Lorentz transformations \cite{ReviewHSZ}. In the relativistic case, these two formulations (metric and generalized flux) are equivalent, up to an identification of parameters and field redefinitions. Specifically, a suitable identification of the double Lorentz parameter with the generalized diffeomorphism parameter, together with a redefinition of the generalized metric should map one formulation into the other.

However, whether this equivalence persists in the NR limit remains an open and interesting question. It is likely that both approaches may yield inequivalent NR limits already at the four-derivative level. This potential non-equivalence could signal the existence of multiple consistent NR limits of heterotic and bosonic string theories, each with distinct symmetry properties: a generalized metric formulation with deformed generalized diffeomorphisms and a generalized flux formulation with deformed Lorentz transformations. This represents a profound insight into the landscape of non-relativistic string theories and their effective descriptions.

\section{Conclusions} \label{Sec:Conclusions}

HSZ theory is a higher-derivative theory of gravity with exact and manifest T-duality invariance. Constructed within the framework of the double geometry, HSZ can be formulated in terms of a generalized metric and a generalized dilaton. Since both fields have a convergent NR expansion \cite{EandD}, this ensures that the HSZ Lagrangian remains finite under a non-relativistic (NR) expansion of its degrees of freedom.  

When we write the theory in the generalized metric approach, the symmetry transformations in HSZ theory arise from a deformed version of generalized diffeomorphisms. The generalized Lie derivative acting on the generalized metric receives a Green–Schwarz–like correction, and consequently, the transformations of its components induce higher-derivative corrections to the symmetry transformations of the fundamental fields in the NR limit, as seen in (\ref{transfzeroa})–(\ref{transfzerob}). While the contributions with $\zeta_{\mu}$ can be eliminated from $\Delta \tau^{\mu \nu}$, $\Delta \tau_{\mu \nu}$, $\Delta h^{\mu \nu}$, $\Delta h_{\mu \nu}$ and $\Delta b_{\mu \nu}$, part of their corrections to the diffeomorphism transformations cannot be removed and become unambiguous in this limit and $\tilde h_{\mu \nu}$, $\tilde \tau^{\mu \nu}$ and $\tilde b_{\mu \nu}$ acquire new higher-derivative boost corrections.   

In the relativistic case, the transformation of the metric tensor can be fully trivialized through field redefinitions, while the transformation of the $B$-field retains an ambiguous term given by (\ref{GS}). However, applying the NR limit directly to this configuration leads to divergences in the effective Lagrangian, necessitating appropriate field redefinitions before taking the NR limit. This provides a key lesson for constructing higher-derivative structures in bosonic or heterotic string theory from a supergravity perspective: while terms like Riem$^2$ may introduce divergences, field redefinitions can play a crucial role in curing potential divergences of the higher-derivative theory.

In this work, we also constructed the four-derivative correction to the action (\ref{DFTaction}), which directly depends on $b_{\mu \nu}$. Since HSZ theory interpolates, order by order, between the heterotic and bosonic strings \cite{EandDiego}, the corrected NR action constructed in this paper contains part of the four-derivative structure of heterotic string theory in the NR regime. 

\noindent {\bf \underline{Acknowledgments:}} The author is very grateful to D. Marques for discussions. This work is supported by the SONATA BIS grant 2021/42/E/ST2/00304 from the National Science Centre (NCN), Poland.

\appendix
\section{HSZ four-derivative Lagrangian in terms of the generalized metric}
\label{AppA}
The compact Lagrangian (\ref{LagHSZ}) can be written explicitly in terms of the generalized metric as
\bea
L^{(4)}_{\textrm{HSZ}} & = & \frac{1}{12} {\cal H}^{P Q} \partial^{J}{{\cal H}^{L S}} \partial_{P Q L}{{\cal H}_{J S}} - \frac{1}{4} {\cal H}^{P Q} \partial_{P}{{\cal H}^{J L}} \partial_{Q J}{}^{S}{{\cal H}_{L S}} - \frac{1}{24} {\cal H}^{P Q} \partial_{P}{}^{J}{{\cal H}^{L S}} \partial_{Q L}{{\cal H}_{J S}} \nn \\ && + \frac{1}{12} \partial^{N}{{\cal H}_{N}{}^{Q}} \partial^{J}{{\cal H}^{L S}} \partial_{Q L}{{\cal H}_{J S}} - \frac{3}{8} \partial^{N}{{\cal H}^{Q J}} \partial_{Q}{{\cal H}^{L S}} \partial_{J L}{{\cal H}_{N S}} - \frac{1}{4} \partial^{N}{{\cal H}^{Q J}} \partial_{Q}{{\cal H}_{N}{}^{L}} \partial_{J}{}^{S}{{\cal H}_{L S}} \nn \\ && - \frac{1}{6} {\cal H}^{P Q} \partial^{J}{{\cal H}^{L S}} \partial_{P}{d} \partial_{Q L}{{\cal H}_{J S}} + \frac{1}{2} {\cal H}^{P Q} \partial_{P}{{\cal H}^{J L}} \partial_{J}{d} \partial_{Q}{}^{S}{{\cal H}_{L S}} \nn \\ &&  + \frac{1}{4} {\cal H}^{K L} {\cal H}^{R S} {\cal H}^{T U} \partial_{K}{{\cal H}_{R}{}^{W}} \partial_{L S T}{{\cal H}_{U W}} + \frac{1}{8} {\cal H}^{K L} {\cal H}^{R S} {\cal H}^{T U} \partial_{K R}{{\cal H}_{T}{}^{W}} \partial_{L U}{{\cal H}_{S W}} \nn \\ && + \frac{1}{8} {\cal H}^{I J} {\cal H}^{K L} \partial_{I}{{\cal H}_{J}{}^{S}} \partial_{K}{{\cal H}_{S}{}^{U}} \partial_{L}{}^{W}{{\cal H}_{U W}} - \frac{1}{4} {\cal H}^{I J} {\cal H}^{K L} \partial_{I}{{\cal H}_{J}{}^{S}} \partial_{K}{{\cal H}^{U W}} \partial_{L U}{{\cal H}_{S W}} \nn \\ && - \frac{1}{4} {\cal H}^{I J} {\cal H}^{K L} \partial_{I}{{\cal H}_{J}{}^{S}} \partial^{U}{{\cal H}_{S}{}^{W}} \partial_{K L}{{\cal H}_{U W}} + \frac{1}{4} {\cal H}^{I J} {\cal H}^{K L} \partial_{I}{{\cal H}_{J}{}^{S}} \partial^{U}{{\cal H}_{S}{}^{W}} \partial_{K W}{{\cal H}_{L U}} \nn \\ && - \frac{1}{4} {\cal H}^{I J} {\cal H}^{K L} \partial_{I}{{\cal H}_{K}{}^{S}} \partial_{J}{{\cal H}_{L}{}^{U}} \partial_{S}{}^{W}{{\cal H}_{U W}} + \frac{1}{8} {\cal H}^{I J} {\cal H}^{K L} \partial_{I}{{\cal H}_{K}{}^{S}} \partial_{L}{{\cal H}^{U W}} \partial_{J U}{{\cal H}_{S W}} \nn \\ && - \frac{1}{8} {\cal H}^{I J} {\cal H}^{K L} \partial_{I}{{\cal H}_{K}{}^{S}} \partial_{L}{{\cal H}^{U W}} \partial_{S U}{{\cal H}_{J W}} + \frac{1}{8} {\cal H}^{I J} {\cal H}^{K L} \partial_{I}{{\cal H}_{K}{}^{S}} \partial^{U}{{\cal H}_{J}{}^{W}} \partial_{L W}{{\cal H}_{S U}} \nn \\ && + \frac{1}{12} {\cal H}^{I J} {\cal H}^{K L} \partial_{I}{{\cal H}_{K}{}^{S}} \partial^{U}{{\cal H}_{L}^{W}} \partial_{J W}{{\cal H}_{S U}} + \frac{1}{4} {\cal H}^{I J} {\cal H}^{K L} \partial_{I}{{\cal H}_{K}{}^{S}} \partial^{U}{{\cal H}_{L}{}^{W}} \partial_{S W}{{\cal H}_{J U}} \nn \\ && + \frac{1}{48} {\cal H}^{I J} {\cal H}^{K L} \partial_{I}{{\cal H}^{S U}} \partial_{J}{{\cal H}_{S}{}^{W}} \partial_{K L}{{\cal H}_{U W}} + \frac{1}{6} {\cal H}^{I J} {\cal H}^{K L} \partial_{I}{{\cal H}^{S U}} \partial_{K}{{\cal H}_{S}{}^{W}} \partial_{J L}{{\cal H}_{U W}} \nn \\ && + \frac{1}{6} {\cal H}^{I J} {\cal H}^{K L} \partial_{I}{{\cal H}^{S U}} \partial_{K}{{\cal H}_{S}{}^{W}} \partial_{J W}{{\cal H}_{L U}} + \frac{1}{4} {\cal H}^{I J} {\cal H}^{K L} \partial_{I}{{\cal H}^{S U}} \partial_{S}{{\cal H}_{J}{}^{W}} \partial_{K U}{{\cal H}_{L W}} \nn \\ && - \frac{1}{8} {\cal H}^{I J} {\cal H}^{K L} \partial_{I}{{\cal H}^{S U}} \partial_{S}{{\cal H}_{K}{}^{W}} \partial_{J L}{{\cal H}_{U W}} + \frac{3}{8} {\cal H}^{I J} {\cal H}^{K L} \partial_{I}{{\cal H}^{S U}} \partial_{S}{{\cal H}_{K}{}^{W}} \partial_{L U}{{\cal H}_{J W}} \nn \\ && - \frac{1}{4} {\cal H}^{I J} {\cal H}^{K L} \partial_{I}{{\cal H}^{S U}} \partial_{S}{{\cal H}_{U}{}^{W}} \partial_{J K}{{\cal H}_{L W}} + \frac{1}{8} {\cal H}^{I J} {\cal H}^{K L} \partial_{I}{{\cal H}^{S U}} \partial^{W}{{\cal H}_{S W}} \partial_{J K}{{\cal H}_{L U}} \nn \\ && - \frac{1}{4} {\cal H}^{I J} {\cal H}^{K L} \partial^{S}{{\cal H}_{I S}} \partial^{U}{{\cal H}_{J}{}^{W}} \partial_{K L}{{\cal H}_{U W}} + \frac{1}{4} {\cal H}^{I J} {\cal H}^{K L} \partial^{S}{{\cal H}_{I S}} \partial^{U}{{\cal H}_{J}{}^{W}} \partial_{K W}{{\cal H}_{L U}} \nn \\ && - \frac{1}{12} {\cal H}^{P Q} \partial^{J}{{\cal H}_{P}{}^{L}} \partial_{L}{{\cal H}_{Q}{}^{S}} \partial^{U}{{\cal H}_{J S}} \partial^{W}{{\cal H}_{U W}} + \frac{1}{4} {\cal H}^{P Q} \partial^{J}{{\cal H}_{P}{}^{L}} \partial_{L}{{\cal H}_{Q}{}^{S}} \partial^{U}{{\cal H}_{J U}} \partial^{W}{{\cal H}_{S W}} \nn \\ && + \frac{1}{8} {\cal H}^{P Q} \partial^{J}{{\cal H}_{P}{}^{L}} \partial_{L}{{\cal H}_{Q}{}^{S}} \partial^{U}{{\cal H}_{J}{}^{W}} \partial_{W}{{\cal H}_{S U}} + \frac{1}{8} {\cal H}^{P Q} \partial_{P}{{\cal H}^{J L}} \partial_{J}{{\cal H}_{L}{}^{S}} \partial_{S}{{\cal H}^{U W}} \partial_{U}{{\cal H}_{Q W}} \nn \\ && + \frac{1}{12} {\cal H}^{P Q} \partial_{P}{{\cal H}^{J L}} \partial_{J}{{\cal H}_{L}{}^{S}} \partial^{U}{{\cal H}_{Q S}} \partial^{W}{{\cal H}_{U W}} + \frac{1}{8} {\cal H}^{P Q} \partial_{P}{{\cal H}^{J L}} \partial_{Q}{{\cal H}^{S U}} \partial_{J}{{\cal H}_{L S}} \partial^{W}{{\cal H}_{U W}} \nn \\ && - \frac{1}{16} {\cal H}^{P Q} \partial_{P}{{\cal H}^{J L}} \partial_{Q}{{\cal H}^{S U}} \partial_{J}{{\cal H}_{L}{}^{W}} \partial_{W}{{\cal H}_{S U}} + \frac{1}{4} {\cal H}^{P Q} \partial_{P}{{\cal H}^{J L}} \partial_{J}{{\cal H}_{Q}{}^{S}} \partial_{L}{{\cal H}_{S}{}^{U}} \partial^{W}{{\cal H}_{U W}} \nn \\ && - \frac{1}{16} {\cal H}^{P Q} \partial_{P}{{\cal H}^{J L}} \partial_{J}{{\cal H}_{Q}{}^{S}} \partial_{L}{{\cal H}^{U W}} \partial_{S}{{\cal H}_{U W}} + \frac{3}{8} {\cal H}^{P Q} \partial_{P}{{\cal H}^{J L}} \partial_{J}{{\cal H}_{Q}{}^{S}} \partial_{S}{{\cal H}^{U W}} \partial_{U}{{\cal H}_{L W}} \nn 
\eea
\bea
&& - \frac{1}{4} {\cal H}^{P Q} \partial_{P}{{\cal H}^{J L}} \partial_{J}{{\cal H}_{Q}{}^{S}} \partial^{U}{{\cal H}_{L U}} \partial^{W}{{\cal H}_{S W}} - \frac{1}{4} {\cal H}^{P Q} \partial_{P}{{\cal H}^{J L}} \partial_{J}{{\cal H}_{Q}{}^{S}} \partial^{U}{{\cal H}_{L}{}^{W}} \partial_{W}{{\cal H}_{S U}} \nn \\ && - \frac{1}{8} {\cal H}^{P Q} \partial_{P}{{\cal H}^{J L}} \partial_{J}{{\cal H}^{S U}} \partial_{L}{{\cal H}_{S}{}^{W}} \partial_{U}{{\cal H}_{Q W}} - \frac{1}{8} {\cal H}^{P Q} \partial_{P}{{\cal H}^{J L}} \partial_{J}{{\cal H}^{S U}} \partial_{S}{{\cal H}_{L}{}^{W}} \partial_{W}{{\cal H}_{Q U}}
\nn \\ && + \frac{1}{4} {\cal H}^{P Q} \partial_{P}{{\cal H}^{J L}} \partial_{J}{{\cal H}^{S U}} \partial_{S}{{\cal H}_{Q}{}^{W}} \partial_{U}{{\cal H}_{L W}} - \frac{1}{8} {\cal H}^{P Q} \partial_{P}{{\cal H}^{J L}} \partial_{J}{{\cal H}^{S U}} \partial_{S}{{\cal H}_{Q}{}^{W}} \partial_{W}{{\cal H}_{L U}} \nn \\ && + \frac{1}{8} {\cal H}^{P Q} \partial_{P}{{\cal H}^{J L}} \partial_{J}{{\cal H}^{S U}} \partial_{S}{{\cal H}_{U}{}^{W}} \partial_{W}{{\cal H}_{Q L}} + \frac{5}{48} {\cal H}^{P Q} \partial_{P}{{\cal H}^{J L}} \partial_{Q}{{\cal H}^{S U}} \partial_{J}{{\cal H}_{S}{}^{W}} \partial_{L}{{\cal H}_{U W}} \nn \\ && + \frac{1}{24} {\cal H}^{P Q} \partial_{P}{{\cal H}^{J L}} \partial_{Q}{{\cal H}^{S U}} \partial_{J}{{\cal H}_{S}{}^{W}} \partial_{W}{{\cal H}_{L U}} + \frac{1}{4} {\cal H}^{P Q} \partial_{P}{{\cal H}_{Q}{}^{J}} \partial^{L}{{\cal H}_{J}{}^{S}} \partial_{S}{{\cal H}_{L}{}^{U}} \partial^{W}{{\cal H}_{U W}} \nn \\ && - \frac{1}{4} {\cal H}^{P Q} \partial_{P}{{\cal H}_{Q}{}^{J}} \partial^{L}{{\cal H}_{J}{}^{S}} \partial^{U}{{\cal H}_{L S}} \partial^{W}{{\cal H}_{U W}} + \frac{1}{4} {\cal H}^{P Q} \partial_{P}{{\cal H}^{J L}} \partial_{Q}{{\cal H}_{J}{}^{S}} \partial_{L}{{\cal H}_{S}{}^{U}} \partial^{W}{{\cal H}_{U W}} \nn \\ && + \frac{1}{32} {\cal H}^{P Q} \partial_{P}{{\cal H}^{J L}} \partial_{Q}{{\cal H}_{J}{}^{S}} \partial_{L}{{\cal H}^{U W}} \partial_{S}{{\cal H}_{U W}} - \frac{1}{4} {\cal H}^{P Q} \partial_{P}{{\cal H}^{J L}} \partial_{Q}{{\cal H}_{J}{}^{S}} \partial_{L}{{\cal H}^{U W}} \partial_{U}{{\cal H}_{S W}} \nn \\ && + \frac{1}{48} {\cal H}^{P Q} \partial_{P}{{\cal H}^{J L}} \partial_{Q}{{\cal H}_{J}{}^{S}} \partial^{U}{{\cal H}_{L S}} \partial^{W}{{\cal H}_{U W}} + \frac{1}{8} {\cal H}^{P Q} \partial_{P}{{\cal H}^{J L}} \partial_{Q}{{\cal H}_{J}{}^{S}} \partial^{U}{{\cal H}_{L U}} \partial^{W}{{\cal H}_{S W}} \nn \\ && + \frac{11}{48} {\cal H}^{P Q} \partial_{P}{{\cal H}^{J L}} \partial_{Q}{{\cal H}_{J}{}^{S}} \partial^{U}{{\cal H}_{L}{}^{W}} \partial_{W}{{\cal H}_{S U}} - \frac{1}{2} {\cal H}^{K L} {\cal H}^{R S} {\cal H}^{T U} \partial_{K}{{\cal H}_{R}{}^{W}} \partial_{S}{d} \partial_{L T}{{\cal H}_{U W}} \nn \\ && + \frac{1}{2} {\cal H}^{I J} {\cal H}^{K L} \partial_{I}{{\cal H}_{J}{}^{S}} \partial_{K}{{\cal H}_{S}{}^{U}} \partial_{U}{{\cal H}_{L}{}^{W}} \partial_{W}{d} + \frac{1}{2} {\cal H}^{I J} {\cal H}^{K L} \partial_{I}{{\cal H}_{K}{}^{S}} \partial_{J}{{\cal H}_{L}{}^{U}} \partial^{W}{{\cal H}_{S W}} \partial_{U}{d} \nn \\ && + \frac{1}{2} {\cal H}^{I J} {\cal H}^{K L} \partial_{I}{{\cal H}_{K}{}^{S}} \partial_{J}{{\cal H}^{U W}} \partial_{L}{{\cal H}_{S U}} \partial_{W}{d} - \frac{1}{2} {\cal H}^{I J} {\cal H}^{K L} \partial_{I}{{\cal H}_{K}{}^{S}} \partial_{J}{{\cal H}^{U W}} \partial_{S}{{\cal H}_{L U}} \partial_{W}{d} \nn \\ && - \frac{1}{2} {\cal H}^{I J} {\cal H}^{K L} \partial_{I}{{\cal H}_{K}{}^{S}} \partial_{J}{{\cal H}^{U W}} \partial_{U}{{\cal H}_{L W}} \partial_{S}{d} - \frac{1}{2} {\cal H}^{I J} {\cal H}^{K L} \partial_{I}{{\cal H}_{K}{}^{S}} \partial_{L}{{\cal H}_{J}{}^{U}} \partial_{S}{{\cal H}_{U}{}^{W}} \partial_{W}{d} \nn \\ && - \frac{1}{2} {\cal H}^{I J} {\cal H}^{K L} \partial_{I}{{\cal H}_{K}{}^{S}} \partial_{L}{{\cal H}_{J}{}^{U}} \partial^{W}{{\cal H}_{S W}} \partial_{U}{d} - \frac{1}{6} {\cal H}^{I J} {\cal H}^{K L} \partial_{I}{{\cal H}_{K}{}^{S}} \partial_{L}{{\cal H}^{U W}} \partial_{U}{{\cal H}_{S W}} \partial_{J}{d} \nn \\ && - \frac{1}{6} {\cal H}^{I J} {\cal H}^{K L} \partial_{I}{{\cal H}_{K}{}^{S}} \partial_{S}{{\cal H}^{U W}} \partial_{U}{{\cal H}_{L W}} \partial_{J}{d} + \frac{1}{2} {\cal H}^{I J} {\cal H}^{K L} \partial_{I}{{\cal H}_{K}{}^{S}} \partial^{U}{{\cal H}_{J}{}^{W}} \partial_{W}{{\cal H}_{L U}} \partial_{S}{d} \nn \\ && - \frac{1}{24} {\cal H}^{I J} {\cal H}^{K L} \partial_{I}{{\cal H}^{S U}} \partial_{J}{{\cal H}_{S}{}^{W}} \partial_{K}{{\cal H}_{U W}} \partial_{L}{d} - \frac{1}{6} {\cal H}^{I J} {\cal H}^{K L} \partial_{I}{{\cal H}^{S U}} \partial_{K}{{\cal H}_{S}{}^{W}} \partial_{U}{{\cal H}_{J W}} \partial_{L}{d} \nn \\ && + \frac{1}{32} {\cal H}^{K L} {\cal H}^{R S} {\cal H}^{T U} \partial_{K}{{\cal H}_{R}{}^{W}} \partial_{L}{{\cal H}_{S}{}^{Y}} \partial_{T}{{\cal H}_{W}{}^{P}} \partial_{U}{{\cal H}_{Y P}} \nn \\ && - \frac{1}{8} {\cal H}^{K L} {\cal H}^{R S} {\cal H}^{T U} \partial_{K}{{\cal H}_{R}{}^{W}} \partial_{L}{{\cal H}_{W}{}^{Y}} \partial_{S}{{\cal H}_{T}{}^{P}} \partial_{P}{{\cal H}_{U Y}} \nn \\ && - \frac{1}{16} {\cal H}^{K L} {\cal H}^{R S} {\cal H}^{T U} \partial_{K}{{\cal H}_{R}{}^{W}} \partial_{S}{{\cal H}_{L}{}^{Y}} \partial_{T}{{\cal H}_{W}{}^{P}} \partial_{U}{{\cal H}_{Y P}} 
\eea

\section{Non-covariant $b_{2}^{\alpha \beta}{}_{\omega[\mu \nu]}$}
Here we report a non-covariant piece of the $\delta_{\xi}b_{\mu \nu}$ transformation,
\bea
&& b_{2}^{\alpha \beta}{}_{\omega[\mu \nu]} = \frac14 \partial_{[\nu} h^{\alpha \beta}  h_{\mu] \omega}   
+ \frac14 (b_{\epsilon [\nu|} h^{\epsilon \rho} + c_{\epsilon [\nu|} \tau^{\epsilon \rho}) \Big[ b_{\pi \omega} h^{\pi \chi} \partial_{\rho} h^{\beta \alpha} 
 - 2 h^{\chi \beta} \partial_{\rho}(b_{\pi \omega} h^{\pi \alpha} + c_{\pi \omega} \tau^{\pi \alpha})
\nn \\ && - 2 h^{\chi \beta} \partial_{\omega}(b_{\pi \rho} h^{\pi \alpha} + c_{\pi \rho} \tau^{\pi \alpha}) \Big] h_{\chi |\mu]} - \frac12 (h_{[\nu|\rho} + b_{\epsilon [\nu|} h^{\epsilon \pi} b_{\pi \rho} + 2 c_{\epsilon ([\nu|} \tau^{\epsilon \pi} b_{\pi \rho)}) h^{\chi \beta} \partial_{\omega} h^{\alpha \rho} h_{\chi |\mu]} 
\nn \\ &&
- \frac14 h^{\chi \rho} (h_{[\nu| \omega} + b_{\epsilon [\nu|} h^{\epsilon \pi} b_{\pi \omega} + 2 c_{\epsilon ([\nu|} \tau^{\epsilon \pi} b_{\pi \omega)}) \partial_{\rho} h^{\beta \alpha} h_{\chi \mu]} 
- \frac14 (b_{\epsilon [\nu|} h^{\epsilon \beta} + c_{\epsilon [\nu|} \tau^{\epsilon \beta}) h^{\chi \rho} \partial_{\rho}(b_{\pi \omega} h^{\pi \alpha} + c_{\pi \omega} \tau^{\pi \alpha}) h_{\chi |\mu]} 
 \nn \\ && 
 - \frac12 (b_{\epsilon [\nu|} h^{\epsilon \alpha} + c_{\epsilon [\nu|} \tau^{\epsilon \alpha}) \Big[ - h^{\chi \rho} \partial_{\omega} (b_{\pi \rho} h^{\pi \beta} + c_{\pi \rho} \tau^{\pi \beta})
+ b_{\pi \rho} h^{\pi \chi} \partial_{\omega}h^{\rho \beta} \Big] h_{\chi |\mu]}  
\nn \\ &&
- b_{\psi [\nu|} \Big[ - \frac12 h^{(\psi| \rho} b_{\gamma \omega} h^{\gamma \chi)} \partial_{\rho} h^{\alpha \beta} - \frac14 h^{\chi \rho} c_{\gamma \omega} \tau^{\gamma \psi} \partial_{\rho} h^{\alpha \beta} + \frac12 h^{(\psi| \rho} h^{|\chi) \beta} \partial_{\rho}(b_{\gamma \omega} h^{\gamma \alpha)} + c_{\gamma \omega} \tau^{\gamma \alpha)}) \nn \\ && - h^{(\psi| \rho} h^{|\chi) \alpha} \partial_{\omega}(b_{\gamma \rho} h^{\gamma \beta} + c_{\gamma \rho} \tau^{\gamma \beta})
- b_{\gamma \rho} h^{\gamma (\psi|} h^{\chi) \alpha} \partial_{\omega} h^{\beta \rho} - \frac12 c_{\gamma \rho} \tau^{\gamma \psi} h^{\chi \alpha} \partial_{\omega} h^{\beta \rho} \Big] h_{\chi |\mu]} 
\nn \\ &&
- c_{\psi [\nu|} \Big[  - \frac14 \tau^{\psi \rho} b_{\epsilon \omega} h^{\epsilon \chi} \partial_{\rho}h^{\alpha \beta} - \frac14 h^{\chi \rho} b_{\pi \omega} \tau^{\pi \psi} \partial_{\rho} h^{\alpha \beta} - \frac14 h^{\chi \rho} c_{\epsilon \omega} \tau^{\epsilon \psi} \partial_{\rho} \tau^{\alpha \beta} 
\nn \\ &&
+ \frac14 \tau^{\psi \rho} h^{\chi \beta} \partial_{\rho} (b_{\epsilon \omega} h^{\epsilon \alpha} + c_{\epsilon \omega} \tau^{\epsilon \alpha})  + \frac14 h^{\chi \rho} \tau^{\psi \beta} \partial_{\rho} (b_{\epsilon \omega} h^{\epsilon \alpha} + c_{\epsilon \omega} \tau^{\epsilon \alpha}) \nn \\ &&  
- \frac12 \tau^{\psi \rho} h^{\chi \alpha} \partial_{\omega} (b_{\epsilon \rho} h^{\epsilon \beta} + c_{\epsilon \rho} \tau^{\epsilon \beta}) - \frac12 h^{\chi \rho} \tau^{\psi \alpha} \partial_{\omega} (b_{\epsilon \rho} h^{\epsilon \beta} + c_{\epsilon \rho} \tau^{\epsilon \beta}) \nn \\ && 
- \frac12 b_{\pi \rho} \tau^{\pi \psi} h^{\chi \alpha} \partial_{\omega} h^{\beta \rho} - \frac12 b_{\epsilon \rho} h^{\epsilon \chi} \tau^{\psi \alpha} \partial_{\omega}h^{\beta \rho} - \frac12 c_{\epsilon \rho} \tau^{\epsilon \psi} h^{\chi \alpha} \partial_{\omega} \tau^{\beta \rho} \Big] h_{\chi |\mu]} 
\nn \\ &&
- \frac14 \Big[ -\, {\partial}_{[\nu}{{\tau}^{\alpha \beta}}\, \, \tau_{\omega \mu]}
+ \, b_{\delta [\nu} \tau^{\delta \sigma} c_{\lambda \omega} \tau^{\lambda \rho} {\partial}_{\sigma}{h}^{\alpha \beta}\, \tau_{\rho \mu]}
 + \, (b_{\lambda [\nu} h^{\lambda \sigma} + c_{\lambda [\nu} \tau^{\lambda \sigma}) \tau^{\delta \rho} b_{\delta \omega} {\partial}_{\sigma}{h}^{\alpha \beta}\,  \tau_{\rho \mu]} \nn \\ && - \, (b_{\lambda [\nu|} h^{\lambda \sigma} + c_{\lambda [\nu|} \tau^{\lambda \sigma}) {\tau}^{\rho \beta} {\partial}_{\sigma}(b_{\epsilon \omega} h^{\epsilon \alpha} + c_{\epsilon \omega} \tau^{\epsilon \alpha})\,  \tau_{\rho |\mu]} \,
+ \, (b_{\lambda [\nu|} h^{\lambda \sigma} + c_{\lambda [\nu|} \tau^{\lambda \sigma}) c_{\epsilon \omega} \tau^{\epsilon \rho} {\partial}_{\sigma}{{\tau}^{\alpha \beta}}\,  \tau_{\rho |\mu]}  \nn \\ &&
+ 2 \, (b_{\lambda [\nu|} h^{\lambda \gamma} + c_{\lambda [\nu|} \tau^{\lambda \gamma}) {\tau}^{\rho \alpha} {\partial}_{\omega}{(b_{\epsilon \gamma} h^{\epsilon \beta} + c_{\epsilon \gamma} \tau^{\epsilon \beta})}\,  \tau_{\rho |\mu]}  \, 
+ 2 \, (h_{[\nu| \gamma} + b_{\lambda [\nu|} h^{\lambda \tau} b_{\tau \gamma} + 2 c_{\lambda ([\nu|} \tau^{\lambda \tau} b_{\tau \gamma)}) {\tau}^{\rho \alpha} {\partial}_{\omega}{{h}^{\beta \gamma}}\,  \tau_{\rho |\mu]} \nn \\ &&
 + \, 
{\tau}^{\rho \sigma} (h_{[\nu| \omega} + b_{\lambda [\nu|} h^{\lambda \tau} b_{\tau \omega} + 2 c_{\lambda ([\nu|} \tau^{\lambda \tau} b_{\tau \omega)}) 
\partial_{\sigma}{h}^{\alpha \beta}\, \tau_{\rho |\mu]}\,
 - \, 
{\tau}^{\rho \sigma} (b_{\lambda [\nu|} h^{\lambda \beta} + c_{\lambda [\nu|} \tau^{\lambda \beta}) 
\partial_{\sigma}(b_{\epsilon \omega} h^{\epsilon \alpha} + c_{\epsilon \omega} \tau^{\epsilon \alpha})\,  \tau_{\rho |\mu]}\,
 \,
 \nn \\ && + 2\, {\tau}^{\rho \gamma} (b_{\lambda [\nu|} h^{\lambda \alpha} + c_{\lambda [\nu|} \tau^{\lambda \alpha}) {\partial}_{\omega}(b_{\epsilon \gamma} h^{\epsilon \beta} + c_{\epsilon \gamma} \tau^{\epsilon \beta})\, \tau_{\rho |\mu]}  + 2\, \tau^{\delta \rho} b_{\delta \gamma} (b_{\lambda [\nu|} h^{\lambda \alpha} + c_{\lambda [\nu|} \tau^{\lambda \alpha}) {\partial}_{\omega}{{h}^{\beta \gamma}}\,  \tau_{\rho |\mu]}
 \nn \\ && + 2\, c_{\lambda \gamma} \tau^{\lambda \rho} b_{\delta [\nu|} \tau^{\delta \alpha} {\partial}_{\omega}{{h}^{\beta \gamma}}\, \tau_{\rho |\mu]}
+ 2 \, c_{\lambda \gamma} \tau^{\lambda \rho} (b_{\lambda [\nu} h^{\lambda \alpha} + c_{\lambda [\nu|} \tau^{\lambda \alpha}) {\partial}_{\omega}{{\tau}^{\beta \gamma}}\,
\, \tau_{\rho |\mu]} \Big] 
\nn \\ && 
- b_{\rho [\nu|} \Big[ - \frac14 \tau^{\sigma \chi} b_{\epsilon \omega} h^{\epsilon \rho} \partial_{\chi}h^{\alpha \beta}  - \frac12 \tau^{(\rho| \chi} c_{\epsilon \omega} \tau^{\epsilon \sigma)} \partial_{\chi}h^{\alpha \beta} - \frac14 h^{\rho \chi} b_{\pi \omega} \tau^{\pi \sigma} \partial_{\chi} h^{\alpha \beta} - \frac14 h^{\rho \chi} c_{\epsilon \omega} \tau^{\epsilon \sigma} \partial_{\chi} \tau^{\alpha \beta}  \nn \\ && + \frac14 \tau^{\sigma \chi} h^{\rho \beta} \partial_{\chi} (b_{\epsilon \omega} h^{\epsilon \alpha} + c_{\epsilon \omega} \tau^{\epsilon \alpha}) + \frac14 h^{\rho \chi} \tau^{\sigma \beta} \partial_{\chi} (b_{\epsilon \omega} h^{\epsilon \alpha} + c_{\epsilon \omega} \tau^{\epsilon \alpha})  - \frac12 \tau^{\sigma \chi} h^{\rho \alpha} \partial_{\omega} (b_{\epsilon \chi} h^{\epsilon \beta} + c_{\epsilon \chi} \tau^{\epsilon \beta}) \nn \\ && - \frac12 h^{\rho \chi} \tau^{\sigma \omega} \partial_{\omega} (b_{\epsilon \chi} h^{\epsilon \beta} + c_{\epsilon \chi} \tau^{\epsilon \beta})  - \frac12 b_{\pi \chi} \tau^{\pi \sigma} h^{\rho \alpha} \partial_{\omega} h^{\beta \chi}  - \frac12 b_{\epsilon \chi} h^{\epsilon \rho} \tau^{\sigma \alpha} \partial_{\omega}h^{\beta \chi} \nn \\ && - c_{\epsilon \chi} \tau^{\epsilon (\rho|} \tau^{\sigma)\alpha} \partial_{\omega}h^{\beta \chi}  - \frac12  c_{\epsilon \chi} \tau^{\epsilon \sigma} h^{\rho \alpha} \partial_{\omega} \tau^{\beta \chi}  \Big] \tau_{\sigma |\mu]}  \, . 
\eea

\section{Contributions to the four-derivative Lagrangian}
Here we report the terms with b-dependence in the four-derivative Lagrangian,

\bea
T^{\mu \nu \rho \sigma \gamma \epsilon \lambda \alpha \beta}(h,\tau) & = & \partial_{\xi}{h_{\pi \psi}} (\frac18h^{\lambda \pi} h^{\mu \sigma} h^{\nu \gamma} h^{\rho \alpha} h^{\epsilon \xi} h^{\beta \psi} + \frac18h^{\lambda \nu} h^{\mu \sigma} h^{\rho \gamma} h^{\epsilon \pi} h^{\alpha \xi} h^{\beta \psi} + \frac18h^{\lambda \xi} h^{\mu \nu} h^{\sigma \gamma} h^{\rho \alpha} h^{\epsilon \pi} h^{\beta \psi} \nn \\ && - \frac18h^{\lambda \xi} h^{\mu \nu} h^{\sigma \gamma} h^{\rho \pi} h^{\epsilon \alpha} h^{\beta \psi} + \frac18h^{\lambda \xi} h^{\mu \nu} h^{\sigma \pi} h^{\rho \gamma} h^{\epsilon \alpha} h^{\beta \psi}) + \partial_{\xi}{h^{\mu \nu}} (\frac{1}{4}h^{\lambda \gamma} h^{\sigma \alpha} h^{\rho \beta} h^{\epsilon \xi} \nn \\ && - \frac{1}{4}h^{\lambda \gamma} h^{\sigma \alpha} h^{\rho \epsilon} h^{\beta \xi} - \frac{1}{4}h^{\lambda \gamma} h^{\sigma \alpha} h^{\rho \xi} h^{\epsilon \beta} + \frac{1}{4}h^{\lambda \gamma} h^{\sigma \rho} h^{\epsilon \alpha} h^{\beta \xi} + \frac{3}{8}h^{\lambda \gamma} h^{\sigma \xi} h^{\rho \alpha} h^{\epsilon \beta} \nn \\ && - \frac{1}{4}h^{\lambda \rho} h^{\sigma \gamma} h^{\epsilon \alpha} h^{\beta \xi} - \frac18h^{\lambda \rho} h^{\sigma \xi} h^{\gamma \alpha} h^{\epsilon \beta} + \frac18h^{\lambda \sigma} h^{\rho \xi} h^{\gamma \alpha} h^{\epsilon \beta} - \frac{1}{4}h^{\lambda \xi} h^{\sigma \gamma} h^{\rho \alpha} h^{\epsilon \beta} \nn \\ && + \frac18h^{\lambda \xi} h^{\sigma \rho} h^{\gamma \alpha} h^{\epsilon \beta}) + \partial_{\xi}{h^{\mu \sigma}} ( - \frac{1}{4}h^{\lambda \nu} h^{\rho \gamma} h^{\epsilon \alpha} h^{\beta \xi} - \frac18h^{\lambda \nu} h^{\rho \xi} h^{\gamma \alpha} h^{\epsilon \beta}) \nn \\ && + \partial_{\xi}{h^{\nu \gamma}} (\frac18h^{\lambda \mu} h^{\sigma \alpha} h^{\rho \beta} h^{\epsilon \xi} + \frac{1}{4}h^{\lambda \rho} h^{\mu \alpha} h^{\sigma \beta} h^{\epsilon \xi}) + \frac{1}{16}h^{\lambda \mu} h^{\sigma \xi} h^{\rho \alpha} h^{\epsilon \pi} h^{\beta \psi} \partial_{\xi}{h^{\nu \gamma}} h_{\pi \psi} \nn \\ && + \frac18h^{\lambda \sigma} h^{\nu \gamma} h^{\rho \alpha} h^{\epsilon \xi} h^{\beta \pi} \partial_{\xi}{h^{\mu \psi}} h_{\psi \pi} + \frac{1}{16}h^{\lambda \xi} h^{\mu \sigma} h^{\rho \gamma} h^{\epsilon \alpha} h^{\beta \pi} \partial_{\xi}{h^{\nu \psi}} h_{\psi \pi} \, ,
\eea
\bea
T^{\mu \nu \rho \sigma \gamma}(h,\tau) & = & \frac{1}{12} h^{\nu \epsilon} h^{\sigma \gamma} \partial_{\epsilon}{h^{\mu \rho}} - \frac14 h^{\nu \sigma} h^{\gamma \epsilon} \partial_{\epsilon}{h^{\mu \rho}} + \frac14 h^{\mu \rho} h^{\nu \lambda} h^{\sigma \epsilon} h^{\gamma \alpha} \partial_{\epsilon}{h_{\lambda \alpha}} \, ,
\eea
\bea
T^{\mu \nu \rho \sigma}(h,\tau) & = & h^{\nu \gamma} ( - \frac{3}{8} \partial_{\epsilon}{h^{\mu \rho}} \partial_{\gamma}{h^{\sigma \epsilon}} + \frac{1}{12} \partial_{\gamma}{h^{\mu \rho}} \partial_{\epsilon}{h^{\sigma \epsilon}}) - \frac14 h^{\nu \sigma} \partial_{\gamma}{h^{\mu \epsilon}} \partial_{\epsilon}{h^{\rho \gamma}} 
\nn \\ && 
+ h^{\sigma \gamma} (\frac{5}{12} \partial_{\epsilon}{h^{\mu \rho}} \partial_{\gamma}{h^{\nu \epsilon}} - \frac14 \partial_{\gamma}{h^{\mu \rho}} \partial_{\epsilon}{h^{\nu \epsilon}}) + \frac14 h^{\mu \sigma} h^{\nu \gamma} \partial_{\gamma}{h^{\rho \epsilon}} \partial_{\lambda}{h^{\lambda \alpha}} h_{\epsilon \alpha} \nn \\ && - \frac14 h^{\nu \gamma} h^{\rho \sigma} \partial_{\gamma}{h^{\mu \epsilon}} \partial_{\lambda}{h^{\lambda \alpha}} h_{\epsilon \alpha} + h^{\nu \gamma} h^{\sigma \epsilon} ( - \frac{1}{24} \partial_{\epsilon}{\tau^{\lambda \alpha}} \partial_{\gamma}{h^{\mu \rho}} \tau_{\lambda \alpha} - \frac{1}{6} \partial_{\epsilon}{\phi} \partial_{\gamma}{h^{\mu \rho}} \nn \\ && - \frac{1}{12} \partial_{\gamma \epsilon}{h^{\mu \rho}} - \frac{1}{24} \partial_{\gamma}{h^{\mu \rho}} \partial_{\epsilon}{h^{\lambda \alpha}} h_{\lambda \alpha} + \frac{1}{12} \partial_{\epsilon}{h^{\mu \lambda}} \partial_{\gamma}{h^{\rho \alpha}} h_{\lambda \alpha}) + h^{\nu \rho} h^{\sigma \gamma} (\frac18 \partial_{\epsilon}{\tau^{\lambda \alpha}} \partial_{\gamma}{h^{\mu \epsilon}} \tau_{\lambda \alpha} \nn \\ && + \frac{1}{2} \partial_{\epsilon}{\phi} \partial_{\gamma}{h^{\mu \epsilon}} + \frac18 \partial_{\gamma}{h^{\mu \epsilon}} \partial_{\epsilon}{h^{\lambda \alpha}} h_{\lambda \alpha}) - \frac14 h^{\nu \sigma} h^{\gamma \epsilon} \partial_{\gamma}{h^{\mu \lambda}} \partial_{\epsilon}{h^{\rho \alpha}} h_{\lambda \alpha} \nn \\ && + h^{\mu \gamma} h^{\nu \epsilon} h^{\sigma \lambda} ( - \frac{3}{8} \partial_{\alpha}{h_{\gamma \lambda}} \partial_{\epsilon}{h^{\rho \alpha}} - \frac{1}{6} \partial_{\lambda}{h_{\alpha \gamma}} \partial_{\epsilon}{h^{\rho \alpha}}) + h^{\mu \rho} h^{\nu \gamma} h^{\sigma \epsilon} ( - \frac14 \partial_{\lambda}{h_{\alpha \gamma}} \partial_{\epsilon}{h^{\lambda \alpha}} \nn \\ && + \frac18 \partial_{\epsilon}{h_{\lambda \gamma}} \partial_{\alpha}{h^{\lambda \alpha}}) + h^{\mu \sigma} h^{\nu \gamma} h^{\epsilon \lambda} (\frac14 \partial_{\epsilon}{h_{\alpha \lambda}} \partial_{\gamma}{h^{\rho \alpha}} + \frac14 \partial_{\alpha}{h_{\gamma \epsilon}} \partial_{\lambda}{h^{\rho \alpha}}) \nn \\ && + h^{\nu \gamma} h^{\rho \epsilon} h^{\sigma \lambda} ( - \frac14 \partial_{\epsilon}{h_{\alpha \lambda}} \partial_{\gamma}{h^{\mu \alpha}} - \frac18 \partial_{\alpha}{h_{\gamma \epsilon}} \partial_{\lambda}{h^{\mu \alpha}} + \frac{1}{3} \partial_{\epsilon}{h_{\alpha \gamma}} \partial_{\lambda}{h^{\mu \alpha}}) \nn \\ && + h^{\nu \gamma} h^{\rho \sigma} h^{\epsilon \lambda} ( - \frac14 \partial_{\epsilon}{h_{\alpha \lambda}} \partial_{\gamma}{h^{\mu \alpha}} + \frac{1}{24} \partial_{\epsilon}{h_{\alpha \gamma}} \partial_{\lambda}{h^{\mu \alpha}}) \nn \\ && + h^{\nu \gamma} h^{\sigma \epsilon} h^{\lambda \alpha} ( - \frac14 \partial_{\lambda}{h_{\gamma \alpha}} \partial_{\epsilon}{h^{\mu \rho}} + \frac18 \partial_{\epsilon}{h_{\gamma \lambda}} \partial_{\alpha}{h^{\mu \rho}}) + h^{\nu \rho} h^{\sigma \gamma} h^{\epsilon \lambda} (\frac18 \partial_{\epsilon}{h_{\alpha \lambda}} \partial_{\gamma}{h^{\mu \alpha}} \nn \\ && - \frac14 \partial_{\gamma}{h_{\alpha \epsilon}} \partial_{\lambda}{h^{\mu \alpha}} - \frac14 \partial_{\epsilon}{h_{\alpha \gamma}} \partial_{\lambda}{h^{\mu \alpha}}) + \frac14 h^{\mu \gamma} h^{\nu \epsilon} h^{\rho \lambda} h^{\sigma \alpha} \partial_{\gamma \lambda}{h_{\epsilon \alpha}} \nn \\ && + h^{\mu \rho} h^{\nu \gamma} h^{\sigma \epsilon} h^{\lambda \alpha} ( - \frac18 \partial_{\lambda}{\tau^{\beta \xi}} \partial_{\epsilon}{h_{\gamma \alpha}} \tau_{\beta \xi} - \frac18 \partial_{\epsilon}{h_{\gamma \lambda}} \partial_{\alpha}{h^{\beta \xi}} h_{\beta \xi} - \frac{1}{2} \partial_{\lambda}{\phi} \partial_{\epsilon}{h_{\gamma \alpha}}) \, ,
\eea

\bea
T^{\mu \nu \rho}(h.\tau) & = & h^{\mu \rho} ( - \frac{1}{4} \partial_{\sigma}{h^{\nu \gamma}} \partial_{\gamma}{h^{\sigma \epsilon}} \partial_{\lambda}{h^{\lambda \alpha}} h_{\epsilon \alpha} + \frac{1}{4} \partial_{\sigma}{h^{\nu \gamma}} \partial_{\epsilon}{h^{\sigma \lambda}} \partial_{\alpha}{h^{\epsilon \alpha}} h_{\gamma \lambda}) + h^{\mu \sigma} (\frac{1}{4} \partial_{\gamma}{h^{\nu \epsilon}} \partial_{\epsilon \sigma}{h^{\rho \gamma}} 
\nn \\ && 
- \frac{1}{8} \partial_{\gamma}{h^{\nu \gamma}} \partial_{\sigma}{h^{\rho \epsilon}} \partial_{\epsilon}{h^{\lambda \alpha}} h_{\lambda \alpha} + \frac{1}{2} \partial_{\gamma}{\phi} \partial_{\sigma}{h^{\nu \gamma}} \partial_{\epsilon}{h^{\rho \epsilon}} - \frac{1}{12} \partial_{\gamma}{h^{\nu \epsilon}} \partial_{\epsilon}{h^{\gamma \lambda}} \partial_{\sigma}{h^{\rho \alpha}} h_{\lambda \alpha} \nn \\ && + \frac{1}{8} \partial_{\gamma}{\tau^{\epsilon \lambda}} \partial_{\sigma}{h^{\nu \gamma}} \partial_{\alpha}{h^{\rho \alpha}} \tau_{\epsilon \lambda} - \frac{1}{4} \partial_{\sigma}{h^{\nu \gamma}} \partial_{\gamma \epsilon}{h^{\rho \epsilon}} + \frac{1}{2} \partial_{\sigma}{h^{\nu \gamma}} \partial_{\gamma}{h^{\rho \epsilon}} \partial_{\lambda}{h^{\lambda \alpha}} h_{\epsilon \alpha}) \nn \\ && + h^{\rho \sigma} ( - \frac{1}{12} \partial_{\gamma}{h^{\mu \epsilon}} \partial_{\lambda}{h^{\nu \gamma}} \partial_{\alpha}{h^{\lambda \alpha}} h_{\epsilon \sigma} + \frac{1}{4} \partial_{\gamma}{h^{\mu \epsilon}} \partial_{\lambda}{h^{\nu \lambda}} \partial_{\alpha}{h^{\gamma \alpha}} h_{\epsilon \sigma} + \frac{1}{8} \partial_{\gamma}{h^{\mu \epsilon}} \partial_{\lambda}{h^{\nu \alpha}} \partial_{\alpha}{h^{\gamma \lambda}} h_{\epsilon \sigma} \nn \\ && + \frac{1}{12} \partial_{\gamma}{h^{\gamma \epsilon}} \partial_{\epsilon \sigma}{h^{\mu \nu}} - \frac{1}{6} \partial_{\gamma}{h^{\mu \epsilon}} \partial_{\epsilon \sigma}{h^{\nu \gamma}} - \frac{1}{4} \partial_{\gamma}{h^{\mu \epsilon}} \partial_{\sigma}{h^{\nu \gamma}} \partial_{\lambda}{h^{\lambda \alpha}} h_{\epsilon \alpha} - \frac{1}{4} \partial_{\gamma}{h^{\mu \epsilon}} \partial_{\lambda}{h^{\nu \gamma}} \partial_{\sigma}{h^{\lambda \alpha}} h_{\epsilon \alpha} \nn \\ && - \frac{1}{12} \partial_{\gamma}{h^{\mu \gamma}} \partial_{\epsilon}{h^{\nu \lambda}} \partial_{\sigma}{h^{\epsilon \alpha}} h_{\lambda \alpha} - \frac{1}{4} \partial_{\sigma}{h^{\mu \gamma}} \partial_{\gamma \epsilon}{h^{\nu \epsilon}} - \frac{1}{4} \partial_{\sigma}{h^{\mu \gamma}} \partial_{\epsilon}{h^{\nu \lambda}} \partial_{\lambda}{h^{\epsilon \alpha}} h_{\gamma \alpha} \nn \\ && + \frac{1}{8} \partial_{\sigma}{h^{\mu \gamma}} \partial_{\gamma}{h^{\epsilon \lambda}} \partial_{\epsilon}{h^{\nu \alpha}} h_{\lambda \alpha}) + h^{\sigma \gamma} ( - \frac{1}{12} \partial_{\epsilon}{h^{\mu \lambda}} \partial_{\sigma}{h^{\nu \epsilon}} \partial_{\gamma}{h^{\rho \alpha}} h_{\lambda \alpha} + \frac{1}{12} \partial_{\epsilon}{h^{\mu \nu}} \partial_{\sigma \gamma}{h^{\rho \epsilon}} \nn \\ && - \frac{1}{24} \partial_{\epsilon}{h^{\mu \nu}} \partial_{\sigma}{h^{\rho \epsilon}} \partial_{\gamma}{h^{\lambda \alpha}} h_{\lambda \alpha} - \frac{1}{2} \partial_{\epsilon}{\phi} \partial_{\sigma}{h^{\mu \nu}} \partial_{\gamma}{h^{\rho \epsilon}} + \frac{1}{4} \partial_{\epsilon}{h^{\mu \nu}} \partial_{\sigma}{h^{\rho \lambda}} \partial_{\gamma}{h^{\epsilon \alpha}} h_{\lambda \alpha} \nn \\ && - \frac{1}{8} \partial_{\epsilon}{\tau^{\lambda \alpha}} \partial_{\sigma}{h^{\mu \nu}} \partial_{\gamma}{h^{\rho \epsilon}} \tau_{\lambda \alpha} - \frac{1}{24} \partial_{\sigma}{\tau^{\epsilon \lambda}} \partial_{\alpha}{h^{\mu \nu}} \partial_{\gamma}{h^{\rho \alpha}} \tau_{\epsilon \lambda} - \frac{1}{4} \partial_{\sigma}{h^{\mu \nu}} \partial_{\epsilon \gamma}{h^{\rho \epsilon}} - \frac{1}{6} \partial_{\sigma}{h^{\nu \epsilon}} \partial_{\epsilon \gamma}{h^{\mu \rho}} \nn \\ && - \frac{1}{6} \partial_{\sigma}{\phi} \partial_{\epsilon}{h^{\mu \nu}} \partial_{\gamma}{h^{\rho \epsilon}} - \frac{1}{8} \partial_{\sigma}{h^{\mu \nu}} \partial_{\gamma}{h^{\rho \epsilon}} \partial_{\epsilon}{h^{\lambda \alpha}} h_{\lambda \alpha} + \frac{1}{4} \partial_{\sigma}{h^{\mu \epsilon}} \partial_{\lambda}{h^{\nu \lambda}} \partial_{\gamma}{h^{\rho \alpha}} h_{\epsilon \alpha}) \nn \\ && - \frac{1}{4} h^{\mu \nu} h^{\rho \sigma} \partial_{\gamma}{h_{\epsilon \sigma}} \partial_{\lambda}{h^{\gamma \epsilon}} \partial_{\alpha}{h^{\lambda \alpha}} + h^{\mu \rho} h^{\sigma \gamma} ( - \frac{1}{4} \partial_{\epsilon}{h_{\lambda \sigma}} \partial_{\gamma}{h^{\nu \epsilon}} \partial_{\alpha}{h^{\lambda \alpha}} + \frac{1}{2} \partial_{\epsilon}{\phi} \partial_{\sigma}{h^{\nu \lambda}} \partial_{\gamma}{h^{\epsilon \alpha}} h_{\lambda \alpha} \nn \\ && + \frac{1}{6} \partial_{\epsilon}{h_{\lambda \alpha}} \partial_{\sigma}{h^{\nu \lambda}} \partial_{\gamma}{h^{\epsilon \alpha}} + \frac{1}{8} \partial_{\epsilon}{\tau^{\lambda \alpha}} \partial_{\sigma}{h^{\nu \beta}} \partial_{\gamma}{h^{\epsilon \xi}} \tau_{\lambda \alpha} h_{\beta \xi} + \frac{1}{4} \partial_{\sigma}{h_{\epsilon \lambda}} \partial_{\alpha}{h^{\nu \epsilon}} \partial_{\gamma}{h^{\lambda \alpha}} \nn \\ && + \frac{1}{8} \partial_{\sigma}{h^{\nu \epsilon}} \partial_{\gamma}{h^{\lambda \alpha}} \partial_{\lambda}{h^{\beta \xi}} h_{\epsilon \alpha} h_{\beta \xi} + \frac{1}{8} \partial_{\sigma}{h_{\epsilon \lambda}} \partial_{\gamma}{h^{\nu \epsilon}} \partial_{\alpha}{h^{\lambda \alpha}}) + h^{\mu \sigma} h^{\gamma \epsilon} ( - \frac{1}{4} \partial_{\gamma}{h_{\lambda \epsilon}} \partial_{\alpha}{h^{\nu \lambda}} \partial_{\sigma}{h^{\rho \alpha}} \nn \\ && - \frac{1}{8} \partial_{\lambda}{h_{\alpha \sigma}} \partial_{\gamma}{h^{\nu \lambda}} \partial_{\epsilon}{h^{\rho \alpha}} - \frac{1}{8} \partial_{\lambda}{h_{\alpha \gamma}} \partial_{\sigma}{h^{\nu \lambda}} \partial_{\epsilon}{h^{\rho \alpha}} + \frac{1}{4} \partial_{\gamma}{h_{\lambda \alpha}} \partial_{\sigma}{h^{\nu \lambda}} \partial_{\epsilon}{h^{\rho \alpha}} + \frac{1}{8} \partial_{\gamma}{h_{\lambda \epsilon}} \partial_{\sigma}{h^{\nu \lambda}} \partial_{\alpha}{h^{\rho \alpha}} \nn \\ && - \frac{1}{8} \partial_{\sigma}{h_{\lambda \gamma}} \partial_{\alpha}{h^{\nu \lambda}} \partial_{\epsilon}{h^{\rho \alpha}}) + h^{\mu \sigma} h^{\rho \gamma} (\frac{1}{8} \partial_{\epsilon}{h_{\lambda \gamma}} \partial_{\alpha}{h^{\nu \alpha}} \partial_{\sigma}{h^{\epsilon \lambda}} + \frac{5}{24} \partial_{\epsilon}{h_{\lambda \gamma}} \partial_{\alpha}{h^{\nu \lambda}} \partial_{\sigma}{h^{\epsilon \alpha}} \nn \\ && + \frac{1}{24} \partial_{\epsilon}{h_{\lambda \gamma}} \partial_{\alpha}{h^{\nu \epsilon}} \partial_{\sigma}{h^{\lambda \alpha}} + \frac{1}{4} \partial_{\epsilon}{h_{\lambda \alpha}} \partial_{\gamma}{h^{\nu \lambda}} \partial_{\sigma}{h^{\epsilon \alpha}} - \frac{1}{16} \partial_{\epsilon}{h_{\lambda \alpha}} \partial_{\gamma}{h^{\nu \epsilon}} \partial_{\sigma}{h^{\lambda \alpha}} + \frac{1}{8} \partial_{\gamma}{h^{\epsilon \lambda}} \partial_{\alpha \sigma}{h^{\nu \alpha}} h_{\epsilon \lambda} \nn \\ && + \frac{1}{8} \partial_{\gamma}{\tau^{\epsilon \lambda}} \partial_{\alpha \sigma}{h^{\nu \alpha}} \tau_{\epsilon \lambda} + \frac{1}{8} \partial_{\epsilon}{\tau^{\lambda \alpha}} \partial_{\beta}{h^{\nu \beta}} \partial_{\sigma}{h^{\epsilon \xi}} \tau_{\lambda \alpha} h_{\xi \gamma} - \frac{1}{8} \partial_{\epsilon}{\tau^{\lambda \alpha}} \partial_{\gamma}{h^{\nu \beta}} \partial_{\sigma}{h^{\epsilon \xi}} \tau_{\lambda \alpha} h_{\beta \xi} \nn \\ && - \frac{1}{8} \partial_{\gamma}{\tau^{\epsilon \lambda}} \partial_{\alpha}{h^{\nu \beta}} \partial_{\sigma}{h^{\alpha \xi}} \tau_{\epsilon \lambda} h_{\beta \xi} - \frac{1}{8} \partial_{\epsilon}{h_{\lambda \sigma}} \partial_{\alpha}{h^{\nu \lambda}} \partial_{\gamma}{h^{\epsilon \alpha}} - \frac{1}{8} \partial_{\epsilon}{h_{\lambda \sigma}} \partial_{\alpha}{h^{\nu \epsilon}} \partial_{\gamma}{h^{\lambda \alpha}} \nn \\ && - \frac{1}{8} \partial_{\epsilon}{h^{\nu \lambda}} \partial_{\sigma}{h^{\epsilon \alpha}} \partial_{\gamma}{h^{\beta \xi}} h_{\lambda \alpha} h_{\beta \xi} + \frac{1}{2} \partial_{\epsilon}{\phi} \partial_{\lambda}{h^{\nu \lambda}} \partial_{\sigma}{h^{\epsilon \alpha}} h_{\alpha \gamma} + \frac{1}{8} \partial_{\epsilon}{h_{\lambda \sigma}} \partial_{\gamma}{h^{\nu \alpha}} \partial_{\alpha}{h^{\epsilon \lambda}} \nn \\ && + \frac{1}{12} \partial_{\epsilon}{h_{\lambda \sigma}} \partial_{\gamma}{h^{\nu \lambda}} \partial_{\alpha}{h^{\epsilon \alpha}} - \frac{1}{4} \partial_{\gamma}{h_{\epsilon \sigma}} \partial_{\lambda}{h^{\nu \lambda}} \partial_{\alpha}{h^{\epsilon \alpha}} - \frac{1}{4} \partial_{\gamma}{h_{\epsilon \sigma}} \partial_{\lambda}{h^{\nu \alpha}} \partial_{\alpha}{h^{\epsilon \lambda}} \nn \\ && - \frac{1}{8} \partial_{\gamma}{h^{\nu \epsilon}} \partial_{\sigma}{h^{\lambda \alpha}} \partial_{\lambda}{h^{\beta \xi}} h_{\epsilon \alpha} h_{\beta \xi} + \frac{1}{2} \partial_{\gamma}{\phi} \partial_{\epsilon \sigma}{h^{\nu \epsilon}} - \frac{1}{4} \partial_{\epsilon \sigma \gamma}{h^{\nu \epsilon}} + \frac{1}{8} \partial_{\epsilon}{h^{\nu \epsilon}} \partial_{\sigma}{h^{\lambda \alpha}} \partial_{\lambda}{h^{\beta \xi}} h_{\alpha \gamma} h_{\beta \xi} \nn \\ && + \frac{1}{12} \partial_{\epsilon}{h^{\lambda \alpha}} \partial_{\lambda \sigma}{h^{\nu \epsilon}} h_{\alpha \gamma} + \frac{1}{16} \partial_{\epsilon}{h_{\lambda \alpha}} \partial_{\sigma}{h^{\nu \epsilon}} \partial_{\gamma}{h^{\lambda \alpha}} - \frac{1}{4} \partial_{\epsilon}{h_{\lambda \alpha}} \partial_{\sigma}{h^{\nu \lambda}} \partial_{\gamma}{h^{\epsilon \alpha}} + \frac{1}{4} \partial_{\epsilon}{h_{\lambda \gamma}} \partial_{\sigma}{h^{\nu \epsilon}} \partial_{\alpha}{h^{\lambda \alpha}} \nn \\ && - \frac{1}{4} \partial_{\epsilon}{h_{\lambda \gamma}} \partial_{\sigma}{h^{\nu \alpha}} \partial_{\alpha}{h^{\epsilon \lambda}} + \frac{1}{24} \partial_{\epsilon}{h_{\lambda \gamma}} \partial_{\sigma}{h^{\nu \lambda}} \partial_{\alpha}{h^{\epsilon \alpha}} - \frac{1}{2} \partial_{\epsilon}{\phi} \partial_{\gamma}{h^{\nu \lambda}} \partial_{\sigma}{h^{\epsilon \alpha}} h_{\lambda \alpha} + \frac{1}{4} \partial_{\gamma}{h_{\epsilon \lambda}} \partial_{\sigma}{h^{\nu \epsilon}} \partial_{\alpha}{h^{\lambda \alpha}} 
\nn
\eea
\bea
&& + \frac{1}{16} \partial_{\gamma}{h_{\epsilon \lambda}} \partial_{\sigma}{h^{\nu \alpha}} \partial_{\alpha}{h^{\epsilon \lambda}} - \frac{1}{2} \partial_{\gamma}{\phi} \partial_{\epsilon}{h^{\nu \lambda}} \partial_{\sigma}{h^{\epsilon \alpha}} h_{\lambda \alpha} - \frac{1}{4} \partial_{\sigma}{h^{\epsilon \lambda}} \partial_{\epsilon \alpha}{h^{\nu \alpha}} h_{\lambda \gamma} - \frac{1}{8} \partial_{\sigma}{h_{\epsilon \lambda}} \partial_{\alpha}{h^{\nu \epsilon}} \partial_{\gamma}{h^{\lambda \alpha}} \nn \\ && + \frac{1}{24} \partial_{\sigma}{h_{\epsilon \gamma}} \partial_{\lambda}{h^{\nu \epsilon}} \partial_{\alpha}{h^{\lambda \alpha}} + \frac{1}{4} \partial_{\sigma}{h_{\epsilon \gamma}} \partial_{\lambda}{h^{\nu \lambda}} \partial_{\alpha}{h^{\epsilon \alpha}} + \frac{11}{24} \partial_{\sigma}{h_{\epsilon \gamma}} \partial_{\lambda}{h^{\nu \alpha}} \partial_{\alpha}{h^{\epsilon \lambda}} - \frac{1}{16} \partial_{\sigma}{h_{\epsilon \lambda}} \partial_{\gamma}{h^{\nu \alpha}} \partial_{\alpha}{h^{\epsilon \lambda}}) \nn \\ && + h^{\nu \sigma} h^{\rho \gamma} ( - \frac{1}{12} \partial_{\epsilon}{h_{\lambda \sigma}} \partial_{\alpha}{h^{\mu \alpha}} \partial_{\gamma}{h^{\epsilon \lambda}} + \frac{1}{8} \partial_{\epsilon}{h_{\lambda \sigma}} \partial_{\alpha}{h^{\mu \lambda}} \partial_{\gamma}{h^{\epsilon \alpha}} - \frac{1}{4} \partial_{\epsilon}{h_{\lambda \sigma}} \partial_{\alpha}{h^{\mu \epsilon}} \partial_{\gamma}{h^{\lambda \alpha}} - \frac{1}{4} \partial_{\epsilon}{h_{\lambda \sigma}} \partial_{\gamma}{h^{\mu \epsilon}} \partial_{\alpha}{h^{\lambda \alpha}} \nn \\ && - \frac{1}{2} \partial_{\epsilon}{\phi} \partial_{\sigma}{h^{\mu \lambda}} \partial_{\gamma}{h^{\epsilon \alpha}} h_{\lambda \alpha} - \frac{1}{16} \partial_{\epsilon}{h_{\lambda \alpha}} \partial_{\sigma}{h^{\mu \epsilon}} \partial_{\gamma}{h^{\lambda \alpha}} + \frac{1}{6} \partial_{\epsilon}{h_{\lambda \alpha}} \partial_{\sigma}{h^{\mu \lambda}} \partial_{\gamma}{h^{\epsilon \alpha}} - \frac{1}{8} \partial_{\epsilon}{\tau^{\lambda \alpha}} \partial_{\sigma}{h^{\mu \beta}} \partial_{\gamma}{h^{\epsilon \xi}} \tau_{\lambda \alpha} h_{\beta \xi} \nn \\ && + \frac{1}{8} \partial_{\sigma}{h_{\epsilon \lambda}} \partial_{\alpha}{h^{\mu \epsilon}} \partial_{\gamma}{h^{\lambda \alpha}} + \frac{1}{12} \partial_{\sigma}{h_{\epsilon \gamma}} \partial_{\lambda}{h^{\mu \epsilon}} \partial_{\alpha}{h^{\lambda \alpha}} - \frac{1}{4} \partial_{\sigma}{h_{\epsilon \gamma}} \partial_{\lambda}{h^{\mu \lambda}} \partial_{\alpha}{h^{\epsilon \alpha}} - \frac{1}{8} \partial_{\sigma}{h_{\epsilon \gamma}} \partial_{\lambda}{h^{\mu \alpha}} \partial_{\alpha}{h^{\epsilon \lambda}} \nn \\ && - \frac{1}{8} \partial_{\sigma}{h^{\mu \epsilon}} \partial_{\gamma}{h^{\lambda \alpha}} \partial_{\lambda}{h^{\beta \xi}} h_{\epsilon \alpha} h_{\beta \xi} + \frac{1}{16} \partial_{\sigma}{h_{\epsilon \lambda}} \partial_{\gamma}{h^{\mu \alpha}} \partial_{\alpha}{h^{\epsilon \lambda}}) + h^{\rho \sigma} h^{\gamma \epsilon} ( - \frac{1}{4} \partial_{\lambda}{h_{\alpha \gamma}} \partial_{\epsilon}{h^{\mu \lambda}} \partial_{\sigma}{h^{\nu \alpha}} \nn \\ && - \frac{1}{4} \partial_{\lambda}{h^{\lambda \alpha}} \partial_{\gamma \epsilon}{h^{\mu \nu}} h_{\alpha \sigma} + \frac{1}{8} \partial_{\lambda}{h_{\alpha \gamma}} \partial_{\epsilon}{h^{\mu \nu}} \partial_{\sigma}{h^{\lambda \alpha}} + \frac{1}{8} \partial_{\lambda}{h_{\alpha \gamma}} \partial_{\epsilon}{h^{\mu \alpha}} \partial_{\sigma}{h^{\nu \lambda}} - \frac{1}{2} \partial_{\lambda}{\phi} \partial_{\gamma}{h^{\mu \nu}} \partial_{\epsilon}{h^{\lambda \alpha}} h_{\alpha \sigma} \nn \\ && - \frac{1}{2} \partial_{\lambda}{\phi} \partial_{\gamma}{h^{\mu \alpha}} \partial_{\epsilon}{h^{\nu \lambda}} h_{\alpha \sigma} - \frac{1}{4} \partial_{\gamma}{h_{\lambda \epsilon}} \partial_{\alpha}{h^{\mu \lambda}} \partial_{\sigma}{h^{\nu \alpha}} - \frac{1}{4} \partial_{\gamma}{h_{\lambda \epsilon}} \partial_{\alpha}{h^{\mu \alpha}} \partial_{\sigma}{h^{\nu \lambda}} - \frac{1}{6} \partial_{\gamma}{\phi} \partial_{\lambda}{h^{\mu \nu}} \partial_{\epsilon}{h^{\lambda \alpha}} h_{\alpha \sigma} \nn \\ && - \frac{1}{16} \partial_{\gamma}{h_{\lambda \alpha}} \partial_{\epsilon}{h^{\mu \nu}} \partial_{\sigma}{h^{\lambda \alpha}} - \frac{1}{4} \partial_{\gamma}{h_{\lambda \alpha}} \partial_{\epsilon}{h^{\mu \lambda}} \partial_{\sigma}{h^{\nu \alpha}} - \frac{1}{24} \partial_{\gamma}{\tau^{\lambda \alpha}} \partial_{\epsilon}{h^{\mu \beta}} \partial_{\sigma}{h^{\nu \xi}} \tau_{\lambda \alpha} h_{\beta \xi} + \frac{5}{24} \partial_{\lambda}{h_{\alpha \sigma}} \partial_{\gamma}{h^{\mu \lambda}} \partial_{\epsilon}{h^{\nu \alpha}} \nn \\ && + \frac{1}{24} \partial_{\gamma}{h_{\lambda \sigma}} \partial_{\alpha}{h^{\mu \lambda}} \partial_{\epsilon}{h^{\nu \alpha}} - \frac{1}{4} \partial_{\gamma}{h_{\lambda \sigma}} \partial_{\alpha}{h^{\mu \nu}} \partial_{\epsilon}{h^{\lambda \alpha}} + \frac{1}{24} \partial_{\gamma}{h_{\lambda \sigma}} \partial_{\alpha}{h^{\mu \alpha}} \partial_{\epsilon}{h^{\nu \lambda}} - \frac{1}{24} \partial_{\gamma}{h^{\mu \lambda}} \partial_{\sigma}{h^{\nu \alpha}} \partial_{\epsilon}{h^{\beta \xi}} h_{\lambda \alpha} h_{\beta \xi} \nn \\ && + \frac{1}{8} \partial_{\gamma}{h_{\lambda \sigma}} \partial_{\epsilon}{h^{\mu \nu}} \partial_{\alpha}{h^{\lambda \alpha}} + \frac{5}{24} \partial_{\gamma}{h_{\lambda \sigma}} \partial_{\epsilon}{h^{\mu \alpha}} \partial_{\alpha}{h^{\nu \lambda}} + \frac{1}{4} \partial_{\gamma}{h_{\lambda \sigma}} \partial_{\epsilon}{h^{\mu \lambda}} \partial_{\alpha}{h^{\nu \alpha}} + \frac{1}{4} \partial_{\lambda}{h_{\sigma \gamma}} \partial_{\alpha}{h^{\mu \lambda}} \partial_{\epsilon}{h^{\nu \alpha}} \nn \\ && + \frac{1}{12} \partial_{\lambda}{h_{\sigma \gamma}} \partial_{\epsilon}{h^{\mu \nu}} \partial_{\alpha}{h^{\lambda \alpha}} + \frac{1}{4} \partial_{\lambda}{h_{\sigma \gamma}} \partial_{\epsilon}{h^{\mu \lambda}} \partial_{\alpha}{h^{\nu \alpha}} - \frac{1}{8} \partial_{\lambda}{h_{\sigma \gamma}} \partial_{\epsilon}{h^{\mu \alpha}} \partial_{\alpha}{h^{\nu \lambda}} - \frac{1}{4} \partial_{\gamma}{h_{\sigma \epsilon}} \partial_{\lambda}{h^{\mu \nu}} \partial_{\alpha}{h^{\lambda \alpha}} \nn \\ && - \frac{1}{6} \partial_{\gamma}{\phi} \partial_{\sigma \epsilon}{h^{\mu \nu}} - \frac{1}{6} \partial_{\lambda}{h_{\alpha \gamma}} \partial_{\sigma}{h^{\mu \lambda}} \partial_{\epsilon}{h^{\nu \alpha}} - \frac{1}{4} \partial_{\lambda}{h_{\alpha \gamma}} \partial_{\sigma}{h^{\mu \alpha}} \partial_{\epsilon}{h^{\nu \lambda}} - \frac{1}{24} \partial_{\gamma}{h^{\lambda \alpha}} \partial_{\sigma \epsilon}{h^{\mu \nu}} h_{\lambda \alpha} \nn \\ && + \frac{7}{24} \partial_{\gamma}{h_{\lambda \alpha}} \partial_{\sigma}{h^{\mu \lambda}} \partial_{\epsilon}{h^{\nu \alpha}} - \frac{1}{24} \partial_{\gamma}{\tau^{\lambda \alpha}} \partial_{\sigma \epsilon}{h^{\mu \nu}} \tau_{\lambda \alpha} + \frac{1}{4} \partial_{\gamma}{h_{\lambda \epsilon}} \partial_{\sigma}{h^{\mu \lambda}} \partial_{\alpha}{h^{\nu \alpha}} - \frac{1}{6} \partial_{\gamma}{\phi} \partial_{\epsilon}{h^{\mu \lambda}} \partial_{\sigma}{h^{\nu \alpha}} h_{\lambda \alpha} \nn \\ && - \frac{1}{8} \partial_{\sigma}{h_{\lambda \gamma}} \partial_{\alpha}{h^{\mu \lambda}} \partial_{\epsilon}{h^{\nu \alpha}} + \frac{1}{8} \partial_{\sigma}{h_{\lambda \gamma}} \partial_{\alpha}{h^{\mu \nu}} \partial_{\epsilon}{h^{\lambda \alpha}} - \frac{1}{8} \partial_{\sigma}{h_{\lambda \gamma}} \partial_{\epsilon}{h^{\mu \alpha}} \partial_{\alpha}{h^{\nu \lambda}} + \frac{1}{12} \partial_{\sigma \gamma \epsilon}{h^{\mu \nu}} \nn \\ && - \frac{1}{16} \partial_{\sigma}{h_{\lambda \alpha}} \partial_{\gamma}{h^{\mu \nu}} \partial_{\epsilon}{h^{\lambda \alpha}} + \frac{1}{24} \partial_{\sigma}{h_{\lambda \alpha}} \partial_{\gamma}{h^{\mu \lambda}} \partial_{\epsilon}{h^{\nu \alpha}} - \frac{1}{24} \partial_{\lambda}{h^{\mu \nu}} \partial_{\gamma}{h^{\lambda \alpha}} \partial_{\epsilon}{h^{\beta \xi}} h_{\alpha \sigma} h_{\beta \xi} \nn \\ && - \frac{1}{8} \partial_{\lambda}{\tau^{\alpha \beta}} \partial_{\gamma}{h^{\mu \nu}} \partial_{\epsilon}{h^{\lambda \xi}} \tau_{\alpha \beta} h_{\xi \sigma} - \frac{1}{8} \partial_{\lambda}{\tau^{\alpha \beta}} \partial_{\gamma}{h^{\mu \xi}} \partial_{\epsilon}{h^{\nu \lambda}} \tau_{\alpha \beta} h_{\xi \sigma} - \frac{1}{24} \partial_{\gamma}{\tau^{\lambda \alpha}} \partial_{\beta}{h^{\mu \nu}} \partial_{\epsilon}{h^{\beta \xi}} \tau_{\lambda \alpha} h_{\xi \sigma} \nn \\ && - \frac{1}{8} \partial_{\gamma}{h^{\mu \lambda}} \partial_{\epsilon}{h^{\nu \alpha}} \partial_{\alpha}{h^{\beta \xi}} h_{\lambda \sigma} h_{\beta \xi} - \frac{1}{8} \partial_{\gamma}{h^{\mu \nu}} \partial_{\epsilon}{h^{\lambda \alpha}} \partial_{\lambda}{h^{\beta \xi}} h_{\alpha \sigma} h_{\beta \xi}) + h^{\sigma \gamma} h^{\epsilon \lambda} (\frac{1}{8} \partial_{\sigma}{h_{\alpha \epsilon}} \partial_{\lambda}{h^{\mu \nu}} \partial_{\gamma}{h^{\rho \alpha}} \nn \\ && - \frac{1}{6} \partial_{\sigma}{h_{\alpha \gamma}} \partial_{\epsilon}{h^{\mu \nu}} \partial_{\lambda}{h^{\rho \alpha}}) + h^{\mu \nu} h^{\rho \sigma} h^{\gamma \epsilon} ( - \frac{1}{4} \partial_{\gamma}{h^{\lambda \alpha}} \partial_{\lambda \epsilon}{h_{\alpha \sigma}} + \frac{1}{8} \partial_{\gamma}{h_{\lambda \sigma}} \partial_{\alpha \epsilon}{h^{\lambda \alpha}}) \nn \\ && + h^{\mu \rho} h^{\sigma \gamma} h^{\epsilon \lambda} (\frac{1}{8} \partial_{\sigma}{\tau^{\alpha \beta}} \partial_{\epsilon}{h_{\xi \gamma}} \partial_{\lambda}{h^{\nu \xi}} \tau_{\alpha \beta} + \frac{1}{8} \partial_{\sigma}{h^{\nu \alpha}} \partial_{\gamma \epsilon}{h_{\alpha \lambda}} - \frac{1}{4} \partial_{\sigma}{h_{\alpha \epsilon}} \partial_{\gamma \lambda}{h^{\nu \alpha}} + \frac{1}{8} \partial_{\sigma}{h_{\alpha \epsilon}} \partial_{\gamma}{h^{\nu \alpha}} \partial_{\lambda}{h^{\beta \xi}} h_{\beta \xi} \nn \\ && + \frac{1}{2} \partial_{\sigma}{\phi} \partial_{\epsilon}{h_{\alpha \gamma}} \partial_{\lambda}{h^{\nu \alpha}}) + h^{\mu \sigma} h^{\nu \gamma} h^{\rho \epsilon} (\partial_{\lambda}{\phi} \partial_{\gamma}{h_{\alpha \epsilon}} \partial_{\sigma}{h^{\lambda \alpha}} - \frac{1}{8} \partial_{\gamma}{h^{\lambda \alpha}} \partial_{\lambda \epsilon}{h_{\alpha \sigma}} + \frac{1}{2} \partial_{\gamma}{\phi} \partial_{\lambda}{h_{\alpha \epsilon}} \partial_{\sigma}{h^{\lambda \alpha}} \nn \\ && - \frac{1}{8} \partial_{\lambda}{\tau^{\alpha \beta}} \partial_{\gamma}{h_{\xi \sigma}} \partial_{\epsilon}{h^{\lambda \xi}} \tau_{\alpha \beta} + \frac{1}{8} \partial_{\gamma}{\tau^{\lambda \alpha}} \partial_{\epsilon}{h_{\beta \sigma}} \partial_{\xi}{h^{\beta \xi}} \tau_{\lambda \alpha} - \frac{1}{8} \partial_{\gamma}{h_{\lambda \sigma}} \partial_{\alpha}{h^{\lambda \alpha}} \partial_{\epsilon}{h^{\beta \xi}} h_{\beta \xi} - \frac{1}{8} \partial_{\gamma}{h_{\lambda \sigma}} \partial_{\epsilon}{h^{\lambda \alpha}} \partial_{\alpha}{h^{\beta \xi}} h_{\beta \xi} \nn \\ && - \frac{1}{2} \partial_{\gamma}{\phi} \partial_{\sigma}{h_{\lambda \epsilon}} \partial_{\alpha}{h^{\lambda \alpha}} - \frac{1}{8} \partial_{\lambda}{h_{\alpha \gamma}} \partial_{\sigma}{h^{\lambda \alpha}} \partial_{\epsilon}{h^{\beta \xi}} h_{\beta \xi} - \frac{1}{2} \partial_{\lambda}{\phi} \partial_{\gamma}{h_{\alpha \sigma}} \partial_{\epsilon}{h^{\lambda \alpha}} + \frac{1}{8} \partial_{\gamma}{h^{\lambda \alpha}} \partial_{\lambda \sigma}{h_{\alpha \epsilon}} \nn \\ && - \frac{1}{8} \partial_{\gamma}{\tau^{\lambda \alpha}} \partial_{\sigma}{h_{\beta \epsilon}} \partial_{\xi}{h^{\beta \xi}} \tau_{\lambda \alpha} + \frac{1}{4} \partial_{\gamma}{h_{\lambda \epsilon}} \partial_{\sigma}{h^{\lambda \alpha}} \partial_{\alpha}{h^{\beta \xi}} h_{\beta \xi} + \frac{1}{2} \partial_{\gamma}{\phi} \partial_{\epsilon}{h_{\lambda \sigma}} \partial_{\alpha}{h^{\lambda \alpha}} - \frac{1}{4} \partial_{\sigma}{h_{\lambda \gamma}} \partial_{\alpha \epsilon}{h^{\lambda \alpha}} \nn 
\eea
\bea
&& + \frac{1}{8} \partial_{\sigma}{h_{\lambda \gamma}} \partial_{\alpha}{h^{\lambda \alpha}} \partial_{\epsilon}{h^{\beta \xi}} h_{\beta \xi} + \frac{1}{8} \partial_{\sigma}{h_{\lambda \gamma}} \partial_{\alpha}{h^{\lambda \beta}} \partial_{\epsilon}{h^{\alpha \xi}} h_{\beta \xi} + \frac{1}{4} \partial_{\lambda}{\tau^{\alpha \beta}} \partial_{\gamma}{h_{\xi \epsilon}} \partial_{\sigma}{h^{\lambda \xi}} \tau_{\alpha \beta} + \frac{1}{8} \partial_{\gamma}{\tau^{\lambda \alpha}} \partial_{\beta}{h_{\xi \epsilon}} \partial_{\sigma}{h^{\beta \xi}} \tau_{\lambda \alpha}) \nn \\ && + h^{\mu \sigma} h^{\rho \gamma} h^{\epsilon \lambda} (\frac{1}{8} \partial_{\alpha}{\tau^{\beta \xi}} \partial_{\epsilon}{h_{\sigma \gamma}} \partial_{\lambda}{h^{\nu \alpha}} \tau_{\beta \xi} - \frac{1}{12} \partial_{\epsilon}{h^{\nu \alpha}} \partial_{\gamma \lambda}{h_{\alpha \sigma}} - \frac{1}{48} \partial_{\epsilon}{h_{\alpha \gamma}} \partial_{\lambda}{h^{\nu \alpha}} \partial_{\sigma}{h^{\beta \xi}} h_{\beta \xi} \nn \\ && - \frac{1}{24} \partial_{\alpha}{h_{\gamma \epsilon}} \partial_{\lambda}{h^{\nu \alpha}} \partial_{\sigma}{h^{\beta \xi}} h_{\beta \xi} + \frac{1}{8} \partial_{\epsilon}{h_{\gamma \lambda}} \partial_{\alpha \sigma}{h^{\nu \alpha}} - \frac{1}{12} \partial_{\epsilon}{\phi} \partial_{\lambda}{h_{\alpha \gamma}} \partial_{\sigma}{h^{\nu \alpha}} + \frac{1}{16} \partial_{\epsilon}{h_{\alpha \beta}} \partial_{\lambda}{h^{\nu \alpha}} \partial_{\sigma}{h^{\beta \xi}} h_{\xi \gamma} \nn \\ && - \frac{1}{24} \partial_{\epsilon}{\tau^{\alpha \beta}} \partial_{\gamma}{h_{\xi \sigma}} \partial_{\lambda}{h^{\nu \xi}} \tau_{\alpha \beta} - \frac{1}{24} \partial_{\epsilon}{\tau^{\alpha \beta}} \partial_{\lambda}{h_{\xi \sigma}} \partial_{\gamma}{h^{\nu \xi}} \tau_{\alpha \beta} + \frac{1}{3} \partial_{\epsilon}{h^{\nu \alpha}} \partial_{\sigma \lambda}{h_{\alpha \gamma}} + \frac{1}{2} \partial_{\alpha}{\phi} \partial_{\epsilon}{h_{\sigma \gamma}} \partial_{\lambda}{h^{\nu \alpha}} \nn \\ && - \frac{1}{8} \partial_{\epsilon}{h_{\alpha \sigma}} \partial_{\gamma \lambda}{h^{\nu \alpha}} - \frac{1}{4} \partial_{\epsilon}{h^{\nu \alpha}} \partial_{\sigma \gamma}{h_{\alpha \lambda}} + \frac{1}{6} \partial_{\epsilon}{h^{\nu \alpha}} \partial_{\alpha \sigma}{h_{\gamma \lambda}} - \frac{1}{24} \partial_{\epsilon}{h_{\alpha \sigma}} \partial_{\gamma}{h^{\nu \alpha}} \partial_{\lambda}{h^{\beta \xi}} h_{\beta \xi} - \frac{1}{4} \partial_{\gamma}{h^{\nu \alpha}} \partial_{\sigma \epsilon}{h_{\alpha \lambda}} \nn \\ && - \frac{1}{24} \partial_{\gamma}{h_{\alpha \sigma}} \partial_{\epsilon}{h^{\nu \alpha}} \partial_{\lambda}{h^{\beta \xi}} h_{\beta \xi} + \frac{1}{2} \partial_{\epsilon}{\phi} \partial_{\sigma}{h_{\alpha \lambda}} \partial_{\gamma}{h^{\nu \alpha}} + \frac{1}{8} \partial_{\epsilon}{h_{\sigma \gamma}} \partial_{\lambda}{h^{\nu \alpha}} \partial_{\alpha}{h^{\beta \xi}} h_{\beta \xi} + \frac{1}{8} \partial_{\alpha}{h^{\nu \alpha}} \partial_{\sigma \epsilon}{h_{\gamma \lambda}} \nn \\ && - \frac{1}{12} \partial_{\epsilon}{\phi} \partial_{\sigma}{h_{\alpha \gamma}} \partial_{\lambda}{h^{\nu \alpha}} + \frac{1}{2} \partial_{\gamma}{\phi} \partial_{\sigma}{h_{\alpha \epsilon}} \partial_{\lambda}{h^{\nu \alpha}} - \frac{1}{4} \partial_{\epsilon}{h_{\alpha \lambda}} \partial_{\sigma \gamma}{h^{\nu \alpha}} - \frac{1}{6} \partial_{\epsilon}{\phi} \partial_{\lambda}{h_{\alpha \sigma}} \partial_{\gamma}{h^{\nu \alpha}} \nn \\ && - \frac{1}{48} \partial_{\epsilon}{\tau^{\alpha \beta}} \partial_{\sigma}{h_{\xi \gamma}} \partial_{\lambda}{h^{\nu \xi}} \tau_{\alpha \beta} + \frac{1}{3} \partial_{\epsilon}{h_{\alpha \gamma}} \partial_{\sigma \lambda}{h^{\nu \alpha}} - \frac{1}{48} \partial_{\epsilon}{h_{\alpha \gamma}} \partial_{\sigma}{h^{\nu \alpha}} \partial_{\lambda}{h^{\beta \xi}} h_{\beta \xi} - \frac{1}{6} \partial_{\epsilon}{\phi} \partial_{\gamma}{h_{\alpha \sigma}} \partial_{\lambda}{h^{\nu \alpha}} \nn \\ && + \frac{1}{8} \partial_{\gamma}{\tau^{\alpha \beta}} \partial_{\sigma}{h_{\xi \epsilon}} \partial_{\lambda}{h^{\nu \xi}} \tau_{\alpha \beta} - \frac{1}{8} \partial_{\gamma}{h_{\alpha \epsilon}} \partial_{\sigma \lambda}{h^{\nu \alpha}} + \frac{1}{8} \partial_{\epsilon}{\tau^{\alpha \beta}} \partial_{\sigma}{h_{\xi \lambda}} \partial_{\gamma}{h^{\nu \xi}} \tau_{\alpha \beta} - \frac{1}{24} \partial_{\sigma}{\tau^{\alpha \beta}} \partial_{\xi}{h_{\gamma \epsilon}} \partial_{\lambda}{h^{\nu \xi}} \tau_{\alpha \beta} \nn \\ && + \frac{1}{24} \partial_{\sigma}{h^{\nu \alpha}} \partial_{\epsilon \lambda}{h_{\alpha \gamma}} + \frac{1}{8} \partial_{\sigma}{h_{\alpha \epsilon}} \partial_{\lambda}{h^{\nu \alpha}} \partial_{\gamma}{h^{\beta \xi}} h_{\beta \xi} - \frac{1}{48} \partial_{\sigma}{\tau^{\alpha \beta}} \partial_{\epsilon}{h_{\xi \gamma}} \partial_{\lambda}{h^{\nu \xi}} \tau_{\alpha \beta} + \frac{1}{24} \partial_{\sigma}{h_{\alpha \gamma}} \partial_{\epsilon \lambda}{h^{\nu \alpha}} \nn \\ &&  - \frac{1}{48} \partial_{\sigma}{h_{\alpha \gamma}} \partial_{\epsilon}{h^{\nu \alpha}} \partial_{\lambda}{h^{\beta \xi}} h_{\beta \xi} - \frac{1}{6} \partial_{\sigma}{\phi} \partial_{\alpha}{h_{\gamma \epsilon}} \partial_{\lambda}{h^{\nu \alpha}} + \frac{1}{16} \partial_{\sigma}{h_{\alpha \gamma}} \partial_{\epsilon}{h^{\nu \beta}} \partial_{\lambda}{h^{\alpha \xi}} h_{\beta \xi} - \frac{1}{8} \partial_{\sigma}{h_{\gamma \epsilon}} \partial_{\alpha}{h^{\nu \beta}} \partial_{\lambda}{h^{\alpha \xi}} h_{\beta \xi} \nn \\ &&  - \frac{1}{4} \partial_{\sigma}{h_{\alpha \epsilon}} \partial_{\gamma \lambda}{h^{\nu \alpha}} + \frac{1}{8} \partial_{\sigma}{h_{\alpha \epsilon}} \partial_{\gamma}{h^{\nu \alpha}} \partial_{\lambda}{h^{\beta \xi}} h_{\beta \xi} - \frac{1}{12} \partial_{\sigma}{\phi} \partial_{\epsilon}{h_{\alpha \gamma}} \partial_{\lambda}{h^{\nu \alpha}} - \frac{1}{48} \partial_{\epsilon}{\tau^{\alpha \beta}} \partial_{\lambda}{h_{\xi \gamma}} \partial_{\sigma}{h^{\nu \xi}} \tau_{\alpha \beta}) \nn \\ && + h^{\nu \sigma} h^{\rho \gamma} h^{\epsilon \lambda} (\frac{1}{24} \partial_{\epsilon}{\tau^{\alpha \beta}} \partial_{\lambda}{h_{\xi \sigma}} \partial_{\gamma}{h^{\mu \xi}} \tau_{\alpha \beta} - \frac{1}{8} \partial_{\epsilon}{h^{\mu \alpha}} \partial_{\sigma \lambda}{h_{\alpha \gamma}} + \frac{1}{12} \partial_{\epsilon}{h_{\alpha \sigma}} \partial_{\gamma \lambda}{h^{\mu \alpha}} + \frac{1}{6} \partial_{\epsilon}{h^{\mu \alpha}} \partial_{\alpha \sigma}{h_{\gamma \lambda}} \nn \\ &&  + \frac{1}{24} \partial_{\epsilon}{h_{\alpha \sigma}} \partial_{\gamma}{h^{\mu \alpha}} \partial_{\lambda}{h^{\beta \xi}} h_{\beta \xi} + \frac{1}{8} \partial_{\epsilon}{h_{\alpha \lambda}} \partial_{\sigma}{h^{\mu \alpha}} \partial_{\gamma}{h^{\beta \xi}} h_{\beta \xi} + \frac{1}{6} \partial_{\epsilon}{\phi} \partial_{\lambda}{h_{\alpha \sigma}} \partial_{\gamma}{h^{\mu \alpha}} - \frac{1}{2} \partial_{\sigma}{\phi} \partial_{\epsilon}{h_{\alpha \lambda}} \partial_{\gamma}{h^{\mu \alpha}} \nn \\ &&  + \frac{1}{8} \partial_{\sigma}{h_{\alpha \epsilon}} \partial_{\gamma \lambda}{h^{\mu \alpha}} + \frac{1}{4} \partial_{\sigma}{h^{\mu \alpha}} \partial_{\alpha \epsilon}{h_{\gamma \lambda}} - \frac{1}{8} \partial_{\sigma}{\tau^{\alpha \beta}} \partial_{\epsilon}{h_{\xi \lambda}} \partial_{\gamma}{h^{\mu \xi}} \tau_{\alpha \beta}) + h^{\rho \sigma} h^{\gamma \epsilon} h^{\lambda \alpha} ( - \frac{1}{24} \partial_{\gamma}{\tau^{\beta \xi}} \partial_{\epsilon}{h_{\sigma \lambda}} \partial_{\alpha}{h^{\mu \nu}} \tau_{\beta \xi} \nn \\ &&  - \frac{1}{6} \partial_{\gamma}{\phi} \partial_{\epsilon}{h_{\sigma \lambda}} \partial_{\alpha}{h^{\mu \nu}} - \frac{1}{24} \partial_{\gamma}{h_{\sigma \lambda}} \partial_{\alpha}{h^{\mu \nu}} \partial_{\epsilon}{h^{\beta \xi}} h_{\beta \xi} - \frac{1}{4} \partial_{\gamma}{h_{\sigma \epsilon}} \partial_{\lambda \alpha}{h^{\mu \nu}} - \frac{1}{4} \partial_{\gamma}{h^{\mu \nu}} \partial_{\epsilon \lambda}{h_{\sigma \alpha}} \nn \\ &&  - \frac{1}{8} \partial_{\gamma}{h_{\beta \lambda}} \partial_{\alpha}{h^{\mu \xi}} \partial_{\epsilon}{h^{\nu \beta}} h_{\xi \sigma}) + h^{\mu \sigma} h^{\nu \gamma} h^{\rho \epsilon} h^{\lambda \alpha} (\frac{1}{8} \partial_{\lambda}{h_{\beta \sigma}} \partial_{\gamma}{h_{\xi \epsilon}} \partial_{\alpha}{h^{\beta \xi}} - \frac{1}{8} \partial_{\gamma}{h_{\beta \sigma}} \partial_{\lambda}{h_{\xi \epsilon}} \partial_{\alpha}{h^{\beta \xi}} - \frac{1}{2} \partial_{\gamma}{\phi} \partial_{\sigma \lambda}{h_{\epsilon \alpha}} \nn \\ &&  - \frac{1}{8} \partial_{\gamma}{h^{\beta \xi}} \partial_{\sigma \lambda}{h_{\epsilon \alpha}} h_{\beta \xi} - \frac{1}{8} \partial_{\gamma}{\tau^{\beta \xi}} \partial_{\sigma \lambda}{h_{\epsilon \alpha}} \tau_{\beta \xi} + \frac{1}{16} \partial_{\sigma}{h_{\beta \gamma}} \partial_{\lambda}{h_{\xi \epsilon}} \partial_{\alpha}{h^{\beta \xi}} + \frac{1}{4} \partial_{\sigma \gamma \lambda}{h_{\epsilon \alpha}}) \nn \\ &&  + h^{\mu \sigma} h^{\rho \gamma} h^{\epsilon \lambda} h^{\alpha \beta} ( - \frac{1}{8} \partial_{\epsilon}{h_{\xi \alpha}} \partial_{\beta}{h_{\gamma \lambda}} \partial_{\sigma}{h^{\nu \xi}} + \frac{1}{16} \partial_{\epsilon}{h_{\xi \alpha}} \partial_{\lambda}{h_{\gamma \beta}} \partial_{\sigma}{h^{\nu \xi}}) + \frac{1}{12} \partial_{\sigma}{h^{\mu \nu}} \partial_{\gamma}{h^{\rho \sigma}} \partial_{\epsilon}{h^{\gamma \epsilon}} \nn \\ &&  - \frac{1}{6} \partial_{\sigma}{h^{\mu \gamma}} \partial_{\gamma}{h^{\nu \epsilon}} \partial_{\epsilon}{h^{\rho \sigma}} - \frac{1}{8} \partial_{\sigma}{h^{\mu \nu}} \partial_{\gamma}{h^{\rho \epsilon}} \partial_{\epsilon}{h^{\sigma \gamma}} - \frac{1}{4} \partial_{\sigma}{h^{\mu \gamma}} \partial_{\gamma}{h^{\nu \sigma}} \partial_{\epsilon}{h^{\rho \epsilon}} \, ,
\eea
where the compact notation $\partial_{\mu \nu}=\partial_{\mu} \partial_{\nu}$ have been used.

\end{document}